\documentclass[useAMS,usenatbib]{mn2e}
\usepackage{graphicx}
\usepackage{times}
\newcounter{subfigure}
\usepackage{colordvi}
\usepackage{epsfig}
\usepackage{appendix}
\title[Dust and abundances in QSO absorption line systems]{ Average
Extinction Curves and Relative Abundances for QSO Absorption Line
Systems at 1$\le z_{abs}< $2} 
\author[D. G. York, P. Khare et al.] {Donald G.
York$^{1,2}$\thanks{E-mail: don@oddjob.uchicago.edu (DGY)}, Pushpa
Khare$^3$, Daniel Vanden Berk$^4$, Varsha P. Kulkarni$^5$,\newauthor
Arlin P. S. Crotts$^6$, James T. Lauroesch$^7$, Gordon T. Richards$^8$,
Donald P. Schneider$^4$,\newauthor Daniel E. Welty$^1$, Yusra
Alsayyad$^1$, Abhishek Kumar$^9$, Britt Lundgren$^{10}$ ,\newauthor
Natela Shanidze$^1$, Tristan Smith$^{11}$, Johnny Vanlandingham$^1$,
Britt Baugher$^{12}$,\newauthor Patrick B. Hall$^{13}$, Edward B.
Jenkins$^{14}$, Brice Menard$^{15}$, Sandhya Rao$^{16}$,\newauthor Jason
Tumlinson$^1$, David Turnshek$^{16}$, Ching-Wa Yip$^{8}$, and  Jon
Brinkmann$^{17}$ \\
$^1$Department of Astronomy and Astrophysics, University of Chicago,
Chicago, IL 60637, USA \\ $^2$Enrico Fermi Institute, University of
Chicago, Chicago, IL 60637, USA \\ $^3$Department of Physics, Utkal
University, Bhubaneswar, 751004, India \\ $^4$Department of Astronomy
and Astrophysics, The Pennsylvania State University, 525 Davey
Laboratory, University Park, PA 16802, USA \\ $^5$Department of Physics
and Astronomy, University of South Carolina, Columbia, SC 29207, USA \\
$^6$Department of Astronomy, Columbia University, New York, NY 10027,
USA \\ $^7$Department of Physics and Astronomy, Northwestern
University, Evanston, IL 60208, USA \\ $^8$Department of Physics and
Astronomy, The Johns Hopkins University, 3400 North Charles Street,
Baltimore, MD 21218, USA\\ $^9$New York University, 4 Washington Place,
New York, NY 10003, USA \\ $^{10}$Department of Astronomy, University
of Illinois at Urbana-Champaign, MC-221 1002 W. Green Street, Urbana,
IL 61801, USA \\ $^{11}$ Physics Department, California Institute of
Technology, Pasadena, California, CA 91125, USA\\ $^{12}$University of
California, Santa Barbara, CA 93101, USA \\  $^{13}$Department of
Physics and Astronomy, York University, 4700 Keele St., Toronto, ON,
M3J 1P3, Canada \\ $^{14}$Princeton University Observatory, Princeton
University, Peyton Hall, Princeton, NJ 08544, USA\\ $^{15}$Institute
for Advanced Study, Princeton, NJ 08544-1001, USA \\ $^{16}$Department
of Physics and Astronomy, University of Pittsburgh, Pittsburgh, PA
15260, USA \\ $^{17}$Apache point Observatory, 2001 Apache Point Road,
Sunspot, New Mexico 88349-0059, USA\\}
\begin{document}
\date{}
\maketitle
\label{firstpage}
\begin{abstract}
We have studied a sample of 809 Mg II absorption systems with 1.0$\le
z_{abs}\le$1.86 in the spectra of Sloan Digital Sky Survey QSOs, with the
aim of understanding the nature and abundance of the dust and the
chemical abundances in the intervening absorbers. Normalized, composite
spectra were derived, for abundance measurements, for the full sample and
several sub-samples, chosen on the basis of the line strengths and other
absorber and QSO properties. Average extinction curves were obtained for
the sub-samples by comparing their geometric mean spectra with those of
matching samples of QSOs without absorbers in their spectra. There is
clear evidence for the presence of dust in the intervening absorbers. The
2175 {\AA} feature is not present in the extinction curves, for any of
the sub-samples. The extinction curves are similar to the SMC extinction
curve with a rising UV extinction below 2200 {\AA}. The absorber rest
frame colour excess, $E(B-V)$, derived from the extinction curves,
depends on the absorber properties and ranges from $<$0.001 to 0.085 for
various sub-samples. The column densities  of Mg II, Al II, Si II, Ca II,
Ti II, Cr II, Mn II, Fe II, Co II, Ni II, and Zn II do not show such a
correspondingly large variation. The overall depletions in the high
$E(B-V)$ samples are consistent with those found for individual damped
Lyman alpha systems, the depletion pattern being similar to halo clouds
in the Galaxy. Assuming an SMC gas-to-dust ratio we find a trend of
increasing abundance with decreasing extinction; systems with N$_{\rm
H\;I} \sim 10^{20}$ cm$^{-2}$ show solar abundance of Zn. The large
velocity spread of strong Mg II systems seems to be mimicked by weak
lines of other elements. The ionization of the absorbers, in general
appears to be low: the ratio of column densities of Al III to Al II
is always less than 1/2. QSOs with absorbers are, in general, at least
three times as likely to have highly reddened spectra as compared to QSOs
without any absorption systems in their spectra.
\end{abstract}
\section{Introduction}
\subsection{The nature of QSO absorbers} Absorption lines in spectra of
quasi-stellar objects have been known since shortly after the discovery
of QSOs (Sandage 1965; Schmidt 1966; Arp, Bolton \& Kinman 1967;
Burbidge, Lynds \& Burbidge 1966; Stockton \& Lynds 1966). While many
of the first detections were close to the redshifts of the QSOs, what are
now termed associated systems, Bahcall, Peterson \& Schmidt (1966) found
one of the earliest systems which could be termed intervening and Bahcall
(1968) established the reality of such systems. Bahcall and Spitzer
(1969) presented plausibility arguments that foreground galaxies with
extended halos could produce some of the intervening systems. If such
were the origin of the QSO absorption line systems (QSOALSs), they would
provide an excellent way to understand various aspects of structure
formation and evolution in the universe, in particular, the buildup of
the elements over cosmic time and evolution of element flows into and out
of galaxies as they form (Hermann et al. 2001; Adelberger et al. 2003,
2005; Tumlinson \& Fang 2005).

There are, indeed, specific examples of QSO absorption lines coming from
specific galaxies (Cohen et al. 1987; Yanny, York \& Gallagher 1989;
Bergeron 1988; Chen, Kennicut \& Rauch 2005). On the other hand, there
are numerous cases in which a QSO sightline passes near a visible galaxy
without producing absorption lines (Bechtold \& Ellingson 1992). Some of
the galaxies are almost certainly dwarf galaxies (York et al. 1986; Rao
et al. 2003; Haehnelt, Stenimetz \& Rauch 1998). Others may have an
outer morphology that resembles the Milky Way (MW) distribution of 21 cm
clouds (York 1982; Wakker \& van Woerden 1997). Such a patchy
configuration would explain why absorbers are sometimes found near
galaxies and why some QSO/near galaxy coincidences do not produce
absorbers. In some cases extended, turbulent regions around star-forming
galaxies (Adelberger et al. 2005) may show up as QSOALSs.

Despite numerous observations of absorption line systems in spectra of QSOs,
there may be some selection effects in the samples. Why, for instance are
QSOALSs with fully or near solar abundances seldom seen (Prochaska et al.
2003a; Khare et al. 2004; Kulkarni et al. 2005a)? Only a few systems with
metallicity above solar are definitely known: SDSSJ1323-0021, $z_{abs}$=0.716
(Khare et al. 2004), Q0058+019, $z_{abs}$=0.612 (Pettini et al. 2000).  There
may be some selection effects that keep such absorbers out of the observed
sample.  For example, dwarf galaxies may constitute a large part of the sample;
these usually have sub-solar abundances. Also, absorbers with solar abundances
may have more dust, making any background objects fainter and keeping those
QSOs out of magnitude-limited samples (e.g. Fall \& Pei 1993; Boisse et al.
1998;  Vladilo \& Peroux 2005). Ostriker \& Heisler (1984) argued that
foreground absorbers were probably absorbing light from background QSOs, making
them redder. The effect is possibly strong enough to eliminate the QSOs
completely from magnitude-limited samples, in some cases, since the attenuation
in the rest frame UV is much stronger than that in the rest frame optical
range. Akerman et al. (2005) find no evidence for such an effect, but
testing on a larger sample could yield different results. 

The presence of and the nature of dust in QSOALSs is therefore of interest, as
it may shed light on the evolution of dust particles in cosmic time, may
allow the evaluation of selection effects and may allow a discrimination
between different kinds of dust in the intergalactic medium and in the
circum-QSO environment. 
\subsection{Inference of dust from gas abundances} Dust has long been
suspected to be a constituent of the QSO absorbers given the general
connections of galaxies with such systems, the presence of dust in known
galaxies and the evidence of depletion of gas phase elements onto dust
grains in the QSOALSs (Meyer, Welty \& York 1989; Khare et al. 2004 and
references therein), similar to that seen in the MW (Hobbs et al. 1993;
Sembach \& Savage 1996; Jenkins 2004), Small Magellanic Cloud (SMC, Welty et
al. 2001) and Large Magellanic Cloud (LMC, Welty et al. 1999a). The signature
of dust depletion is often characterized by the ratio of column densities of
dominant ionization stages of a refractory (depleted) species to that of a
volatile (weakly depleted) species. Examples include N$_{\rm Cr\;II}$/N$_{\rm
Zn\;II}$, N$_{\rm Fe\;II}$/N$_{\rm Zn\;II}$ (Khare et al. 2004 and references
therein), or N$_{\rm Fe\;II}$/N$_{\rm Si\;II}$ (Prochaska et al. 2003b), N
being the column density, in units of number of ions per square centimeter.
More quantitatively, N$_{\rm Cr\;II}$/N$_{\rm Zn\;II}$ and N$_{\rm
Fe\;II}$/N$_{\rm Zn\;II}$ are smaller than their cosmic abundance ratios by
factors of 2-5 in QSOALSs, whereas, in the Galaxy, the values range from 2-100
depending on the kind of gas. Akerman et al. (2005) do not find
statistically significant difference in Zn metallicity as well as depletions
between their sample of 15 DLAs in the complete optical and radio absorption
line systems and the values in optically selected sample published in the
literature. 
\subsection{Extinction in QSO absorption line systems} Attempts to
quantitatively detect dust by its differential extinction of a given object
have been difficult. Pei, Fall \& Bechtold (1991) found higher extinction in
a sample of 13 QSOs pre-selected to have damped Ly\, $\alpha$ systems
(DLAs), compared to a sample of 15 QSOs that had no DLAs and concluded that
extinction existed in the first sample, but not the second: a dust-to-gas
ratio of 10\% of the local interstellar medium was found with the
absorber rest frame colour excess, hereafter, $E(B-V), < $0.03. Ellison,
Hall \& Lira (2005) similarly find a slight reddening ($E(B-V) <$
0.04) in 14 out of 42 QSOs with DLAs. The Sloan Digital Sky Survey (SDSS)
archive has been used for several studies of extinction in sightlines to
QSOs. Richards et al. (2003) showed that an observer frame colour excess,
hereafter, $\Delta(g-i)$, which is the difference between the actual colours
of a QSO and the median colours of QSOs at that redshift, could be defined
for the SDSS QSO spectra and could be used to form templates of objects with
apparent degrees of extinction. However, they could not discern if the
extinction so detected was from the QSO itself or from intervening systems.
Hopkins et al.  (2004) used composite QSO spectra to show that the
extinction towards QSOs is dominated by SMC-like extinction, which they
argued was predominantly located at QSO redshifts.  Recently, Wild \& Hewett
(2005) have found evidence of dust in QSOALSs with detected Ca II lines, 
with $E(B-V)$=0.06, while Wild, Hewitt \& Pettini (2005) find strong Ca II
absorbers to have $E(B-V)>0.1$. Murphy \& Liske (2004) studied the spectral
indices of QSOs in SDSS Data Release 2 and found no evidence for the
presence of dust in DLAs at a redshift $\sim$ 3. They derived an upper limit
of 0.02 on $E(B-V)$.
\subsection{Searches for the 2175 {\AA} bump} The 2175 {\AA} bump is a
feature of the extinction curve of gas in the Local Group. It was
discovered in the 1960s using sounding rockets (Stecher 1965, 1969). The
origin of the feature is unknown, but it is thought to be made by
carbonaceous material in dust or large molecules (Draine 2003). It is
well correlated with the total density of hydrogen (atomic and molecular)
and is, by inference, an unsaturated signature of material in dust grains
or molecules. Since it is known to vary in strength with the steepness of
the general extinction (Valencic, Clayton \& Gordon 2005), knowledge of
its behavior is an indicator of different sizes and types of grains. Its
absence indicates, in models, a dearth of certain types or sizes of
grains in a given locale. Understanding its behavior in other galaxies at
different times or under different conditions reveals what is generic and
what is particular about the dust in space.

Malhotra (1997) reported detection of the 2175 {\AA} bump, in the
extinction curves of QSOALSs, by averaging 92 QSO spectra in the absorber
rest frames (See section 9.2 of this paper for a discussion of this
result).  Ostriker, Vogeley \& York (1990) showed a possible detection
of the 2175 {\AA} bump in a single object.  Recently, the presence of the
bump has been recognized in a few QSOALSs (Motta et al. 2002; Wang et al.
2004; Junkkarinen et al. 2004).  Yet another ISM constituent, the diffuse
interstellar bands (Herbig 1995) are seen in one system (AO 0235+164,
Junkkarinen et al. 2004). 
\subsection{Overview of the paper} In this paper, we proceed to use all
three techniques noted above to understand dust in QSOALSs: looking for
the presence of reddening, searching for the 2175 {\AA} bump,  and
studying the relative abundances of elements in a large and homogeneous
sample of QSOs provided by the SDSS (York et al. 2000). It is of
particular interest to derive the relationship between these three
aspects of dust discernment, to test the universality of the
characteristics of dust throughout the Universe. It is also of interest
to distinguish dust associated with QSOALSs from the dust near QSOs.
Since the presence of dust may affect the selection of QSOs for such a
study, it is important to use samples with well-defined selection
criteria that can be modeled for the purpose of generalizing conclusions.
Finally, by surveying well-defined samples of large numbers of absorbers,
albeit at 170 km s$^{-1}$ resolution, we can hope to separate true trends
from selection effects that may arise in small samples by accident.

The contents of the paper, by section, are briefly outlined here. Section
2 includes a description of the SDSS survey and spectra, as they relate to the
present paper. The selection of 809 absorption systems for the study of the
2175 {\AA} bump is outlined in section 3. The same sample is used for our
derivation of extinction and of line strengths. Averaging procedures used to
obtain composite spectra, along with the line identifications, using the full
sample as an example, are given in section 4. Two different procedures for
finding average extinction, one using QSO spectra and the other using the SDSS
colours are given in section 5. Definition of selection of sub-samples, based
on entities that can be measured in individual QSOs, along with the physical
motivation of those criteria appear in section 6. Table 1, found in section 6,
summarizes the sub-samples and their properties, including the color excesses
determined as described in section 4. Section 7 contains the central
observational results of this study: the effects of extinction are correlated
with the absorption line strengths and the derived column densities; the
strength of the Mg II absorption lines is defined with regard to extinction and
shown not to be a dominant factor; and control samples are used to show that
there are only small effects (on colour excess) in redshift, i-magnitude and
beta. The extinction is shown to be most like the SMC extinction, with no bump
and no inflection below 2000 {\AA}. The abundances and depletions for our
sub-samples are given in section 8, where we identify our highest extinction
samples with traditional DLAs (sample of $\sim$ 100 absorbers) and suggest,
assuming an SMC extinction curve, that the largest sub-sample (698 absorbers)
has, on average, a sub-DLA (systems with N$_{\rm H\;I}$ between 10$^{19}$ to
2$\times 10^{20}$ cm$^{-2}$) value of N$_{\rm H\;I}\sim 10^{20}$ cm$^{-2}$.
With these assumptions, the latter sample shows solar metallicity in the
absorbers.  Uncertainties in column densities, abundances and extinction values
are explored in section 9. Our conclusions are summarized in section 10.
\section{A large catalog of QSO absorbers}
\subsection{The SDSS spectra} We have undertaken a large survey of QSOALSs,
using the QSO spectra from the SDSS. The SDSS project uses a 2.5 meter,
dedicated telescope at Apache Point Observatory (Gunn et al.  2005). The SDSS
survey targets QSOs based on colour selection of objects to an $i$ magnitude of
19.1 for $z_{em}<$3.0 and 20.2 for higher redshifts, based on 5-filter scans of
the northern Galactic cap of the MW (Gunn et al. 1998). This choice of sky
coverage minimizes the effects of Galactic reddening. Based on precise absolute
flux calibration (Hogg et al. 2001; Smith et al. 2002; Ivezic et al. 2004) of
the object images in the SDSS filter system (Fukugita et al. 1996), QSOs can be
reliably separated from most types of stars, and reasonably accurate
photometric redshifts can be estimated for a large fraction of the quasar
candidates, using only colour data (Richards et al. 2004; Weinstein et al.
2004). The spectra are obtained using fibers plugged into pre-drilled plates,
using coordinates known to $<$ 0.1 arcsec (Pier et al 2003). Empirical
testing of the technique shows that the end-to-end completeness of the QSO
spectral survey itself is approximately 90{\%}, for reasons discussed by Vanden
Berk et al. (2005). The completeness is achieved by adaptive tiling of
selected candidates (Blanton et al. 2003).

Selection of objects as candidates for taking of SDSS spectra occurs in
two ways: (a) via a set of rules relating to colour space for the purpose of
reducing the number of false positives (e.g. white dwarfs, A stars) (Richards
et al.  2002a) and (b) via the inclusion of all known X-ray and radio sources
of SDSS objects (Anderson et al. 2003; Stoughton et al. 2002). Objects selected
by the first method are brighter than $i$ magnitude 19.1. Objects selected in the
second way are generally fainter than that magnitude. 

QSOs found using these techniques are entered into QSO catalogs. The
catalogs include spectroscopically confirmed QSOs from among the candidates
described above as well as any other QSOs found spectroscopically from among
any other targeted objects (serendipitous objects, picked as being candidates
for some other class of point source). The redshifts in the catalogue account
for the fact that the actual centers of the C IV and Mg II emission lines, for
example, show systematic shifts with respect to [O III] lines, which presumably
are at the systemic redshift of the QSO (Gaskell 1982; Tytler \& Fan, 1992;
Richards et al. 2002b).

The SDSS spectra thus obtained have wavelength coverage between 3800 {\AA} to
9200 {\AA}. The resulting spectra have a resolving power of about 1800 (170 km
s$^{-1}$) with two-pixel sampling. However, the resolution is a complicated
function of wavelength, so that lines in this paper were observed with a
resolution of 80-190 km s$^{-1}$. The Mg II lines fall in the region of nominal
resolution. 

The $S/N$ of the spectra recorded for SDSS targets is set by the procedure that
15 minute exposures are recorded until the sum of spectra of a single QSO at a
$g$ magnitude 20.2 gives $(S/N)^2 \sim $7. Spectra admitted to the public
database have $S/N$ $\ge$ 4 at a fiber magnitude of 19.9 in the red SDSS
spectrum and a $g$ magnitude 20.2 in the blue spectra and a fiber magnitude of
19.9 in the red spectrum (Adelman-McCarthy et al. 2005). 

The specific SDSS data products are described by Stoughton et al. (2002) and
Abazajian et al. (2003, 2004, 2005). The data used here are publicly available
from the SDSS website, \underline {http://www.sdss.org}. SDSS spectra of QSOs
with absorption lines have been published in conjunction with broad absorption
line QSOs (Hall et al. 2002a; Menou et al. 2001; Tolea, Krolik \& Tsvetanov
2002; Reichard et al. 2003a) and lensed QSOs (Inada et al. 2003; Hall et al.
2002b). All spectra and the photometric data used in this paper have been
corrected for Galactic reddening, using the extinction maps of Schlegel,
Finkbeiner, \& Davis (1998). The QSOs are published in lists, by Data Release
number (EDR: Schneider et al. 2001; DR1: Schneider et al. 2003; DR3: Schneider
et al. 2005). Further details of the observing and catalog building
process may be found in those references. 
\subsection{The SDSS absorption line catalog} Several of us have developed a
separate pipeline (Vanden Berk et al. 1999; Richards 2001; York et al. 2001;
York et al. 2005) to isolate absorption lines and to identify systems of lines.
The systems are organized into a format modeled on the catalog of York et al.
(1991).  The completeness of the lists of absorption line systems is above
95{\%}, as judged by comparing automated lists with those made by hand. The
incompleteness is associated with the key Mg II lines being blended with night
sky emission ($\lambda\lambda$5596, 5890, 5897 and 6300 and numerous lines
above 7000 {\AA}), blended with some other line of another system or blended
together into a single, very strong line. Visual inspection was done to verify
the present work is unaffected by such occurrences. 

The main catalog presents a list of significant absorption lines, equivalent
widths, various line parameters and possible redshifts. These lines are
automatically assigned to discrete systems by matching lines with candidate
redshifts that agree to within a range that depends on the $S/N$ of the
spectrum. Weak lines are often matched to $\pm$30 km s$^{-1}$ in the best
spectra, whereas strong lines in low $S/N$ spectra may be matched to within
$\pm$300 km s$^{-1}$. The systems are graded, depending on the number and
quality of the individual line detections available. Class A systems are made
of at least four, unblended lines with detection significance above 4 $\sigma$
that match in redshift, are free of observational artifacts and are not in the
Ly\, $\alpha$ forest.  In this paper, we deal only with class A systems. Each
system has a mean redshift, determined from the average of the most significant
lines. The redshifts of individual lines are retained. Our list of class A
systems is complete to the limiting equivalent width allowed by each spectrum,
for QSOs that meet our selection criteria. 

The QSOALS catalog based on DR1 spectra (Abazajian et al. 2003; Schneider et
al. 2003) is used for the current paper, though the data are processed with the
DR3 software (Abazajian et al. 2005).  
\section{Sample selection} From a list of $\sim $7,000 grade A absorption
line systems extracted from 16,713 QSOs in DR1, we used our database to
select systems with (1) W$_{\rm Mg\; II \lambda2796}$, the rest
equivalent width of Mg II $\lambda$2796 $\ge$ 0.3 {\AA}; (2) 1.0 $\le
z_{abs} \le $ 1.86; and (3) an apparent velocity with respect to the QSO
of $>$3000 km s$^{-1}$. The Mg II line strengths were used as a first
guess to define systems that might have large column densities in H I
(Rao and Turnshek 2000). The second criterion, the redshift range of the
absorbers, is to allow only objects for which the 2175 {\AA} feature lies
completely within the SDSS spectrograph wavelength range and to ensure
that the Mg II lines are at $\lambda<$8000 {\AA} to avoid the regions of
the SDSS spectra that are contaminated by strong night sky emission
lines. The third criterion rejects associated QSOALSs (Foltz et al.
1986), defined as narrow absorption lines within 3000 km s$^{-1}$ of the
QSO redshift. A separate study is underway for associated systems. We
also imposed an upper limit of 1.95 on the emission redshift of the QSOs
to keep the Ly\, $\alpha$ forest from appearing in the spectra. 

These cuts left us with about 1200 systems. For multiple systems in the
spectra of a single QSO, which are sufficiently close in redshift so that
their 2175 {\AA} features fall within $\sim$ 100 {\AA} of one another, we
only kept the strongest of the systems: this choice maximizes our chances
of finding the 2175 {\AA} feature in our composite spectra without
smearing the feature. 

We then examined each of the 1200 spectra by eye. QSOs were eliminated if
broad absorption line systems (BALs) are present; this examination is
required to minimize the effects of a possible origin of some narrow
absorption lines in relativistic outflows from the QSO as well as the
possible presence of dust and hence reddening in the BALs (Januzzi et al.
1996; Barlow \& Sargent 1997). BALs were identified using the criteria of
Weymann et al. (1991). In addition, we rejected objects with strong
absorption systems with relative velocity with respect to the QSO $<$
3000 km s$^{-1}$.
  
We were finally left with 813 QSOALSs, out of which we used 809 systems
(for reasons stated  in section 5), which we denote as sample 1. A
complete list of the 809 systems with relevant data appears in the
appendix (Tables A1a and A1b). Listed are the QSO (plate, fiber and MJD);
the coordinates; the QSO redshift; the absorption system redshift; $i$
magnitude; $\beta$; $\Delta(g-i)$ and $E(B-V)$ obtained from it (as
explained in section 5); and the equivalent widths, from the main
absorption line catalog, of Mg II $\lambda\lambda$2796, 2803, Mg I
$\lambda$2852, Al II $\lambda$1670, and Fe II $\lambda$2382. These
particular lines are among those used in constructing the sub-samples, as
described later in the paper. 

We assume the spectra are free from contamination of light from the host
galaxy (Vanden Berk et al. 2005) because we only consider QSOs with
$z_{em}>1.0$. Contamination by the spectrum of the host galaxy is well
correlated with the presence of an extended morphology. We checked that
only three of the objects were resolved: two appear to be blends with
stars and the third, SDSS J024634.09-082536.1, is a known, gravitational
lens.

The combination of a nearly fixed signal to noise for each magnitude
(section 2.1) and the selection of systems with line strengths at or
above 4 $\sigma$ in significance leaves us with a limiting equivalent
width that increases with QSO magnitude. We define below a selection
criterion that W$_{\rm Mg\;II\lambda2796}>$0.3 {\AA}, 4 $\sigma$. This
limit is actually reached only in the brightest QSOs. For the faint half
of our sample, the mean $i$ band magnitude is 19.49; for the bright half,
it is 18.59. Histograms of the Mg II $\lambda$2796 line strength show
that the completeness limit is at an equivalent width of 0.9 {\AA} in the
former case and 0.6 {\AA} in the latter case. We will discuss the effect
of this limit below.
\section{Generation of composite spectra for absorption line studies} 
Absorption spectra were combined into composites by normalizing each
spectrum, shifting to the absorber rest frame, masking various pixels,
weighting, and averaging the spectra.  The spectra (and associated error
arrays) were normalized by reconstructions of the QSO continua, using
the first 30 QSO eigenspectra derived by Yip et al. (2004). The
reconstructed QSO continua were formed as a linear combination of the
eigenspectra in the QSO frame -- a technique that has the advantage of
very closely approximating the QSO spectrum while almost entirely
ignoring the narrow absorption features.  After normalizing a spectrum,
it was shifted to the absorber rest frame, and re-sampled onto a common
pixel to wavelength scale, which is the same logarithmic scale used by
the SDSS, but extended to shorter wavelengths. 
\begin{figure*}
\includegraphics[width=7.00in,height=8.8in]{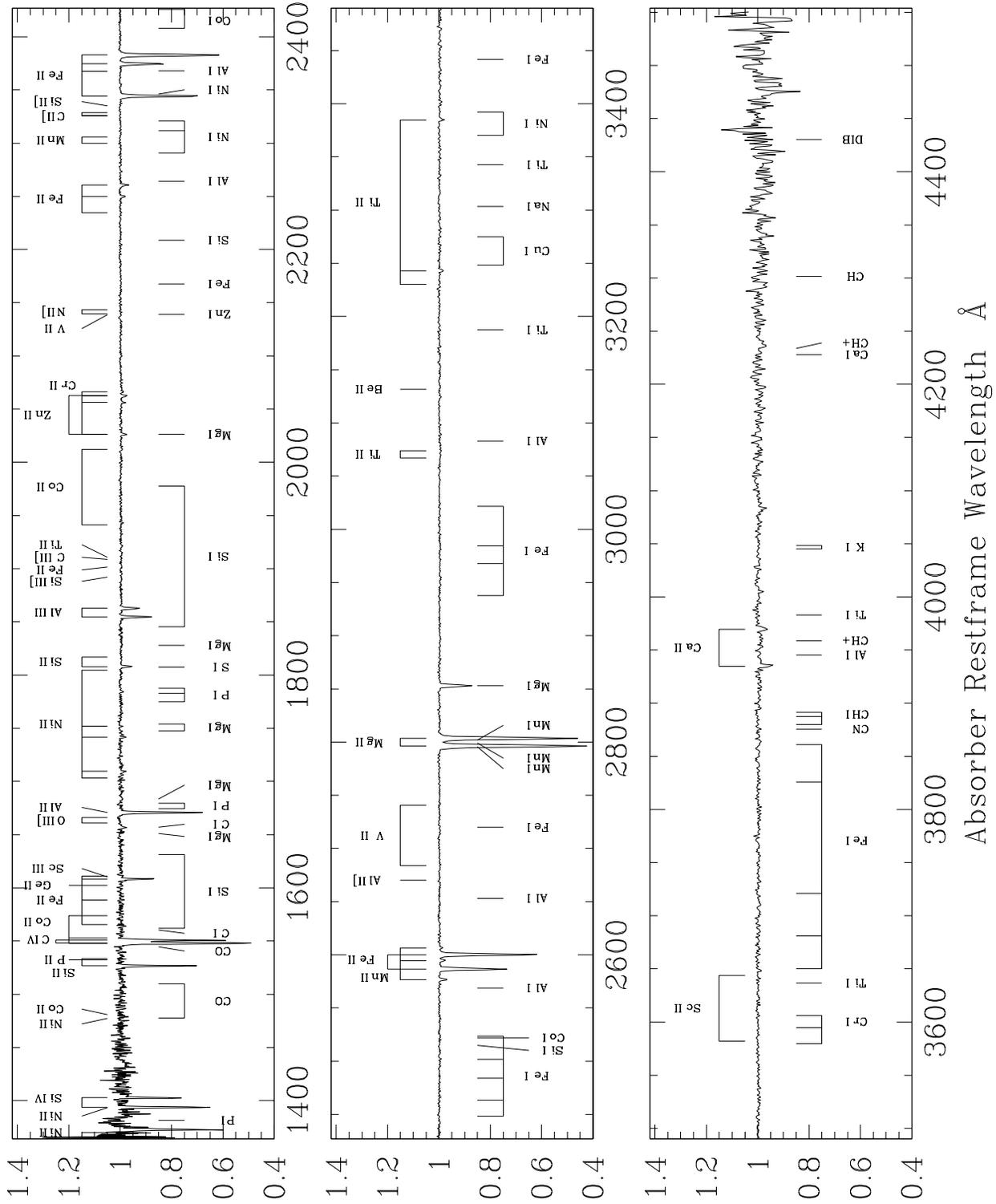}
\caption{The composite spectrum (arithmetic mean) of the systems in the
full sample (sample 1) in the absorber rest frame. Individual spectra
have been normalized using the first 30 eigenspectra of the principle
components of the QSO spectra and weighted by inverse variance. All
spectra contribute to the region between 1900 and 3200 {\AA} which,
therefore, has high signal-to-noise. Locations of lines of atomic and
molecular species are labeled below the spectrum while locations of lines
of ionic species are labeled above the spectrum. See text for the choice
of lines which are labeled. The same figure is shown in the Appendix
(Figure A1) with a smaller range in flux to highlight the weaker lines.}
\end{figure*}
The absorber redshifts used were those based on the individual line (e.g.
Mg\,{\sc ii}\,$\lambda2796$) that was primarily used to select the
sub-sample. As noted in section 2.2, those may differ from the mean
redshift by as much as $\Delta z=0.001$ ($300 {\rm \;km\;s^{-1}}$), but
the differences are often as low as $30 {\rm \,km\,s^{-1}}$. 

In several  cases, individual pixels were masked before summing the
spectra. Pixels flagged by the spectroscopic pipeline as possibly bad in
some way (Stoughton et al. 2002) were masked and not used in constructing
the composites. Also masked were pixels within 5 {\AA} of the expected
line positions of detected absorption systems unrelated to the target
system. The positions of the 18 typically strongest transitions were
masked, for each system, regardless of whether they were actually
detected in the spectrum or not. This helps prevent strong, unrelated
features from ``printing through'' to the final composite, and weaker
features from artificially enhancing the signal from lines of interest.
Such effects would otherwise set the noise limits of the composites,
especially for the smaller sub-samples. 

After masking pixels, the normalized flux density in each remaining pixel
was weighted by the inverse of the associated variance, and the weighted
arithmetic mean of all contributing spectra was calculated for each
pixel.  Error estimates were calculated by propagating the weighted
errors at each pixel, and by measuring the rms scatter about the mean
spectrum; both methods resulted in similar error estimates.

Given our choice of redshift range for searching for the 2175 {\AA}
feature and including Mg II doublet near 2800 {\AA}, the region of
maximum sensitivity for absorption line detection in the co-added spectra
is 1900 {\AA} to 3200 {\AA} in the rest frame. Beyond that range, to
shorter and longer wavelengths, there are fewer spectra being averaged,
the noise is larger and the lines (e.g. C IV $\lambda\lambda$1548, 1550
and Ca II $\lambda\lambda$3934, 3969) no longer correspond to the average
of exactly the same systems as found in the main spectral range.

Figure 1 shows the region of the composite spectra of 809 QSOALSs (sample 1)
from 1500 {\AA} to 4400 {\AA}, in the absorber rest frame. An expanded version
of this figure is shown in Figure A1 in the appendix A in order to highlight the
weak lines. Lines of molecules and neutral atoms are labeled below the spectra.
Ions are labeled above the spectra. We used a complete list of lines in this
region from Morton (2003) as a reference list. To this we added CO lines from
Morton and Noreau (1994). The modified list includes ample weak lines to
confirm detection of each species and to find lines of each species that are on
the linear portion of the curve of growth (lines with strengths $\propto
f\lambda^2$, $f$ being the oscillator strength). Only Mg II and Al II, in this
spectral region, do not have such lines, for our low to medium reddened
samples. 

A complete list of lines thus derived is in the appendix (Table A2). As
the SDSS absorption line catalog grows, weaker and weaker lines will be
discernible in the composite spectra. While many of the lines are not now
detected in QSOALSs, most have been detected in the local interstellar
medium. The list given here provides a minimal set of lines needed to
verify identifications in these improved spectra. All wavelengths quoted
in the paper are vacuum wavelengths: when four-digit wavelengths are used
in discussion, they are truncated values of the wavelengths in Table A2.

In summary, we are able to co-add many SDSS spectra and obtain high
signal-to-noise, composite spectra for our selected objects between 1900 {\AA}
and 3200 {\AA} in the absorber rest frame. At lower $S/N$, the spectra extend
to 1500 {\AA} and to 4500 {\AA}. The spectra show interstellar lines from
typical, intervening QSOALS. The techniques described here are used later to
derive spectra of various sub-samples of our 809 absorbers.
\section{Procedures for determining Extinction} 
\subsection{Using composite spectra} There are numerous ways to define the
unreddened continuum for derivation of an extinction curve, but division by a
QSO spectrum with no absorbers is the one we chose, to avoid selection effects
due to any possible systematic dependence of QSO spectral shape on emission
redshift or brightness. For the full sample, which we will henceforth refer to
as the absorber sample, we constructed a matching sample of QSOs, also referred
to as the non-absorber sample, having no QSOALSs (of any grade) in their
spectra. Thus for every QSO in the absorber sample there is a QSO in the
non-absorber sample which has nearly the same $i$-band magnitude and nearly the
same emission redshift (hence, the same absolute magnitude). Because the
exposure levels are regulated at the telescope, and because the exposure levels
are checked before a spectrum is admitted to the database (see section 2.1),
the matched, non-absorber QSO is sensitive to the same level of equivalent
widths as the QSO with the absorber: systems too weak to be detected in either
sample could be present in both QSOs of a matched pair, and the effects on
extinction due to these systems would thus cancel in the derivation of the
composite extinction curve. 

It is also possible that the QSO spectra are reddened by circum-QSO dust.
We have not selected the QSOs by any  feature related to possible
extinction other than absorption lines with $z_{abs}<z_{em}$, so
extinction that is circum-QSO should also cancel in the composite
spectra. 

\subsubsection{Selection of the non-absorber sample} The appropriateness of a
QSO with no absorber as a match for a QSO with an absorber was determined by
minimizing $({\Delta z_{em}\over{R_{z_{em}}}})^2+({\Delta i\over{R_i}})^2$, R
being the rms.  Most matches are within $\Delta z_{em}<0.05$ and $\Delta
i<0.1$.  Gravitational lensing is not expected to significantly affect these
matchings. Lensing by metal absorbers (Menard 2005) and cosmic magnification
(Scranton et al 2005) induce, on average, magnitude changes that are smaller
than our matching range. 

For four of the absorption systems, in three QSOs, it was not possible to
find a good match (the combined $\Delta z_{em} - \Delta i$ radius is
greater than 0.5). The three QSOs are bright ($i$ band magnitudes of 17)
and are in $z_{em}-i$ regions with a low density of QSOs. The objects are
SDSS J123602.34-033129.9, LBQS0055+0055 (SDSS J005824.75+004113.3) and HS
0037+1351 (SDSS J004023.76+140807.3), with emission line redshifts of
1.815, 1.92 and 1.87, respectively, and absorber redshifts of 1.2689,
1.0719 and (in the last object) 1.1254 and 1.6209. 

The sample was thus reduced to 809 systems. We used 174 non-absorber QSOs
twice: none were used three times. The list of the matched QSOs along
with their $i$ magnitude and $z_{em}$ is given in Table A1a (plate,
fiber, MJD, emission redshift, $i$ magnitude, and match radius) and Table
A1b (plate, fiber, MJD, and $\Delta(g-i)$). The quality of the match
between the two samples is shown in Figure A2, where the difference in
$i$ magnitudes for each pair and the difference in emission redshifts for
each pair are shown. 

Our intent was to form multiple extinction curves using different,
non-absorbed QSOs in each case. There are not enough such objects in DR1,
but it should be possible to use three or four, independent, matched
pairs with subsequent SDSS data releases for absorber samples of the
present size. It will also be possible to use more sophisticated criteria
for matching the QSOs, for instance, based on emission line
characteristics or the nature of the radio flux.

\subsubsection{Derivation of the extinction curves} Geometric mean spectra
were generated from the absorber and matching non-absorber spectra.  The
geometric mean of a set of power law spectra preserves the average power law
index of the spectra, so it is the appropriate statistic to use to determine
the characteristic extinction law (likely to be approximated by a power law)
of the absorber sample.  The geometric mean composite spectra were generated
in much the same way as the normalized composites described in section 4,
except that the spectra were not normalized, the pixels were not weighted, and
the geometric (rather than arithmetic) mean of the spectra was calculated.
Weighting the pixels can have the effect of changing the final index of the
geometric mean composite, depending on the way the $S/N$ varies with
wavelength; we are more interested here in the correct shape of the overall
continuum than in maximizing the $S/N$.  The geometric mean composite spectra
were generated for the absorber and non-absorber spectra in the same way (both
sets of spectra were shifted to the rest frames of the detected absorbers).
As long as the spectroscopic properties of the QSOs in the absorber and
non-absorber samples are statistically equivalent, except for the effects of
the absorption systems, dividing the absorber composite by the non-absorber
composite will yield the relative extinction curve in the absorber frame.  We
use the same procedure to generate extinction curves for specific sub-samples
described in section 6.

The geometric mean composite spectra of the 809 absorber spectra and the
corresponding non-absorber spectra are shown in Fig.~2.  The composite
spectra are smeared out versions of the mean spectra as they would appear
in the QSO frame. For example, very broad features corresponding to the
strong C\,{\sc iv}, C\,{\sc iii}], and Mg\,{\sc ii} emission lines,
smoothed and distorted by the distribution of QSO vs.\ absorber redshift
differences, are evident in  the top panel of Fig.\,2.  Those features
are remarkably similar in both the absorber and non-absorber spectra,
which validates the assumption that the two samples are statistically
equivalent, at least on scales as large as the emission lines.  The
narrow absorption lines are easily seen in the absorber spectrum, and are
absent in the non-absorber spectrum.  The non-absorber spectrum has been
scaled so that the average flux density in the reddest $200$ {\AA} of
both spectra are equal.  It is clear from the figure that the absorber
composite spectrum for our full sample is relatively fainter at the blue
end.

\subsubsection{Nature of the mean extinction curve of the full sample} The
bottom panel of Fig.~2 shows the ratio of the absorber to the non-absorber
composite spectra.  Overlaid are extinction curves based on the standard MW
curve (Fitzpatrick 1999), and an SMC curve (Pei 1992).  The SMC curve
describes the ratio spectrum quite well with a best fit (excluding 5 {\AA}
around each of the narrow absorption features) $E(B-V)$=0.013.  The
statistical error in the value of $E(B-V)$ is less than $0.001$, and was
determined by measuring the rms dispersion of $E(B-V)$ values of a large
sample of ratio spectra generated by adding simulated noise to the original
ratio spectrum. The errors are likely to be larger than this because of the
fact that the non-absorber sub-samples are not always large enough to be
completely free of any biases. For example, for the sub-samples based on
W$_{\rm Mg\;II\lambda2796}$ values (as described in the next section), we
expect the $\Delta(g-i)$ values to be the same for all the non-absorber
samples but they differ by up to 0.03. From the relation between
$<\Delta(g-i)>$ and $E(B-V)$ obtained in section 9.1, we estimate that the
errors in the derived $E(B-V)$ values due to this in 18 out of 27 samples will
be smaller than 0.002. For 7 other sub-samples the error will be smaller than
0.004 and only in one case (sub-sample 22) it could be as large as 0.009. The
MW curve also fits the ratio spectrum well using the same value of $E(B-V)$,
but only at wavelengths longer than about $2500$ {\AA}. 

There is no evidence of the $2175$ {\AA} absorption bump in the ratio
spectrum, and the extinction increases at wavelengths shortward of $2000$
{\AA}; both facts are contrary to what is expected for the MW extinction
curve.
\begin{figure}
\includegraphics[width=3.15in,height=4.0in]{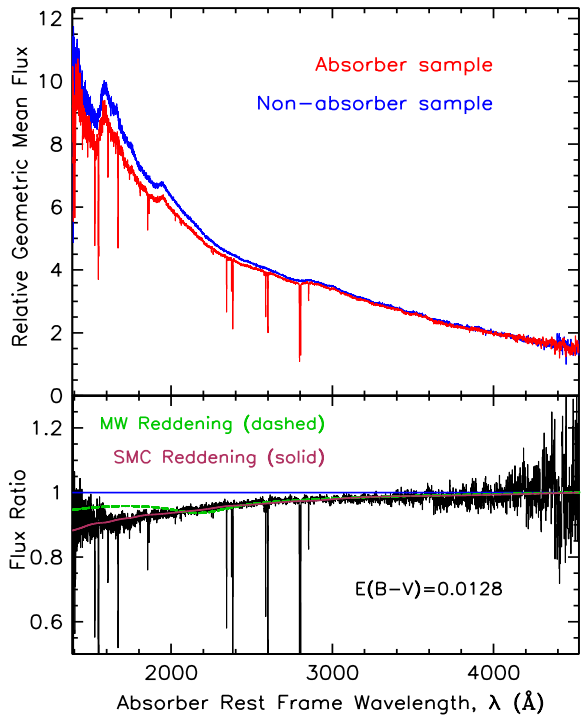}
\caption{The top panel shows the composite spectrum (geometric mean) of
the absorbers, in their rest-frames, in red, and that of the
corresponding non-absorbers, in blue. Note that smeared out emission
lines do appear in the absorber rest-frame composite spectrum due to the
small range in emission and absorption redshifts considered here. The
absorption lines in the red spectra are the obvious, detected absorption
lines seen in Figure 1. The bottom panel shows the ratio of the two
spectra along with the best fit SMC extinction curve in red. The derived
absorber rest frame $E(B-V)$ value is given. The MW extinction curve for
the derived SMC $E(B-V)$ value is shown in green. Extinction curves for
sub-samples with comparable and higher values of $E(B-V)$ are given, in
this same format, in Figure 5.}
\end{figure}
We experimented with the generation of composite spectra after excising a
region up to 8000 km s$^{-1}$ around the major emission lines. We found
no effect on the derived $E(B-V)$ values. Similarly, using a given
matching QSO only once in the non-absorber sample by increasing the
$\Delta z_{em}\;-\;\Delta i$ match radius also has no effect on the
derived $E(B-V)$ values. We note that out of all the  QSOs in our
absorber sample, 65 have two QSOALSs while six have three QSOALSs in
their spectra. It is possible that the reddening is overestimated by the
inclusion of these sightlines with multiple QSOALSs. To test this we
generated composite spectra excluding these QSOs. The resulting spectrum
is almost indistinguishable (except for somewhat higher noise) from the
composite of the whole sample. As will be seen below, the systems with
W$_{\rm Mg\;II\lambda2796}<$1.5 {\AA} do not, on average produce
significant reddening. Out of the 71 sightlines having multiple
absorbers, only 13 have two systems with W$_{\rm Mg\;II\lambda2796}>$ 1.5
{\AA}. Thus our results are not affected by the multiple absorber
sightlines. Some of our QSOs, picked for having grade A systems, also
have grade B (3 detected, 4 $\sigma$ lines) and grade C (two lines) or
grade D (one line) systems. We verified that the grade B systems are
subordinate to the grade A systems in the strength of Mg II (usually,
grade B systems have the two Mg II lines and one Fe II line, only) and
that the grade C systems are mainly C IV-only doublets, with no
detectable Mg II. The grade D systems are isolated, single lines that
could plausibly be one member of the Mg II doublet, but the absence of
the second member makes them insignificant for our study. The effect of
the presence of systems of lower grades on the results is discussed in
section 9.1.
 
\subsection{Using SDSS colours for individual QSOs} A
comparison of the observer frame extinction measure, $\Delta(g-i)$, for the
absorber and non-absorber samples is shown in
Figure 3a as histograms. It is clear that the sample with absorption line
systems as defined herein has more outliers to the right than the non-absorber
sample and that the extinction curves we measure mainly refer to gas in the
intervening absorbers. The nature of the few non-absorber QSOs having high
values of $\Delta(g-i)$ is the topic of a separate paper. These could be due
to statistical errors in the colours or could involve circum-QSO dust or dust
in intervening galaxies in an environment with a high dust-to-gas ratio or a
high dust-to-H I ratio (i.e., hot gas, with dust, Ferland et al. 2003) or
could be due to intervening absorbers at $z_{abs}<$0.5 (which can not be
easily detected). They could also involve peculiar energy distributions
intrinsic to the QSO itself. Figure 3b shows the same histograms for the
absorber and non-absorber samples but additionally shows the distribution of
$\Delta(g-i)$ as a function of $z_{em}$. The data points are colour coded by
Mg II line strength. It is clear that the strongest Mg II lines can occur in
systems of any colour excess, but that the highest relative colour excess
objects generally have the highest Mg II absorption line strengths as well
(the reverse is not true). Out of 111 systems in the absorber sample having
$\Delta(g-i)>$0.2, 105 have W$_{\rm Mg\;II\lambda2796}$ $>$1.0 {\AA}. 
\renewcommand{\thefigure}{\arabic{figure}\alph{subfigure}}
\setcounter{subfigure}{1}
\begin{figure}
\includegraphics[width=3.15in,height=4.0in]{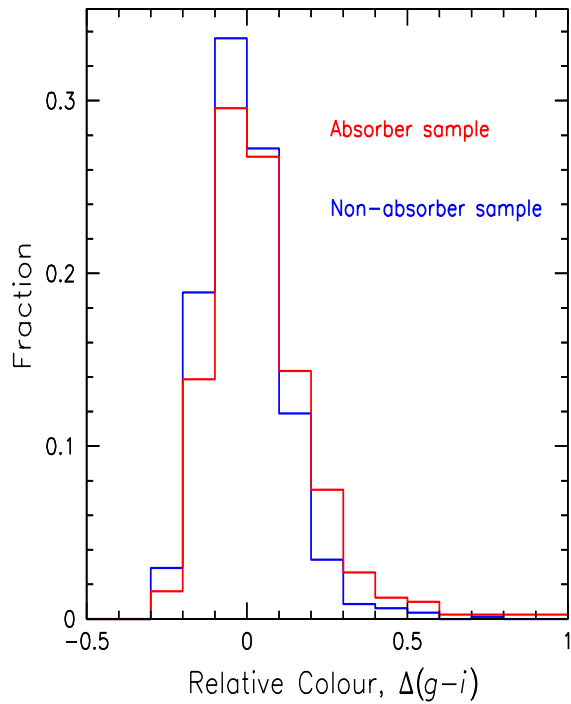}
\caption{Histogram of observer frame colour excess $\Delta(g-i)$ for
the full absorber sample is shown in red and for the non-absorber
sample is shown in blue. The fractional number in each bin is plotted.
Note the excess of QSOs with high values of $\Delta(g-i)$ in the
absorber sample. The average values of $\Delta(g-i)$ for the absorber and 
non-absorber samples and full DR1 are 0.013, -0.013 and 0.003 respectively.
Note that the average for full DR1 is close to zero as expected from
the definition of $\Delta(g-i)$. The K-S probability for the
$\Delta(g-i)$ values for our absorber and the non-absorber samples to
be taken from the same distribution is 1.03$\times10^{-5}$, while that
for the absorber sample and the full DR1 is 0.126, and for the
non-absorber sample and DR1 is 1.15$\times10^{-6}$. The histograms
show the clear, red tail of our sample of QSO absorbers.}
\end{figure}
\addtocounter{figure}{-1}
\addtocounter{subfigure}{1}
\begin{figure}
\includegraphics[width=3.15in,height=4.0in]{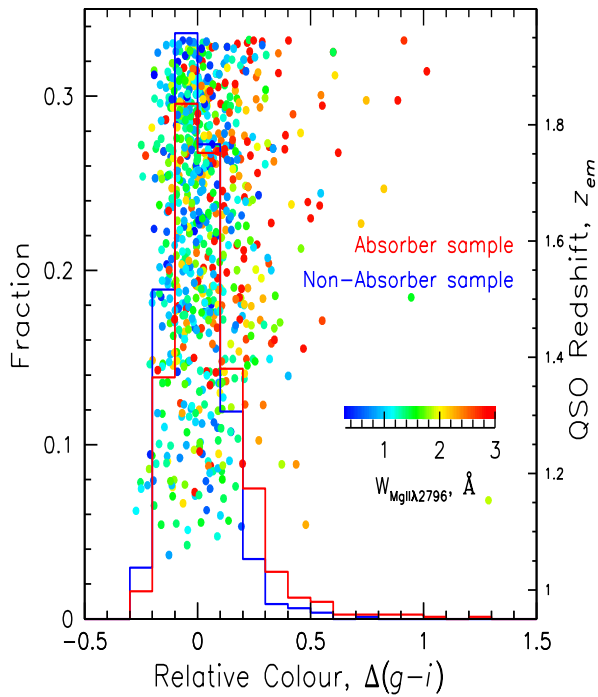}
\caption{Histogram of observer frame colour excess $\Delta(g-i)$ for the full
absorber and non-absorber samples are shown in red and blue respectively, as in
Fig. 3a. Overplotted are the $\Delta(g-i)$ values as a function of emission
redshift (labeled to the right) for individual QSOs whose spectra contain
absorption systems.  The points are colour coded to indicate the strength of
the Mg II $\lambda2796$ line. Note that the highest values of W$_{\rm
Mg\;II\lambda2796}$ occur for all values of $\Delta(g-i)$.}
\end{figure}
\addtocounter{figure}{-1}
\addtocounter{subfigure}{1}
\begin{figure}
\includegraphics[width=3.15in,height=4.0in]{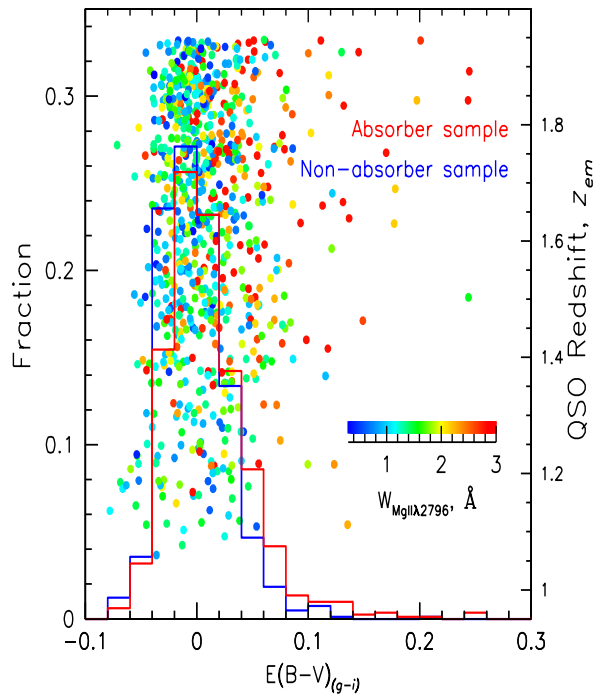}
\caption{Histogram of absorber rest frame $E(B-V)$ values for
individual absorbers as obtained from observed $\Delta(g-i)$ values
using the SMC extinction curve for the full sample is shown in red. For
QSOs in the non-absorber sample the absorption redshifts
for the corresponding absorber QSOs are used to derive the rest-frame
$E(B-V)$ values. The histogram for the non-absorber QSOs is shown in
blue. Note the excess of absorber QSOs with high $E(B-V)_{(g-i)}$
values as compared to the non-absorber QSOs. Overplotted are the
$E(B-V)_{(g-i)}$ as a function of emission redshift (labeled to the
right) for individual QSOs whose spectra contain absorption systems.
The points are colour coded to indicate the strength of the Mg II
$\lambda2796$ line.}
\end{figure}
\addtocounter{figure}{-1}
\addtocounter{subfigure}{1}
\begin{figure}
\includegraphics[width=3.15in,height=4.0in]{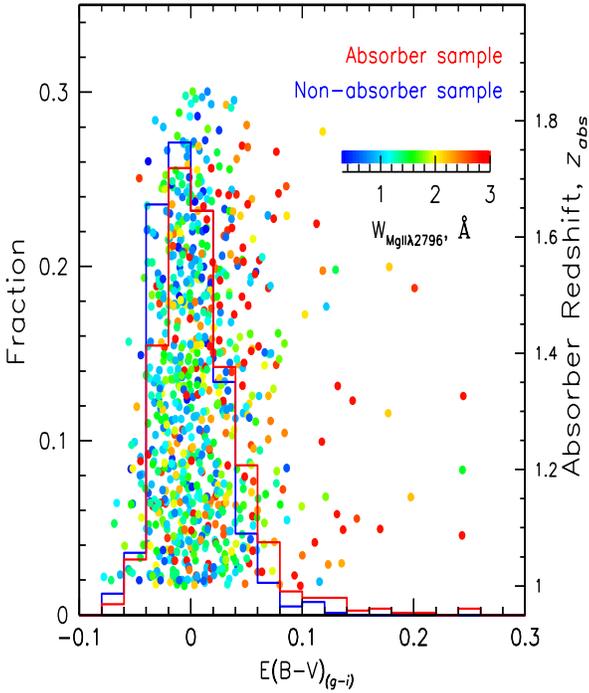}
\caption{Same as Figure 3c except that the $E(B-V)_{(g-i)}$ values for
individual QSOs are plotted as a function of absorption redshift (labeled to
the right). Note that the distribution of the highest extinction QSOs
is much more uniform as compared to that in Figure 3d.}
\end{figure}
\addtocounter{figure}{-1}
\addtocounter{subfigure}{1}
\begin{figure}
\includegraphics[width=3.15in,height=4.0in]{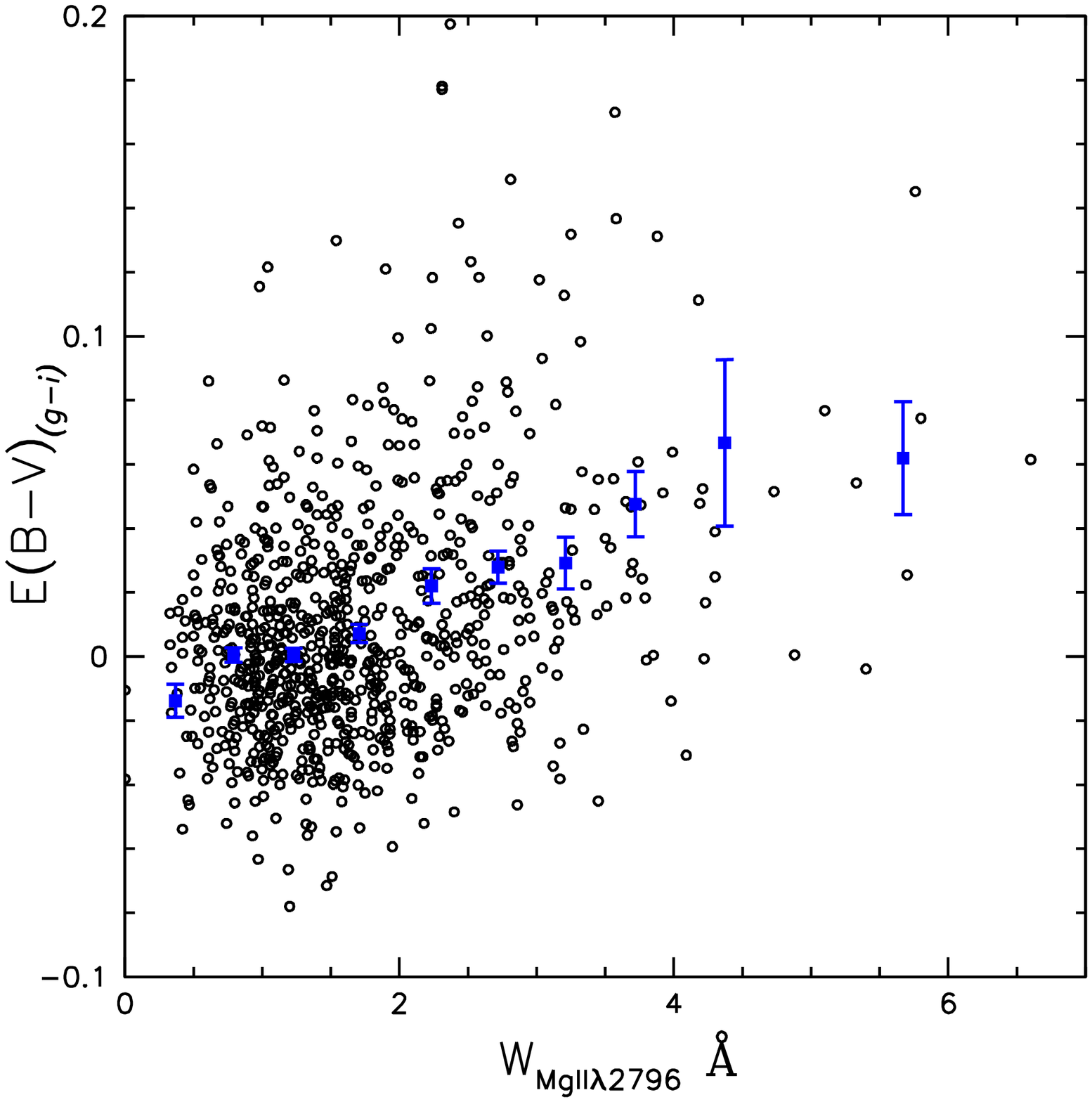}
\caption{Plot showing $E(B-V)_{(g-i)}$, the rest frame colour excess as
obtained from photometric data, for individual absorber QSOs as a
function of the absorber rest frame equivalent width of the Mg II
$\lambda2796$ line. The mean values of $E(B-V)_{(g-i)}$ with error bars
(rms values divided by the square roots of the number of points in each
of the bins) in W$_{\rm Mg\;II}$ bins are plotted which show the trend
of increasing $E(B-V)_{(g-i)}$ with W$_{\rm Mg\;II}$.}
\end{figure}
\renewcommand{\thefigure}{\arabic{figure}}

High $\Delta(g-i)$ values appear selectively at high emission redshifts
(Figure 3b). This is possible as high $z_{abs}$ systems are expected to
be present in the spectra of QSOs with high $z_{em}$ and the more
reddened (UV) part of the spectrum gets shifted to the SDSS filters
giving higher values of $\Delta(g-i)$, as $\Delta(g-i)$ is the colour
excess in the observer frame. We check this as follows. Assuming an SMC
reddening law, A$_\lambda$ = 1.39$\lambda^{-1.2}E(B-V)$ (Prevot et al.
1984) we determine the rest frame colour excess, $E(B-V)_{(g-i)}$ from
the observed $\Delta(g-i)$ values (assuming they are due to the
intervening absorber). Taking $\lambda_g$ and $\lambda_i$ to be 4657.98
and 7461.01 {\AA} respectively, this gives $E(B-V)_{(g-i)}$ = $\Delta
(g-i)$(1+$z_{abs})^{-1.2}$/1.506. In Fig. 3c we have plotted the
histograms for $E(B-V)_{(g-i)}$ for the absorber and the non-absorber
samples. These are similar to the histograms for $\Delta(g-i)$ (Fig. 3a)
and show that the absorber QSOs have higher rest frame $E(B-V)_{(g-i)}$
values as compared to the non-absorber QSOs. Values of $E(B-V)_{(g-i)}$
for individual systems, colour coded by the strength of Mg
II lines are overplotted as a function of emission redshift. As is the
case with $\Delta (g-i)$, there is still a tendency for higher
$E(B-V)_{(g-i)}$ with increasing $z_{em}$. This could then be possibly 
caused by the presence of circum-QSO dust in some QSOs as discussed later
in the paper. 

Figure 3d shows the same histograms with the absorption line redshift
plotted instead of the QSO emission redshift. Now, it is seen that the
distribution of the highest extinction QSOs, in the rest frame, is much more
uniform. In the last column of Table 1 (below in section 6), we give the
average $E(B-V)_{(g-i)}$ for the QSOs in each of our subsamples. Comparing the
color excesses derived using our matched pairs technique with the means from
the SDSS color technique, it is clear that the spectroscopic and photometric
measures of extinction are very similar. We can use the photometric extinction
to define sub-samples of objects with similar, high or low extinction, which
we do in section 6, below. 

In Figure 3e we have plotted the $E(B-V)_{(g-i)}$ values as a function of
W$_{\rm Mg\;II\lambda2796}$. Though the scatter is very large, there is a weak
correlation between $E(B-V)_{(g-i)}$ and W$_{\rm Mg\;II\lambda2796}$. The best
fit straight line is $E(B-V)$=0.016 W$_{\rm Mg\;II\lambda2796}$-0.02 with
rms=0.043. The general rise is consistent with the values of $E(B-V)$ in Table
1 for the sub-samples defined by ever increasing values of W$_{\rm
Mg\;II\lambda2796}$, but the scatter of individual points is so large that
the weak relationship can only be applied to very large samples. 

\subsection{Summary}In this section, we have described our matched pair
technique for the derivation of extinction curves in general and for the full
sample, in particular; described the selection and uncertainties in the
matching process; and compared the resulting values with color excess derived
from SDSS colors for individual objects. The full extinction curve for our
absorption line selected sample of 809 QSOALS has no 2175 {\AA} bump and no
leveling off of extinction shortward of the bump. 

We can now proceed to derive spectroscopic extinctions and absorption
line spectra for sub-samples defined using photometric measures of
extinction in each system, coupled with strengths of absorption lines
in each system.
\section{Sub-sample definitions} Using the equivalent widths of the
lines discussed above, and other parameters, the QSOs were sorted into
various subgroups to search for any systematic effects in the strengths
of lines or extinction detected in the composite spectra. Table 1
includes the the sample number; derived $E(B-V)$ (section 7.1); the
selection criteria for objects in the sample; the sub-section number
describing the sample; the number of systems in the sample and the mean
values of $z_{abs}$, $\beta$, $i$ magnitude and $E(B-V)_{(g-i)}$ for
the sample.

For reference, an SMC extinction curve has $E(B-V)$ = 0.01 for
log(N$_{\rm H\;I}$) = 20.6 (Bouchet et al. 1985). 
\subsection{Sub-samples
in Mg II $\lambda$2796 equivalent width bins} The sample of Mg II lines
was divided into six ranges of W$_{\rm Mg \;II\lambda2796}$, each with a
similar number of absorbers. Recalling that the full sample is called
sample 1, we number these six sub-samples from 2 to 7: sub-sample 2, 0.3
$\le$ W$_{\rm Mg\; II\lambda2796}<$0.93; sub-sample 3, 0.93 $\le$ W$_{\rm
Mg \;II\lambda2796} <$1.2; sub-sample 4, 1.2 $\le$ W$_{\rm Mg\;
II\lambda2796}<$1.53; sub-sample 5, 1.53 $\le$ W$_{\rm Mg\;
II\lambda2796}<$1.91; sub-sample 6, 1.91$\le$ W$_{\rm Mg\;
II\lambda2796}<$2.52; sub-sample 7, 2.52$\le$ W$_{\rm Mg\;
II\lambda2796}<$ 5.00 {\AA}. Sub-sample 8 has all systems with W$_{\rm
Mg\; II\lambda2796}>$2.0 {\AA}. These groups were defined to explore the
relation between Mg II strength and the dust content and thereby the
total gas content if the dust-to-gas ratio is constant.
\subsection{Sub-samples in redshift, magnitude and $\beta$} The full sample was
divided in two equal parts for several parameters: absorbers with
$z_{abs}$ below and above the median value of 1.3127 (sub-samples 9 and 10,
respectively); absorbers in QSOs brighter than or fainter than the median value
in the $i$ band magnitude 19.12 (sub-samples 11 and 12); and absorbers at
$\beta$ below or above the median $\beta$, 0.103 ($\sim $30,000 km s$^{-1}$)
(sub-samples 13 and 14). These sub-samples were defined, respectively, to look
for systematic effects in line strength and reddening with redshift, QSO
brightness, or relative velocity with respect to the QSO.
\subsection{Sub-samples defined to test Routly-Spitzer effect and to
search for dense clouds} Similarly, sub-samples were defined by dividing
the absorbers into two equal sub-sets by line strength: the ratio of
equivalent width of Fe II $\lambda$2382 to that of Mg II $\lambda$2796,
with a median value of 0.577 (sub-samples 15 and 16), and the ratio of
equivalent widths of Al II $\lambda$1670 to Mg I $\lambda$2852, with a
median value of 1.538 (sub-samples 17 and 18). The first set was designed
to see if the Routly-Spitzer effect (Routly \& Spitzer 1952; Silk \&
Siluk 1972) is at play, wherein the Fe in dust grains might be released
in the high velocity clouds and strengthen the wings of Fe II relative to
Mg II , in a manner similar to the strengthening of Ca II with respect to
Na I in the original effect. The second set was made to look for systems
that were highly depleted (missing Al II) and of high density (hence, Mg
II recombined to Mg I).
\begin{table*}
\centering
\begin{minipage}{140mm}
\caption{Samples: definitions and properties}
\small
\begin{tabular}
{|r|r|l|r|r|r|r|r|r|}
\hline
Sample &$E(B-V)$&Selection Criterion$^a$&Defining &Number&$<z_{abs}>$&$<\beta>$&$<i>$&
$<E(B-V)_{(g-i)}>$\\
number&(SMC)&&section&of systems&&&&\\
\hline
1&    0.013& Full sample &3& 809   & 1.333 & 0.120 & 19.04 &  0.010 \\
2& $<$0.001&0.3$\le$W$_{\rm Mg\;II}^b<$0.93& 6.1& 129  & 1.375 & 0.115 & 18.66 &  0.000 \\
3& $<$0.001&0.93$\le$W$_{\rm Mg\;II}<$1.2& 6.1& 135   & 1.318 & 0.116 & 18.94 &  0.002 \\
4& $<$0.001&1.2$\le$W$_{\rm Mg\;II}<$1.53& 6.1& 133   & 1.320 & 0.116 & 19.06 & -0.004 \\
5&    0.007&1.53$\le$W$_{\rm Mg\;II}<$1.91& 6.1& 139   & 1.332 & 0.115 & 19.10 &  0.004 \\
6&    0.018& 1.91$\le$W$_{\rm Mg\;II}<$2.52& 6.1& 132  & 1.316 & 0.123 & 19.17 &  0.021 \\
7&    0.031& 2.52$\le$W$_{\rm Mg\;II}<$5.0 & 6.1&134  & 1.327 & 0.134 & 19.30 &  0.035   \\
8&    0.032& W$_{\rm Mg\;II}\ge$2.0&6.1,6.5&251  & 1.328 & 0.130 & 19.26 &  0.030   \\
9&    0.006&$z_{abs}<$1.3127&6.2& 404  & 1.142 & 0.158 & 19.02 &  0.010 \\
10&   0.012& $z_{\rm abs}\ge$1.3127&6.2& 405  & 1.523 & 0.081 & 19.07 &  0.010 \\
11&   0.009& i$<$19.12&6.2& 398    & 1.319 & 0.119 & 18.59 &  0.008 \\
12&   0.011& i$\ge$19.12&6.2& 411   & 1.346 & 0.121 & 19.49 &  0.012 \\
13&   0.010& $\beta<$0.103&6.2& 405  & 1.431 & 0.052 & 19.05 &  0.008 \\
14&   0.011& $\beta\ge$0.103&6.2& 404  & 1.233 & 0.187 & 19.04 &  0.012 \\
15&   0.009& W$_{\rm Fe\;II}^c$/W$_{\rm Mg\;II}<$0.577& 6.3&368 & 1.328 & 0.122 & 18.96 &  0.009 \\
16&   0.014& W$_{\rm Fe\;II}$/W$_{\rm Mg\;II}\ge$0.577& 6.3& 369  & 1.299 & 0.125 & 19.12 &  0.013 \\
17&   0.034& W$_{\rm Al\;II}$/W$_{\rm Mg\;I}<$1.538& 6.3&85  & 1.523 & 0.079 & 19.04 &  0.023 \\
18&   0.024& W$_{\rm Al\;II}^d$/W$_{\rm Mg\;I}^e\ge$1.538 & 6.3&85& 1.469 & 0.091 & 18.89 &  0.011 \\
19&   0.034& Fe II $\lambda$2260 present&6.4&58  & 1.308 & 0.126 & 19.02 &  0.022 \\
20&   0.019& Fe II $\lambda$2374 present &6.4&392  & 1.300 & 0.123 & 19.02 &  0.014 \\
21&   0.036& Zn II-Mg I $\lambda$2026 present&6.4&83  & 1.305 & 0.132 & 18.98 &  0.028 \\
22&   0.058& Zn II-Cr II $\lambda$2062 present&6.4&31 & 1.310 & 0.136 & 18.76 &  0.038 \\
23&   0.003& W$_{\rm Mg\;II}<$2.0&6.1,6.5& 558   & 1.335 & 0.115 & 18.95 &  0.001 \\
24&   0.002& $\Delta(g-i)<$0.2&6.5& 698     & 1.331 & 0.119 & 19.02 & -0.002 \\
25&   0.081& $\Delta(g-i)\ge$0.2&6.5&111 & 1.343 & 0.125 & 19.19 &  0.086\\
26&   0.012& W$_{\rm Mg\;II}\ge$2.5, $\Delta(g-i)<$ 0.2&6.5& 97 & 1.316 & 0.131 & 19.34 &  0.010   \\
27&   0.085& W$_{\rm Mg\;II}\ge$2.5,$\Delta(g-i)\ge$0.2&6.5&48  & 1.366 & 0.135 & 19.27 &  0.090  \\
\hline
\end{tabular}

$^a$ For selection from the full sample (sample 1)\hfill\break
$^b$ Equivalent width of Mg II $\lambda$2796 in {\AA}, used throughout in
tables\hfill\break
$^c$ Equivalent width of Fe II $\lambda$2382 in {\AA}, used throughout in
tables\hfill\break
$^d$ Equivalent width of Al II $\lambda$1670 in {\AA}, used throughout in
tables\hfill\break
$^e$ Equivalent width of Mg I $\lambda$2852 in {\AA}, used throughout in
tables\hfill\break
\end{minipage}
\end{table*}
\subsection{Sub-samples defined by detected, but intrinsically weak,
lines of heavy elements} In addition, we defined sub-samples of the
systems with detected (4 $\sigma$) lines of the following transitions: Fe
II $\lambda$2260 (sub-sample 19); Fe II $\lambda$2374 (sub-sample 20),
the blend of Zn II and Mg I $\lambda$2026 (sub-sample 21); and the blend
of Zn II and Cr II $\lambda$2062 (sub-sample 22).  These were defined to
test if depletion effects could be seen. Fe II might, for instance,
become weaker as Zn II strengthens, or both might grow, albeit at
different rates, with the total amount of dust. The Fe II $\lambda$2374
line is stronger than Fe II $\lambda$2260 line, so yields larger
sub-samples to work with. 
\subsection{Sub-samples defined to separate Mg II strength from
$\Delta(g-i)$} After examining the results from samples 1 to 22, a set of
sub-samples was defined based on a combination of Mg II line strengths
and $\Delta(g-i)$. We first complemented sub-sample 8 with sub-sample 23
(W$_{\rm Mg\;II2796}<2$ {\AA}). Then, we divided the full sample 1 in two
parts based on whether the $\Delta(g-i)$ was below or above 0.2
(sub-samples 24 and 25). Finally, we divided the sub-sample of all
systems with high Mg II equivalent width ($>$2.5 {\AA}) in two parts,
according to  $\Delta(g-i)<0.2$ or $\Delta(g-i)>0.2$ (sub-samples 26 and
27), chosen to sample the extreme reddening values in $\Delta(g-i)$ in
the histogram of Figure 3a.  These sub-samples were defined to check if
the effects of Mg II equivalent widths and dust could be separated, by
noticing how the weaker lines in the co-added spectra depend on the two
quantities.
\section{Absorption lines for different samples}
Average spectra were generated for all 27 samples defined above (following the
protocols used for Figure 1 (section 4)) and the equivalent widths of lines
were measured. Then, extinction curves, and the corresponding values of
$E(B-V)$, were derived for each sub-sample, as for Figure 2 (section 5). In
section 7.1, we deal with the equivalent widths of the lines for a selected
set of samples and their relation to the derived values of $E(B-V)$. In
section 7.2, we discuss the relation of the line strengths to other
parameters, and in section 7.3, we derive column densities and abundances from
the line strengths.
\begin{figure*}
\includegraphics[width=6.5in,height=5.0in]{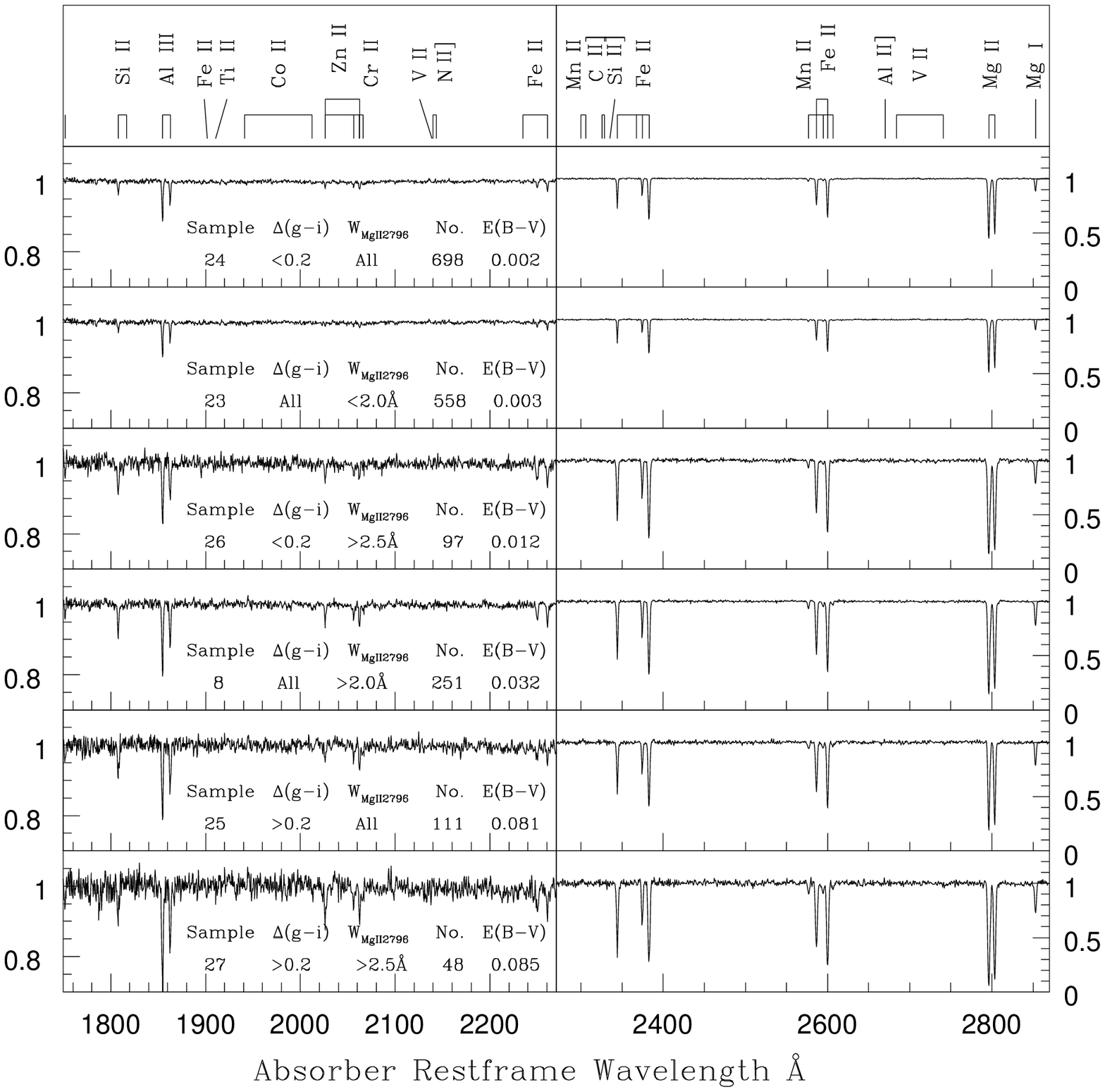}
\caption{Absorber rest frame arithmetic mean composite spectra for
sub-samples selected by ranges in $\Delta(g-i)$ and W$_{\rm Mg
\;II\lambda2796}$. The panels from top to bottom show the spectra for
sub-samples 24, 23, 26, 8, 25 and 27, respectively, in order of
increasing $E(B-V)$ values. The selection criteria [range of W$_{\rm Mg
II\lambda2796}$, range of $\Delta(g-i)$], the number of spectra averaged,
the derived value of $E(B-V)$ and the sub-sample number are listed in the
lower left of each frame. The differing signal-to-noise ratios are due to
the number of spectra added together. Only the wavelength of complete
overlap in our sample of 809 systems is plotted; the region below 2300
{\AA} is shown on an expanded flux scale to improve the visibility of
weak lines in this region. Lines labeled by wavelength are as for the
same portion of Figure 1.}
\end{figure*}
\subsection{Relation of line strengths to extinction properties} 
\subsubsection{Separating extinction and line strength effects} To
investigate the systematics that relate extinction to absorption line
strengths, for our sub-samples, consider the  spectra of sub-samples 8
and 23-27 plotted in Figure 4. These sub-samples were defined to
distinguish the effects of large Mg II equivalent width from the effects
of the SDSS, observer frame reddening, $\Delta(g-i)$. The six panels of
Figure 4 are arranged in the order of increasing, rest frame $E(B-V)$,
derived as discussed above in conjunction with Figure 2 and as listed in
Table 1.

The absorber, rest frame equivalent widths are listed in Table 2, for the
samples of Figure 4. The header information is the same as that on the panels
of Figure 4. It is apparent that the Mg II equivalent widths, while well
correlated with the strengths of the strong lines of Fe II, do not track well
with the spectroscopically derived color excesses. The weaker lines, those of
Cr II, Mn II and the two weakest lines of Fe II, increase monotonically with
the color excesses. Note the very slow change of the strengths of the corrected
Zn II equivalent widths (see appendix B) with color excess.
\begin{table*}
\centering
\begin{minipage}{140mm}
\small
\caption{Equivalent widths for Mg II and relative $\Delta(g-i)$ selected
sub-samples}
\begin{tabular}
{|l|l|r|r|r|r|r|r|r|}
\cline{1-9} 
\multicolumn{3}{|c|}{Sub-sample number}&24& 23 & 26& 8& 25& 27 \\
\hline
\multicolumn{3}{|c|}{Number of systems}& 698& 558& 97& 251& 111& 48 \\
\hline
\multicolumn{3}{|c|}{Selection Criterion W$_{\rm Mg\;II}$}&all&$<$2.0&$>$2.5&$>$2.0&all&$>$2.5 \\
\hline
\multicolumn{3}{|c|}{Selection Criterion $\Delta(g-i)$}&$<$0.2&all &$<0.2$& all& $>$0.2 &$>0.2$ \\
\hline
\multicolumn{3}{|c|}{$E(B-V)$}&0.002&0.003 &0.012&0.032& 0.081&0.085\\
\hline
\hline
$\lambda $& Species& $f\lambda^2$/$10^4 $&\multicolumn{6}{|c|}{Equivalent
width with 1 $\sigma$ errors, in m{\AA}}\\
\cline{1-9} 
2852.96& Mg I& 1489& 258$\pm$4& 192$\pm$5& 596$\pm$16& 585$\pm$10& 536$\pm$15& 795$\pm$26 
\\
2026.48$^a$&Mg I& 46.4& $<$14 ($>$0.5)& $<$5 ($>$8)& $<$86 ($>$24)& $<$79 ($>$20)&
$<$48 ($>$21)& $<$240($>$32) \\
1827.94& Mg I&8.4 & $<$5& $<$5& $<$19& $<$11& $<$24& $<$37 
\\
2796.35& Mg II& 482& 1362$\pm$4& 1076$\pm$4& 3035$\pm$17& 2674$\pm$10& 2198$\pm$15& 3210$\pm$31 
\\
2803.53& Mg II& 241& 1192$\pm$4& 927$\pm$4& 2712$\pm$17& 2400$\pm$10& 2038$\pm$15& 2877$\pm$30 
\\
1670.79& Al II& 486& 441$\pm$6& 330$\pm$6& 1033$\pm$24& 1029$\pm$14& 943$\pm$22& 1346$\pm$35 
\\
1854.72& Al III& 192& 168$\pm$4& 142$\pm$4& 332$\pm$18& 347$\pm$10& 395$\pm$18& 622$\pm$32 
\\
1862.79& Al III& 97& 106$\pm$4& 85$\pm$4& 198$\pm$16& 221$\pm$10& 246$\pm$18& 365$\pm$33 
\\
1808.00& Si II$^b$& 0.69& 62$\pm$5& 46$\pm$5& 187$\pm$19& 167$\pm$11& 175$\pm$24& 212$\pm$37 
\\
2056.26& Cr II&43.6 & 35$\pm$5& 23$\pm$5& 93$\pm$16& 101$\pm$10& 77$\pm$13& 154$\pm$27  
\\
2062.24$^a$&Cr II& 32.4& 23& 17& 65& 76& 75& 131  
\\
2066.16& Cr II& 21.7& 12$\pm$4& 12$\pm$4& 28$\pm$9& 54$\pm$10& 72$\pm$18& 109$\pm$27  
\\
2576.88& Mn II& 240& 47$\pm$4& 32$\pm$4& 143$\pm$14& 140$\pm$9& 156$\pm$17& 237$\pm$26 
\\
2594.50& Mn II& 188& 42$\pm$4& 32$\pm$4& 108$\pm$22& 129$\pm$11& 162$\pm$19& 277$\pm$32 
\\
2606.46& Mn II& 135& 21$\pm$4& 15$\pm$3& 60$\pm$18& 79$\pm$10& 96$\pm$18& 147$\pm$30 
\\
2382.77& Fe II& 182& 720$\pm$4& 540$\pm$4& 1786$\pm$15& 1600$\pm$9& 1359$\pm$17& 1975$\pm$32 
\\
2600.17& Fe II& 162 &723$\pm$4&543$\pm$4&1890$\pm$17&1672$\pm$10&1516$\pm$16&2293$\pm$30\\
2344.21& Fe II& 62.6&513$\pm$4& 367$\pm$4&1351$\pm$14&1202$\pm$9&1049$\pm$16&1658$\pm$29\\
2586.65& Fe II& 46& 462$\pm$4& 333$\pm$4& 1171$\pm$15& 1136$\pm$9& 1144$\pm$17& 1712$\pm$29 
\\
2374.46& Fe II& 17.6& 274$\pm$4& 195$\pm$4& 761$\pm$14& 702$\pm$9& 624$\pm$16& 958$\pm$28 
\\
2260.78& Fe II& 1.23& 43$\pm$4& 31$\pm$3& 137$\pm$16& 113$\pm$9& 90$\pm$14& 145$\pm$18 
\\
2249.88& Fe II& 0.91& 31$\pm$4& 19$\pm$3& 123$\pm$16& 105$\pm$10& 113$\pm$20& 215$\pm$32 
\\
2367.59& Fe II& 0.012& $<$9& $<$4& $<$26& $<$16& $<$30& $<$50 
\\
2012.17& Co II& 15& 13$\pm$4& 9$\pm$3&$<$26 & 56$\pm$13&$<$30 & $<$47 
\\
1941.29& Co II& 12.8& $<$4& $<$4& $<$14& $<$8& $<$15& $<$29 
\\
2026.14$^a$&Zn II& 205& (43)& (30)& (52)& (77)& (84)& (57):$^c$ 
\\
2062.66$^a$&Zn II& 105& 21& 15& 26& 38& 47& 27:$^c$\\
2026.3$^a$&Zn-Mg&bl & 35$\pm$4& 20$\pm$4& 112$\pm$14& 111$\pm$8& 95$\pm$15& 266$\pm$29 \\
2062.4$^a$&Cr-Zn&bl & 45$\pm$4& 32$\pm$5& 91$\pm$13& 114$\pm$8& 122$\pm$14& 158$\pm$20 
\\
\cline{1-9} 
\end{tabular}
\label{tab2}
$^a$ Lines so marked are involved in a correction scheme for blends,
based on the last two lines of the table, which are the measured values.
The equivalent widths of the other four lines marked are estimated as
described in the text.\\
$^b$ The Si II $\lambda$1808 may be contaminated with S I $\lambda$1807
line. However, Jenkins and Tripp (2001) have shown that the ratio of
column densities of C I and S I is likely to be between 16 and 24 for
region with A$_{\rm V}$ between zero and 1.0 in the Galaxy. The upper
limits and values of equivalent widths of C I lines in our sub-samples
indicate that the contamination of Si II $\lambda$1808 by S I
$\lambda$1807 line is negligible.\\ 
$^c$ The Cr II lines are saturated, so the Zn II and Cr II cannot be
disentangled due to component saturation. The values marked are probably
extreme lower limits. See text.\\
\end{minipage}
\end{table*}

As is well known, the effects of wide spread, multiple components are
strong in Mg II and strong Fe II lines in QSOALS (Nestor et al. 2003).
However, the effect seems only slightly related to extinction. Compare, for
instance the two samples with Mg II equivalent widths above 2.5 {\AA}, one
with lower reddening in the absorbers (sample 26, column 6 in Table 2), the
other with higher levels of extinction (sample 27, column 9). The spectroscopic
values of $E(B-V)$ show that the extinction differs by a factor 7, though both
samples have very strong Mg II lines. The extinction curves for both samples
are SMC extinction curves. If the SMC dust to gas ratio applies to both, then
the column densities of H I are not closely related to the line strengths of Mg
II, consistent with the findings of Rao, Turnshek \& Nestor (2006) (see their
Figures 2 and 7).

In constructing Table 2, and the later tables, and in anticipation of
using equivalent widths to derive column densities, we have had to face the
critical questions of line saturation and component blending, especially for
the Zn II/Mg I/Cr II blend at $\lambda$2026 and the Zn II/Cr II blend at
$\lambda$2062.  We have used a procedure similar to that of Khare et al. (2004)
to achieve the separation of these species. The detailed process along with the
error propagation are given in the appendix B, in conjunction with the
discussion of column densities of our 27 samples. In the process, we have
discovered that the Mg I $\lambda$2852 line is unsaturated, for equivalent
widths below 600 m{\AA}, for our samples.

A second result of the analysis given in the appendix C is that the ionization
correction is small in all of our samples, as the line strengths of Al II and
Al III track each other closely, the former line implying a larger column
density than the latter.
\subsubsection{Extinction and the 2175 {\AA} bump} Following from the
discussion in the last section, we can use the SDSS photometric colors to
isolate the most reddened QSOALS in our samples. We now return to our attempt
to detect the 2175 {\AA} bump, which failed for our full sample (Figure 2).
Figure 5 shows three extinction curves, for sub-samples 26, 8 and 25 of Figure
4, for $E(B-V)$ values of 0.012, 0.032 and 0.081. The results are in the
format of Figure 2. It is remarkable that the typical extinction of the MW is
so unique compared to our relatively large sample. However, we note that
virtually all of the known MW extinction curves are for relatively large
values of $E(B-V)$ (Valencic et al. 2004) and that for some very low
extinction sightlines in the MW, the bump is weak or missing (Clayton, Gordon
\& Wolff 2000). In other words, along the very low reddening sightlines
between the clouds of the MW for which $E(B-V)$ is typically measured,
there are regions with SMC extinction curves. We have tried to make a rough
estimate of the fraction of lines of sight that could have MW type of
extinction curves, by assuming the $E(B-V)$ to be same for the SMC and MW
lines of sight. We estimate that a maximum of 30\% of the lines of sight could
have MW type of extinction curves, the bump being masked by the 70\% of the
SMC type extinction curves.

If we assume that the curves are SMC curves, and that the dust-to-gas
ratio is the same as in the SMC, then N$_{\rm
H\;I}$/$E(B-V)$=(4.4$\pm{0.7}$)x10$^{22}$ cm$^{-2}$ (Bouchet et al.
1985). Then, in order of the panels in Figure 4 (top to bottom) and the
columns in Table 2 (left to right), the sub-samples have characteristic
values of N$_{\rm H\;I}$/10$^{20 }$ (the $E(B-V)$ of each sub-sample is
in parentheses) of 0.9 (0.002), 1.4 (0.003), 5.4 (0.012), 14 (0.032), 36
(0.081) and 37 (0.083). According to traditional definition, the last
four values correspond to DLAs and the first two, to sub-DLAs.
\subsection{Line strength as a function of other parameters}
\subsubsection{Line strengths for ranges in Mg II equivalent width} We
have now shown that there is a plausible relationship between the
$E(B-V)$ of an SMC extinction curve for the absorbers and the combination
of absorption line strengths and observer frame extinction
($\Delta(g-i)$). We turn now to samples selected by other criteria as out
lined in section 6. Table 3 includes the equivalent widths for samples
with six
different ranges of Mg II equivalent width, for selected lines of C IV,
Mg I, Mg II, Al II, Al III, Si II, Mn II and Fe II. For these
sub-samples, the strengths of Cr II and other, weak lines often give only
upper limits for equivalent widths, so we have chosen a particular set of
lines for which good data can be obtained in all six Mg II sub-samples. 
\begin{table*}
\centering
\begin{minipage}{140mm}
\caption{Equivalent widths for bins in Mg II equivalent widths}
\begin{tabular}{|l|l|l|l|l|l|l|l|l|}
\hline
\multicolumn{3}{|c|}{Sub-sample Number} & 2& 3& 4& 5& 6& 7\\
\hline
\multicolumn{3}{|c|}{Number of systems} &129&135&133&139&132&134\\
\hline
\multicolumn{3}{|c|}{Range of W$_{\rm Mg\;II}$}& 0.3-0.93& 0.93-1.2& 
1.2-1.53& 1.53-1.91& 1.91-2.52& 2.52-5.0\\
\hline
\multicolumn{3}{|c|}{$E(B-V)$}& $<$0.001& $<$0.001& $<$0.001& 0.007& 0.018& 0.031\\
\hline
\hline
$\lambda$& Species& $f\lambda^2$/10$^4$&\multicolumn{6}{|c|}{Equivalent width in
m{\AA}}\\
\hline 1548.20& C IV& 45.5& 928$\pm$48& 670$\pm$72& 789$\pm$64& 1277$\pm$48$^a$& 
1108$\pm$78$^a$& 1311$\pm$90$^a$ \\
1550.78& C IV& 22.8& 604$\pm$43& 543$\pm$74& 690$\pm$59& 944$\pm$51& 818$\pm$68& 
1154$\pm$125 \\
2852.96& Mg I&1490 & 107$\pm$10& 171$\pm$22& 281$\pm$16& 315$\pm$26& 491$\pm$20& 
629$\pm$34 \\
2796.35& Mg II& 482& 630$\pm$19& 1022$\pm$22& 1391$\pm$20& 1672$\pm$25& 2238$\pm$22& 
3041$\pm$30 \\
2803.53& Mg II& 241& 544$\pm$18& 867$\pm$26& 1167$\pm$18& 1455$\pm$25& 1979$\pm$20& 
2852$\pm$35 \\
1670.79& Al II& 486& 242$\pm$40& 285$\pm$32& 486$\pm$45& 664$\pm$47& 909$\pm$77& 
1024$\pm$48 \\
1854.72& Al III& 192& 85$\pm$16& 128$\pm$16& 169$\pm$15& 209$\pm$24& 337$\pm$31& 
407$\pm$35 \\
1862.79& Al III& 96.5& 43$\pm$13& 77$\pm$14& 111$\pm$15& 136$\pm$24& 214$\pm$33& 
236$\pm$27 \\
1526.71& Si II& 31& 180$\pm$23& 246$\pm$35& 319$\pm$61& 614$\pm$62& 928$\pm$73& 
1165$\pm$126 \\
1808.00& Si II& 0.69& 27$\pm$13& 63$\pm$18& 79$\pm$18& 52$\pm$16& 147$\pm$19& 
177$\pm$22 \\
2576.88& Mn II& 240& $<$19& 27$\pm$11& 61$\pm$14& 58$\pm$14& 113$\pm$20& 
159$\pm$19 \\
2594.50& Mn II& 188& $<$19& 39$\pm$14& 43$\pm$13& 23$\pm$10 & 104$\pm$20& 
130$\pm$19 \\
2606.46& Mn II& 135& $<$19& $<$23& 24$\pm$11& $<$24& 58$\pm$16& 
93$\pm$19 \\
2382.77& 
Fe II& 182& 265$\pm$10& 514$\pm$19& 714$\pm$14& 910$\pm$25& 1330$\pm$21& 1728$\pm$37 \\
2600.17& Fe
II&162&251$\pm$11&524$\pm22$&775$\pm$17&868$\pm$17&1374$\pm$25&1838$\pm$25\\
2344.21&Fe
II&62.6&181$\pm$11&347$\pm$18&521$\pm$22&651$\pm$18&&1249$\pm$25\\
2586.65& Fe II& 46& 138$\pm$12& 284$\pm$15& 517$\pm$21& 564$\pm$15& 949$\pm$26& 
1210$\pm$28 \\
2374.46& Fe II& 17.6& 87$\pm$11& 180$\pm$18& 281$\pm$15& 335$\pm$20& 559$\pm$17& 
679$\pm$23 \\
2260.78& Fe II& 1.23& 25$\pm$10& $<$22& 58$\pm$16& 56$\pm$12& 96$\pm$16& 
176$\pm$28 \\
2249.88& Fe II& 0.91& $<$17& $<$22& 36$\pm$11 & 34$\pm$12 & 55$\pm$14& 
120$\pm$26 \\
\hline
\end{tabular}
\label{tab3}

$^a$ The strong, blended lines of the C IV doublet had been approximately
separated into two lines. The equivalent widths are only rough estimates,
as the detailed profile is not known.\\
\end{minipage}
\end{table*}
We include the C IV measurements here, as a matter of curiosity, even
though we can not measure the lines for the lower redshift QSOALSs,
meaning that the averaging here is not over exactly the same systems as
for the Mg II. 

The strong lines of first ions are all saturated (Mg II, Al II, one line
of Si II, the first three lines of Fe II) and increase in equivalent
width as the mean Mg II equivalent widths increase and as the mean
extinction increases. Equivalent widths of lines of Al II, Al III and Si
II $\lambda$1526 vary over a wider range (a factor of five as compared to
less than a factor of two for Mg II and C IV), but show a monotonic
increase with W$_{\rm Mg\;II\lambda2796}$. The weak lines vary over the
largest range (a factor of ten) and show monotonic increase with W$_{\rm
Mg\;II\lambda2796}$ except for a decrease for the sub-sample with W$_{\rm
Mg\;II\lambda2796}$ between 1.53 to 1.91 (sub-sample 5). The weakest
lines vary by a factor of three while the extinction varies by a factor
of 30 (compare sub-samples 4 and 7).  Evidently, the extinction is
correlated with the equivalent widths of heavy element lines but
increasing depletion and increasing saturation may preclude finding a
perfectly linear relationship. 

\begin{table*}
\centering
\begin{minipage}{140mm}
\caption{Equivalent widths for split sample: redshift, $\beta$, i
magnitude}
\begin{tabular}{|l|l|l|l|l|l|l|l|l|}
\hline
\multicolumn{3}{|c|}{Sub-sample Number}& 9&10$^a$&11$^{a,b}$&12$^b$& 13$^a$&14\\
\hline
\multicolumn{3}{|c|}{Number of systems}&404&405&398&411&405&404\\
\hline
\multicolumn{3}{|c|}{Selection criterion}& $z_{\rm abs}<$1.31& $z_{\rm abs}>$
1.31& i$<$19.1& i$>$19.1& $\beta<$0.1& $\beta>$0.1\\
\hline
\multicolumn{3}{|c|}{$E(B-V)$}&0.006&0.012 &0.009&0.011&0.010&0.011\\
\hline
\hline
$\lambda$&Species&$f\lambda^2$/10$^4$&\multicolumn{6}{|c|}{Equivalent width in
m{\AA}}\\
\hline
1548.20& C IV& 45.5& NA& 971$\pm$23$^c$&993$\pm$28$^c$&948$\pm$58$^c$&
980$\pm$27$^c$&834$\pm$88$^c$ \\
1550.78& C IV& 22.8& NA& 727$\pm$24& 761$\pm$27& 578$\pm$39& 
732$\pm$28& 675$\pm$83 \\
2852.96& Mg I& 1490& 341$\pm$20& 300$\pm$17& 298$\pm$13& 379$\pm$21& 
279$\pm$15& 331$\pm$12 \\
2796.35& Mg II& 482& 1516$\pm$14& 1534$\pm$14& 1471$\pm$13& 1771$\pm$17& 
1444$\pm$14& 1594$\pm$13 \\
2803.53& Mg II& 241& 1366$\pm$13& 1366$\pm$16& 1326$\pm$12& 1569$\pm$20& 
1252$\pm$14& 1455$\pm$13 \\
1670.79& Al II& 486& 524$\pm$117& 442$\pm$21& 401$\pm$26& 522$\pm$33& 
428$\pm$24& 440$\pm$32 \\
1854.72& Al III& 192& 190$\pm$13& 177$\pm$9& 164$\pm$9& 250$\pm$17& 
179$\pm$9& 189$\pm$14 \\
1862.79& Al III& 96.5& 118$\pm$15& 103$\pm$10& 95$\pm$8& 156$\pm$18& 
102$\pm$9& 137$\pm$17 \\
1526.71& Si II& 31& NA& 324$\pm$22& 305$\pm$25& 397$\pm$46& 
337$\pm$23& 167$\pm$55 \\
1808.00& Si II& 0.69& 76$\pm$14& 58$\pm$7& 59$\pm$7& 83$\pm$24& 
59$\pm$7& 88$\pm$19 \\
2576.88& Mn II& 240& 61$\pm$8& 53$\pm$9& 51$\pm$6& 76$\pm$17& 
43$\pm$7& 76$\pm$12 \\
2594.50& Mn II& 188& 54$\pm$7& 42$\pm$9& 54$\pm$9& 60$\pm$12& 
47$\pm$8& 54$\pm$10 \\
2606.46& Mn II& 135& 24$\pm$7& $<$14& 23$\pm$8& $<$23& 
15$\pm$7& 29$\pm$10 \\
2382.77& Fe II& 182& 802$\pm$10& 710$\pm$15& 713$\pm$11& 979$\pm$20& 
695$\pm$12& 858$\pm$13 \\
2600.17& Fe II&
162&860$\pm$11&742$\pm$14&765$\pm$11&1000$\pm$21&758$\pm$13&884$\pm$16\\
2344.21&Fe II& 62.6&
569$\pm$10&488$\pm$10&489$\pm$8&734$\pm$17&486$\pm$10&624$\pm$16\\
2586.65& Fe II& 46& 585$\pm$11& 477$\pm$14& 494$\pm$12& 705$\pm$23& 
494$\pm$13& 583$\pm$14 \\
2374.46& Fe II& 17.6& 307$\pm$7& 253$\pm$11& 252$\pm$8& 416$\pm$18& 
264$\pm$9& 313$\pm$9 \\
2260.78& Fe II& 1.23& 61$\pm$11& 44$\pm$10& 48$\pm$7& 59$\pm$17& 
53$\pm$10& 42$\pm$7 \\
2249.88& Fe II& 0.91& 42$\pm$8& 19$\pm$9& 32$\pm$7& 39$\pm$12& 
30$\pm$8& 39$\pm$11 \\
\hline
\end{tabular}
\label{tab4}

$^a$ The numbers of objects included in sub-samples 10, 11 and 13 
were larger by 2, 1 and 1, respectively, for the line measurements 
than for the measurements of extinction. The numbers for the number 
of spectra included and the mean extinction are from Table 1.\\
$^b$ Note that the lines are stronger in the sample of faint QSOs (sample
12) as the equivalent width limit for detection is higher in this sample
as compared to the sample of bright QSOs (sample 11)\\
$^c$ Same as footnote a in Table 3.\\
\end{minipage}
\end{table*}

While we cannot deal with C IV here, as the sub-samples do not include
the same objects as does the Mg II sub-samples, it is interesting to note
that C IV behaves differently than the first ions, being much less
variable across the full range of sub-samples. This may be an indication
that, in particular, C IV is in a different gas than the Al II, and also
a different gas than the Al III, since Al II and Al III track each other.
This result is not inconsistent with the literature, but is much more
general, because it covers so many objects. High resolution data also
seems to indicate that Al III follows lower ions, not C IV and Si IV.
\subsubsection{Split samples: $i$ magnitude, redshift and $\beta$} As seen
in the last section line strengths appear to increase with extinction (Table 2)
as well as with Mg II equivalent width (Table 3).  To test whether there is any
dependence on other properties of QSOALS or QSOs we split the full sample in
halves: in absorption redshift, at 1.3127; in i-magnitude, at 19.1; and in
relative velocity with respect to the QSO, at 30,000 km s$^{-1}$. The results
of the line measurements are listed in Table 4, using the same transitions that
are listed in Table 3.  The line strengths are remarkably uniform for samples
split in $z_{abs}$ and $\beta$, implying that they are not closely related to
the absorption redshift and implying that there is no evidence for systems with
smaller relative velocity with respect to the QSO being different (which could
have been the case if they were produced in outflows from the QSO) from the
systems with larger relative velocities (which in all probability are produced
by intervening absorbers).

There is a trend for all lines to be stronger in the faint sample than in
the bright sample. The extinction is the same, however, in both samples
(11 and 12, see Table 4). Recall that the magnitude that splits the full
sample of 809 QSOALS in half is 19.1. As mentioned in section 2.1,
brighter QSOs are drawn from the SDSS color selected sample while fainter
QSOs are mainly those chosen for spectroscopic follow up because they are
X-ray sources, radio sources, or serendipitous sources. The bright sample
is therefore subject to systematic effects of extinction, the faint sample is
not. There appear, therefore, to be no strong selection effects that hide
a large number of QSOs due to large extinction, in the SDSS sample.   

The lines are stronger in the faint sample because the weak systems that are
possibly present in that sample are selectively not detected at 4 $\sigma$
significance, our cutoff for inclusion. Thus inclusion of the faint QSOs may
enhance the equivalent widths more than the $E(B-V)$ values and therefore may
yield somewhat higher column densities and abundances. This will affect the
column densities and abundances measured below, for most other sub-samples 
(which should have an equal distribution of bright and faint QSOs) by less
than 25\% (as seen from the differences in the line strengths of samples 11 and
12).  Also, the trend in the abundance variation as a function of $E(B-V)$ that
we determine in this paper should be free from this effect. Measurements
of color excess are evidently not affected by the magnitudes of the QSOs. We
note, from Table 3, that samples of equivalent widths of Mg II $\lambda2796 <$
1.53 {\AA} (samples 2, 3 and 4) do not have discernible extinction.

A remaining anomaly from Table 3 is that the derived, spectroscopic extinction
for sample 9 (low $z_{abs}$, $E(B-V)$=0.006) is one-half the extinction for
sample 10 (high $z_{abs}$, $E(B-V)=$0.012). Both extinction values are low. The
difference could be significant and might be verified by using a larger sample.
However, the mean extinction is affected by a few high values and many low
values. Small number statistics in the former is probably responsible for the
apparent difference.

In summary, the trends of increasing line strengths with extinction (section
7.1.1) and Mg II line strength (section 7.2.1) are highly significant compared
to only weak trends found by splitting the main sample by redshift, $\beta$, or
apparent magnitude.
\subsubsection{Other sub-samples}A final set of sub-samples appears in
Table 5, those made to study the extremes of the sample from the point of
view of the absorption lines, as described in sections 6.3 and 6.4. These
sub-samples provide better limits on the weaker lines and are indicative
of a wider range of constraints on the QSOALS.

\begin{table*}
\centering
\begin{minipage}{160mm}
\footnotesize
\caption{Equivalent widths for the full sample and for selected
sub-samples}
\begin{tabular}
{|l|l|l|l|l|l|l|l|}
\hline
\multicolumn{3}{|c|}{Sample Number}&16& 1& 17& 19& 21\\
\hline
\multicolumn{3}{|c|}{Number of systems}& 369& 809& 85& 58& 83 \\
\hline
\multicolumn{3}{|c|}{Selection criterion}& W$_{\rm
Fe\;II}$/W$_{\rm Mg\;II}$& Full&W$_{\rm Al\;II}$/W$_{\rm Mg\;I}$& Fe II $\lambda2260$&
Zn II $\lambda2026$\\
\multicolumn{3}{|c|}{}  & $>$0.58 &sample & $<$1.54&present& present\\
\hline
\multicolumn{3}{|c|}{$E(B-V)$}& 0.014& 0.013& 0.034& 0.034& 0.036\\
\hline
\hline
$\lambda$&Species&$f\lambda^2$/10$^4$&\multicolumn{5}{|c|}{Equivalent width in
m{\AA}}\\
\hline
1656.93& C I& 40.9& 46$\pm$13& 19$\pm$4& 79$\pm$21&$<$57 & 88$\pm$15 \\
1548.20& C IV& 45.5& 762$\pm$36& 839$\pm$10& 825$\pm$44& 970$\pm$59& 822$\pm$47 \\
1550.78& C IV& 22.8& 602$\pm$37& 580$\pm$9& 718$\pm$45& 706$\pm$62& 551$\pm$48 \\
2852.96& Mg I& 1490& 397$\pm$10& 282$\pm$4& 691$\pm$35& 689$\pm$16& 636$\pm$15 \\
2026.48$^a$& Mg I& 46.4& $<$34($>$16)& $<$32($>$11)&
$<$37($>$26)&$<74$($>28$)
& $<$140($>$25) \\
1827.94& Mg I& 8.4& $<$10& $<$ 5& $<$22& $<$15& $<$15 \\
2796.35& Mg II& 482& 1819$\pm$12& 1432$\pm$4& 1893$\pm$28& 2454$\pm$16& 2508$\pm$15
\\
2803.53& Mg II& 241& 1636$\pm$12& 1261$\pm$4& 1751$\pm$33& 2255$\pm$15& 2312$\pm$15
\\
1670.79& Al II& 486& 716$\pm$18& 471$\pm$6& 662$\pm$29& 958$\pm$26& 936$\pm$22 \\
1854.72& Al III& 192& 253$\pm$13& 183$\pm$4& 245$\pm$24& 353$\pm$16& 345$\pm$17 \\
1862.79& Al III& 97& 168$\pm$14& 112$\pm$4& 172$\pm$22& 190$\pm$15& 202$\pm$17 \\
1526.71& Si II& 31& 599$\pm$31& 383$\pm$10& 510$\pm$30& 814$\pm$67& 619$\pm$55 \\
1808.00& Si II$^b$& 0.69& 121$\pm$10& 68$\pm$5& 145$\pm$22& 240$\pm$15& 224$\pm$14 \\
3934.78& Ca II& 970& 254$\pm$64& 110$\pm$16& NA& 254$\pm$47& 370$\pm$56 \\
3969.59& Ca II& 491& 156$\pm$42& 57$\pm$13& NA& 141$\pm$36& 154$\pm$35 \\
3384.73& Ti II& 410& 67$\pm$16& 31$\pm$5& $<$130& 160$\pm$16& 143$\pm$22 \\
3242.92& Ti II& 244& 49$\pm$10& 32$\pm$5& $<$94& 120$\pm$16& 64$\pm$12 \\
3073.86& Ti II& 114&$<$15 & 20$\pm$6&$<$46 & 43$\pm$12& $<$44 \\
3230.12& Ti II& 72&$<17$ &$<$12 &$<$61 &$<$32 & $<$47 \\
1910.85& Ti II$^c$&36.5& $<$9& $<$14& $<$19& 32$\pm$10& 34$\pm$9 \\
2056.26& Cr II& 43.6& 72$\pm$11& 39$\pm$4& 89$\pm$17& 147$\pm$14& 115$\pm$12 \\
2062.24$^a$& Cr II& 32.4& 56& 40& 70& 123& 90 \\
2066.16& Cr II& 21.7& 39$\pm$10& 14$\pm$4& 50$\pm$16& 100$\pm$13& 65$\pm$13 \\
2576.88& Mn II& 240& 114$\pm$11& 54$\pm$4& 146$\pm$21& 231$\pm$14& 183$\pm$13 \\
2594.50& Mn II& 188& 78$\pm$10& 50$\pm$4& 103$\pm$16& 167$\pm$12& 148$\pm$13 \\
2606.46& Mn II& 135& 49$\pm$9& 26$\pm$4& 64$\pm$14& 142$\pm$13& 130$\pm$14 \\
2382.77& Fe II& 182& 1138$\pm$14& 761$\pm$4& 1007$\pm$32& 1566$\pm$15& 1581$\pm$15 \\
2600.17&Fe II&162&1149$\pm$15&783$\pm$4&1135$\pm$23&1657$\pm$16&1666$\pm$15\\
2344.21&Fe II&62.6&843$\pm$12&550$\pm$4&840$\pm$22&1331$\pm$14&1295$\pm$14\\
2586.65& Fe II& 46&837$\pm$15 & 509$\pm$4& 845$\pm$21& 1447$\pm$17& 1290$\pm$16 \\
2374.46& Fe II& 17.6 &505$\pm$12 & 298$\pm$4& 476$\pm$31& 894$\pm$14& 835$\pm$14 \\
2260.78& Fe II& 1.23 & 99$\pm$11& 46$\pm$4& 125$\pm$23& 280$\pm$10& 147$\pm$12 \\
2249.88& Fe II& 0.91 & 63$\pm$10& 35$\pm$4& 58$\pm$17& 163$\pm$15& 110$\pm$14 \\
2367.59& Fe II& 0.012 & $<$11~& $<$4& ~$<$17& $<$14& $<$14 \\
2012.17& Co II&15 &$<$11 & 15$\pm$4& $<$20& 57$\pm$17& 28$\pm$8 \\
1941.29& Co II& 12.8&$<$10 &$<$8 &$<19$ &$<$27 &$<30$ \\
1741.55& Ni II& 13.0& 84$\pm$24& 28$\pm$4& 80$\pm$20& 124$\pm$16& 118$\pm$15 \\
1709.60& Ni II& 9.35& 23$\pm$6& 22$\pm$5&$<$29 & 108$\pm$22& 65$\pm$15 \\
1751.92& Ni II& 8.58&$<$16 & $<$11& 48$\pm$20& $<$48& 153$\pm$20 \\
2026.14$^a$& Zn II& 205& (42)& (16)& (62)& $<$148& $<$204 \\
2062.64$^a$& Zn II& 105& 21& 8& 31& 102& 101 \\
2026.1$^a$& Zn II-Mg I&bl& 55$\pm$8& 40$\pm$4& 68$\pm$17& 176$\pm$14& 239$\pm$9 \\
2062.6$^a$& Cr II-Zn II&bl & 77$\pm$10& 48$\pm$4& 101$\pm$15& 225$\pm$14& 189$\pm$12 \\
\hline
\end{tabular}
\label{tab5}

$^{a,b}$ are same as for Table 2\\
$^c$ Ti II, $\lambda$1920.6123, f=0.103 and $\lambda$1910.9538, f=0.098, here
combined by averaging the wavelengths and adding the f-values.
\end{minipage}
\end{table*}

Table 5 aims to show the weakest lines of each element and includes species
not represented in Tables 2, 3 and 4: C I, Ca II, Ti II, Co II and Ni II. We
have also added some strong limits for undetected species.

First, consider Mg I. We have corrected the blends of Mg I, Zn II and Cr
II as described in appendix B. Here we add the limit for Mg I
$\lambda$1827, which has an $f$ value 20{\%} of that for Mg I
$\lambda$2026. The measured limits are consistent with our previous
conclusion that Mg I is on the linear portion of the curve of growth: for
most sub-samples we can use $\lambda1827$ to show that the $\lambda2852$
line is not badly saturated. The case is, however, particularly
strengthened for sample 1 (the sample of all 809 systems) and for
sub-sample 21 (the Zn II selected sub-sample). The latter has among the
strongest Mg I $\lambda$2852 lines of all the sub-samples, 636 m{\AA}, a
level that for any other line in our study would be strongly saturated by
a factor of 25. For Mg I, W$_{\rm Mg\;I\lambda1827}$ constrains the
saturation to be less than a factor of 3.
\renewcommand{\thefigure}{\arabic{figure}\alph{subfigure}}
\setcounter{subfigure}{1}
\begin{figure}
\includegraphics[width=3.15in,height=4.0in]{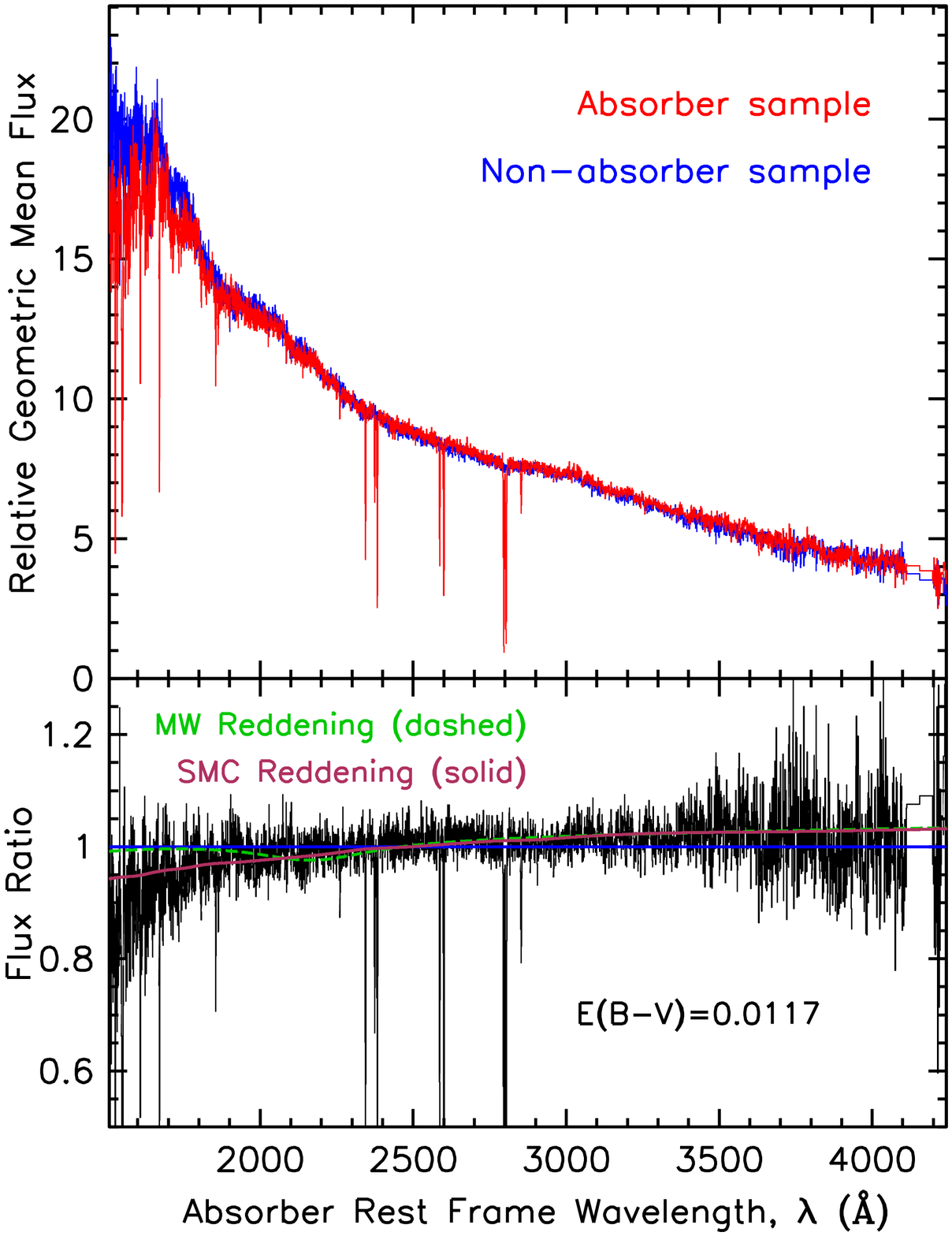}
\caption{Extinction curve (same format as Fig. 2) for sub-sample 26
(normalized spectrum which was plotted in panel 3 of Figure 4). The
sub-sample consists of QSOALS with W$_{\rm Mg\;II\lambda2796}>2.5$ {\AA}
and $\Delta(g-i)<0.2$. Note that this sub-sample contains only 97 systems
as compared to 809 in the full sample plotted in Figure 2. Both samples
have comparable derived $E(B-V)$ values. }
\end{figure}
\addtocounter{figure}{-1}
\addtocounter{subfigure}{1}
\begin{figure}
\includegraphics[width=3.15in,height=4.0in]{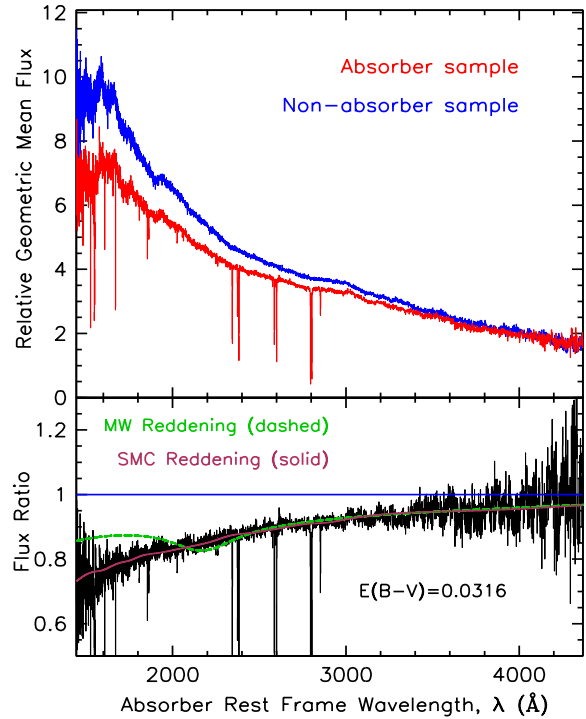}
\caption{Extinction curve (same format as Fig. 2) for sub-sample 8
(normalized spectrum which was plotted in panel 4 of Figure 4). The
sub-sample consists of 251 QSOALS with W$_{\rm Mg\;II\lambda2796}>2.5$
{\AA} (same as for sub-sample 26, plotted in Figure 5a) irrespective of
their $\Delta(g-i)$ values. The inclusion of systems with higher
reddening is clearly indicated in the higher mean extinction as compared
to that for sub-sample 26.}
\end{figure}
\addtocounter{figure}{-1}
\addtocounter{subfigure}{1}
\begin{figure}
\includegraphics[width=3.15in,height=4.0in]{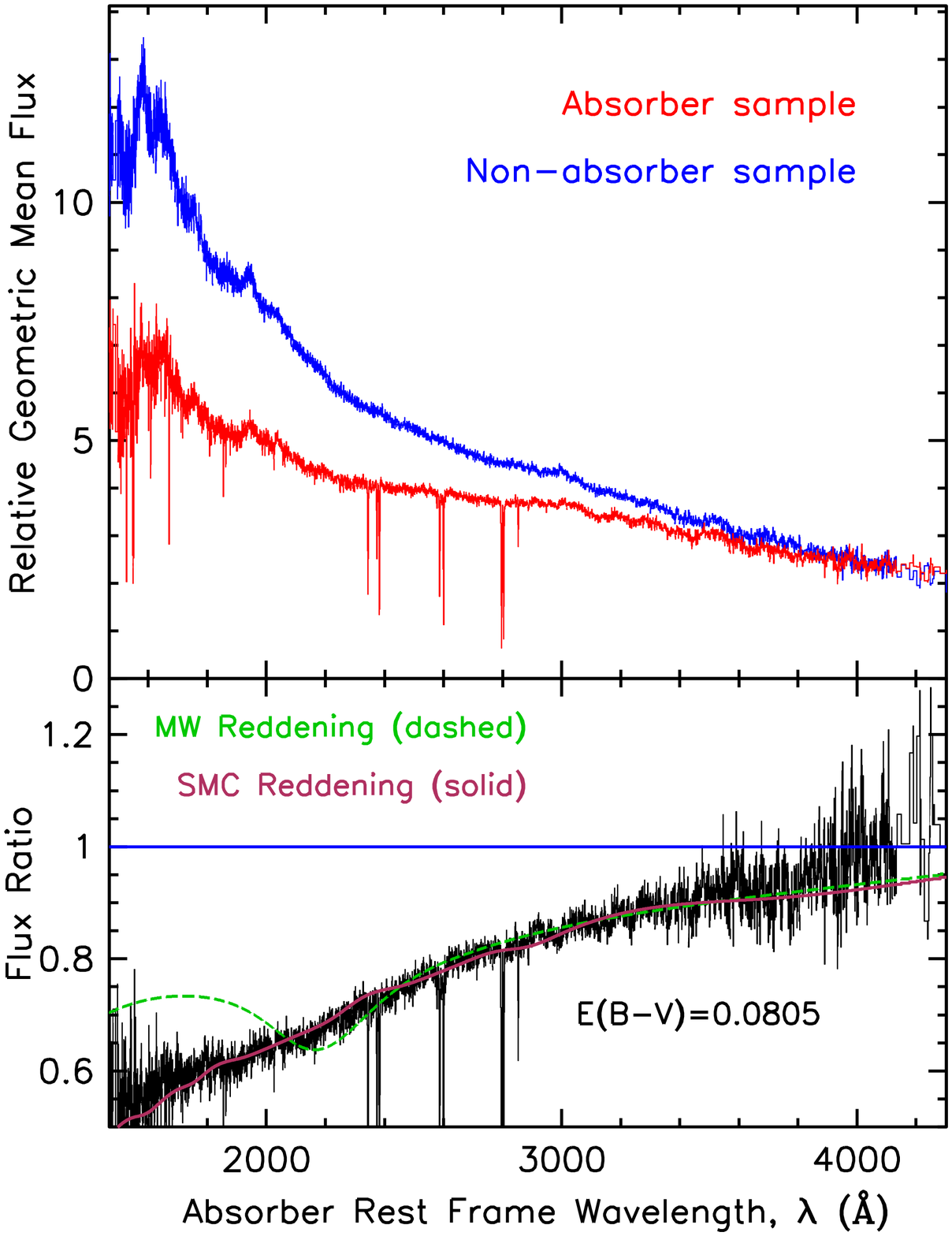}
\caption{Extinction curve (same format as Fig. 2) for sub-sample 25
(normalized spectrum which was plotted in panel 5 of Figure 4). The
sub-sample consists of 111 QSOALS with $\Delta(g-i)>0.2$, irrespective of
their W$_{\rm Mg\;II\lambda2796}$ values. The higher reddening in
individual systems leads to large derived value of $E(B-V)$. Note the
small differences in the strong (saturated) absorption lines in panels 3,
4 and 5 of figure 4, but the increasing strengths of weak lines (towards
the left in figure 4) as the mean extinction increases.}
\end{figure}
\renewcommand{\thefigure}{\arabic{figure}}

Next, consider C I. While we have only one line in our spectral region,
it is detected at the 3.5-4 $\sigma$ level in four separate sub-samples
and is thus real.  The abundance of Mg is about 1/10 that of C in most
known types of interstellar gas (see the later discussion of abundances).
Considering the relative abundances of C and Mg and values of
$f\lambda^2$ in column 3 of Table 5, an origin of Mg I and C I in the
same gas would lead to ratios of the strongest shown lines of the two
species of three, Mg I $\lambda$2852 being stronger.  The observed ratio
ranges from 8 to 15 when both lines are detected, an amount accounted for
by the higher photoionization rate of C I compared to Mg I in the
interstellar medium. On the other hand the recombination coefficient for
Mg$^+$ compared to C$^+$ is much higher in warm gas than in gas with
T$<$3000 K (York \& Kinahan 1979). If Mg I and C I are from the same gas,
the gas in the four samples that contain C I in Table 5 (16, 1, 17, 21)
has a temperature lower than 3000 K. Thus, it is consistent that the Mg I
and C I arise in the same gas. High resolution profiles of each could
immediately confirm this result, a situation that seems to be true within
the SMC as well (Welty et al. 2005, in preparation). The importance of
this conclusion is that the amount of ionized gas in H II regions
(temperature $>$ 8000 K) is small and the ionization corrections to
abundances drawn from first ions is small, as is indicated by the
constant, low ratio of N$_{\rm Al\; III}$ to N$_{\rm Al\; II}$.

To comment briefly on the strong lines, the C IV lines are remarkably
uniform across these sub-samples, contrary to their variation by a factor
of two in Table 3. The lines of Al II and Al III show a pattern similar
to that noticed above; they closely track each other. Al II does not
track the Fe II $\lambda$2374 as well as in other sub-samples. 

We turn now to the weak lines and the saturation properties of those lines.
The last three columns contain the highest average values of $E(B-V)$ and there
we expect the largest saturation, given our analysis of Table 2.  However,
visual examination of the trends and relative strengths in these columns
indicates that if we take the weakest line in each case, we should get accurate
integrated column densities. The exception is actually the full sample
(sample 1), the one most dominated by the numerous weak systems, where lines of
Cr II have equivalent widths $\sim$ 40 m{\AA}, instead of the ratios expected
from the relative $f$-values of the three lines. However, the sizes of the
errors for the two, unblended Cr II lines allow the possibility that the Cr II
lines are unsaturated. Note the apparent unsaturated nature of Ca II, in which
case the equivalent widths of the two lines are similar to the ratio of their
$f$-values, even though the total equivalent widths are very large. These
lines also have an appearance of being broad as do the weakest lines of Mn II
and Fe II. They may have a wide dispersion of components as in the Mg I and Mg
II lines discussed in section 9.6, an effect that alleviates saturation
somewhat.  
\subsection{Column densities} 
\subsubsection{General comments on column density derivations}
We turn now to deriving column densities and abundances from the equivalent
widths. For the low resolution of the SDSS spectrographs, blends of saturated
components are a major concern, as discussed earlier and in more detail in the
appendix. Jenkins (1986) has demonstrated that in large samples, integrated
equivalent widths can be used to derive column densities. Furthermore, if
multiple lines of a species can be shown to be on the linear portion of the
curve of growth, the column densities can be directly derived. In our best
spectra, we can perform this test for lines of Zn II, Cr II, Mn II and Fe II.
In the extended spectra (1800 {\AA}$<\lambda$ and $\lambda>$3100 {\AA}) lines
of Ti II, Co II, Ni II and Ca II are available. Of course, when no line is
detected, the upper limit on equivalent width can reasonably be used to produce
upper limits on the column densities, provided the data are of high $S/N$. For
the weakest lines, the degree of saturation can be assessed if an intrinsically
weaker line is not detected.
\subsubsection{Species with weak lines} We conclude that for our data
set, we can derive approximate column density averages from Mg I
$\lambda$2852, Cr II $\lambda$2066, Mn II $\lambda$2606, Fe II
$\lambda$2249 and Zn II $\lambda$2062 (after correction for Cr II, as
in appendix B) by assuming these particular lines are on the linear portion
of the curve of growth. Lines of Ti II, Ca II and Ni II can also be used
in the manner and we assume the same is true for Si II $\lambda$1808. For
cases with evidence of saturation, we treat the derived values as lower
limits.

We collect in Table A3, a wide range of sub-samples with fairly complete
abundance information and with coverage of a full range of $E(B-V)$. The
table includes equivalent widths for 16 sub-samples that range in derived
$E(B-V)$ from $<$0.001 to 0.085. We have purposefully included some
sub-samples with similar $E(B-V)$ values, to document the dispersion of
equivalent widths in differently defined sub-samples. 
\begin{table}

\tiny
\caption{Column densities}
\vspace{0.1in}
\begin{tabular}
{|c|c|c|c|c|}
\hline
&  SMC & Sub-sample 8& Sub-sample 25  & DLA$^b$ 
\\ &Sk155  & W$_{\rm Mg\;II}>$ 2.0 {\AA}&$\Delta(g-i)>$0.2&\\
\hline
$E(B-V)$ & &0.032 & 0.081 & $<$0.04/0.09 \\
\hline
\hline
Species&\multicolumn{4}{|c|}{Log(N$_{\rm X}$)}\\
\hline
H I$^a$ & 21.4 &21.2 & 21.6 & 21.4\\ 
C I & 13.4 &13.1& 13.6: & -- \\
Mg II & 16.1$^c$ &15.0:& 14.9: &-- \\
Al I & -- &$<$12.2& $<$12.4 & --  \\
Al II & -- &13.9:& 13.8: & 13.2 \\ 
Al III & 13.0 &13.4& 13.5 & 12.8 \\
Si I &--  &$<$12.5 & $<$12.3 & -- \\ 
Si II & 16.2 &15.4& 15.5 & 15.6 \\
Ca II & 12.7 & 12.4 &12.6 & -- \\
Ti II & 12.6 &12.4& 12.3 & $<$12.9 \\
Cr II & 13.4 & 13.3&13.6 & 13.4 \\
Mn II & 13.0 & 12.9&13.0 & --- \\
Fe I & $<$11.6 &$<$11.8& $<$12.0 & --\\
Fe II & 15.2 & 15.1&$>$14.9 & 14.5 \\
Co II & $<$12.7 &$<$13.1& $<$13.4 & $<$13.0  \\
Ni II & 13.7 & 13.9&$>$14.1 & 14.0 \\
Zn II & 13.2 & 12.6&$>$12.7 & 12.7 \\
\hline
\end{tabular}

\noindent $^a$ For the two sub-samples the H I column densities were
estimated from the derived reddening; see section 8 for details. For Sk
155 the H I is from a profile fit that ignores MW gas, from HST and FUSE
data. The column densities of heavy elements for Sk 155 refer only to SMC
gas, as well. For DLA the H I is from profile fit to Lyman $\alpha$ line
from HST data.\\
\noindent $^b$ SDSS1727+5302, $z_{abs}=$1.0311\\ 
\noindent $^c$ See footnote f to table 7\\
\end{table}
Table A3 includes the weakest lines of each detected species to use for
abundance estimates. 
\subsubsection{Species with strong lines}
\renewcommand{\thefigure}{\arabic{figure}\alph{subfigure}}
\setcounter{subfigure}{1}
\begin{figure}
\includegraphics[width=3.15in,height=4.0in]{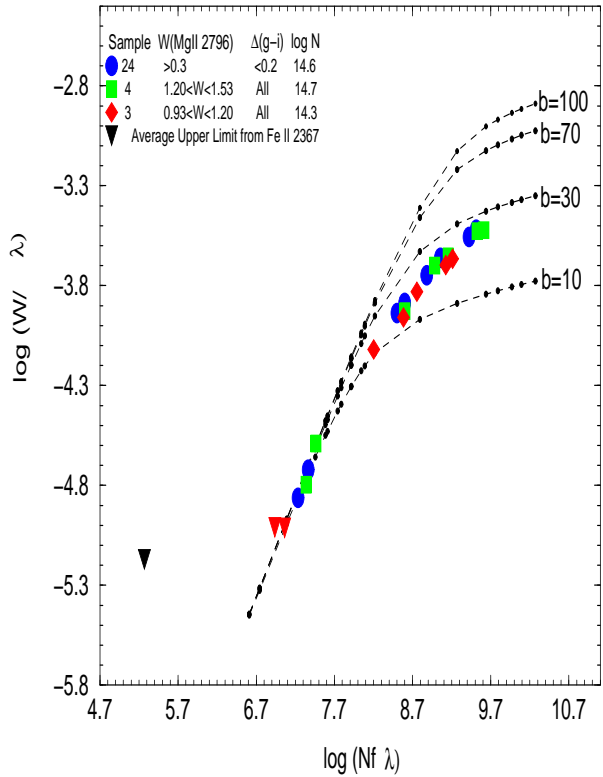}
\caption{Empirical curve of growth for Fe II for selected sub-samples
with low ($\le$0.002) $E(B-V)$ values. Column densities are from weak Fe
II lines (Table A4) and equivalent widths are from Table A3. The curves
deviate from single component curves of growth because the systems have
many components of differing column densities. These curves can be used
to estimate the column densities for systems with only saturated lines
available (Al II, Mg II, in our case).}
\end{figure}
\addtocounter{figure}{-1}
\addtocounter{subfigure}{1}
\begin{figure}
\includegraphics[width=3.15in,height=4.0in]{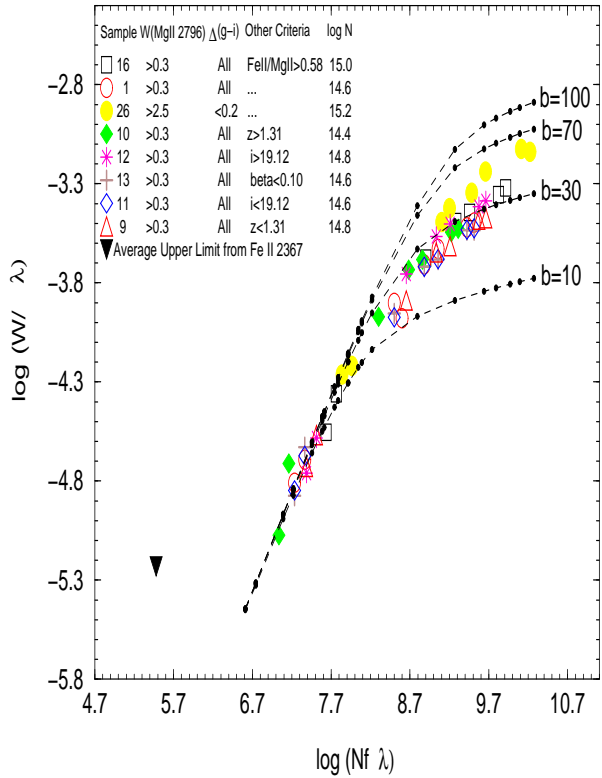}
\caption{Empirical curve of growth for Fe II for selected sub-samples
with $E(B-V)$ values $\sim$ 0.01. The legend gives the symbols in order
of decreasing $E(B-V)$, the values of which range from 0.014 to 0.006, as
listed in Table 1. Column densities and equivalent widths for Fe II are
derived as for Figure 6a, and the use of the curve of growth is the
same.}
\end{figure}
\addtocounter{figure}{-1}
\addtocounter{subfigure}{1}
\begin{figure}
\includegraphics[width=3.15in,height=4.0in]{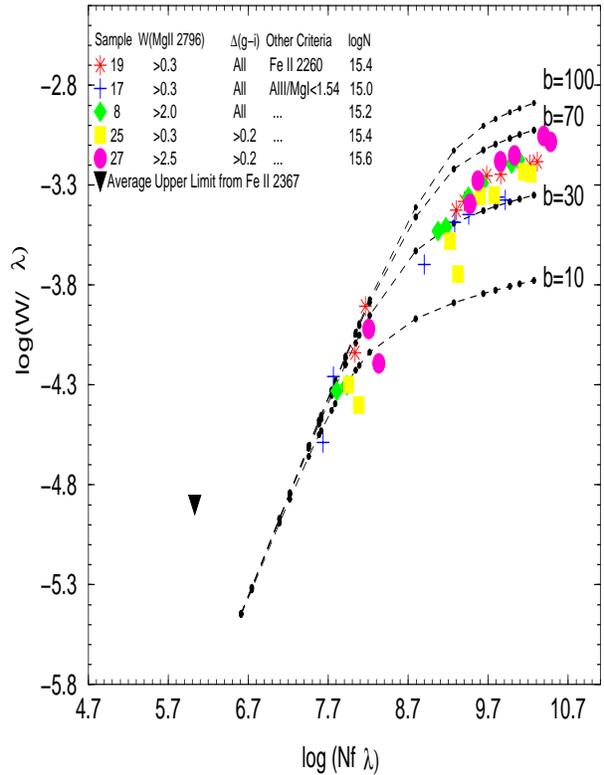}
\caption{Empirical curve of growth for Fe II for selected sub-samples
with high ($\ge$ 0.03) $E(B-V)$ values. Column densities and equivalent
widths for Fe II are derived as for Figure 6a, and the use of the curve
of growth is the same.}
\end{figure}
\renewcommand{\thefigure}{\arabic{figure}}
For Al II and Mg II, other techniques must be applied. Figures 6a, 6b, 6c
give curves of growth based on column densities for Fe II for the 16
sub-samples of Table A3 with low (Fig. 6a), medium (Fig. 6b) and high
(Fig. 6c) values of $E(B-V)$ respectively. These curves can be used to
tie together the weak and strong lines of Fe II in an empirical curve of
growth. By assuming that the Mg II and Al II lines follow the same type
of curve (i.e., that the components match one to one and have the same
relative abundances), the column densities of the Al II and Mg II can be
estimated. These curves of growth have a more linear shape, as saturation
is approached from the left, than do single component curves of growth.
The well known effect occurs because lower column density components,
outside the core absorption, do not affect the integrated strengths of
the weakest lines, yet add strength to the strong, saturating lines. Thus
the weak outliers seem to extend the linear portion of the curve of
growth. We have built such curves of growth for the line of sight to
SN1987A (Welty et al. 1999a) and to Sk 155 (Welty et al. 2001), using
weak, ultraviolet transitions observed with IUE and HST, respectively at
high resolution. By adding the column density in all components and
integrating the entire equivalent widths (ignoring the components
attributable to the MW), we find that the total strengths are similar to
those of the DLAs and that the integrated curve of growth for the
sightline is similar to those in Figure 6.
\subsubsection{An overview of column densities for several sub-samples}
The column densities along with the estimated errors for the sub-samples
in Table A3 are given in Table A4. The samples documented in this table
are chosen to be those that have fairly complete sets of abundances and
to have non-limit values of $E(B-V)$ over the widest possible range,
0.002 to 0.085, and to include redundancy in $E(B-V)$ to measure the
scatter from different sub-samples. When there is any indication of
saturation, we quote the formal error based on the assumption that the
weakest line of the species is on the linear portion of the curve of
growth. When several lines of a given species indicate strong saturation,
we enter values calculated from the weakest lines as lower limits to the
column density. When there is no detection, we give the upper limit to
the average column density that could be present. For Al II and Mg II the
errors are uncertain because of our approximate curve of growth
corrections. For Si II, we use the values obtained from $\lambda1808$
line only, taken as unsaturated. (The Si II$\lambda$1526 line gives
column densities too low by a factor of 10 on the same assumption).
Errors have been propagated and indicated by a, b, c, d, e,.., each
representing an increment of 0.05 dex: a implies $<$0.049 dex, b implies
$<$0.099, etc. 

Table 6 includes column densities for two of the sub-samples (8 and 25 which
are, evidently, typical of the DLA sample in metallicity, depletion and H I
column density), taken from Table A4. Also given are column densities for Sk
155, the star with the most extensive abundance information available in the
SMC and for the $z_{abs}$ = 1.0311 DLA system toward SDSS1727+0048 (Khare et
al. 2004).  All column densities tabulated are averages over all components,
which may differ significantly from one component to the other in abundances.
For Sk 155, the average is taken after the column densities are derived from
higher resolution data. For the other columns, the data are averages based on
lower resolution data, as described in this paper. The entries in Table 6 show
that sub-samples 8 and 25, selected by high W$_{\rm Mg\;II\lambda2796}$ and
high $\Delta(g-i)$ values, respectively, have the same column densities,
approximately as the two chosen companion objects, except that N$_{\rm Al\;II}$
is higher in SDSS1727+0048 and N$_{\rm Si\;II}$ is higher in Sk155. 

The column densities for various species from Table A4 are plotted as a
function of $E(B-V)$ in Figure 7. The general trend of increase of column
densities with $E(B-V)$ is present but the relative increase is small.  The
increase in column densities is clearly not linear in $E(B-V)$, since the
single element values range over less than a factor of 10, while the $E(B-V)$
values range over a factor of 80. It is also clear that the scatter in
column density between different sub-samples with the same
value of $E(B-V)$ is small, even though the sub-samples were chosen in very
different ways and $E(B-V)$ was derived, spectroscopically, after the
fact, in all cases. We note that an increasing Zn Column density with
increasing extinction has also been noted by Vladilo \& Peroux (2005). 

In summary, using equivalent width measurements discussed in earlier sections,
we elaborate on our technique of finding column densities, for elements with
multiple weak lines and for Mg II and Al II, with strong lines. We find that
the column densities follow a smooth pattern with our $E(B-V)$ values, with
little scatter between different samples with similar $E(B-V)$ values (Figure
7). We show that our highest extinction sub-samples have the same column
density patterns as the SMC and a DLA observed at higher resolution.

\renewcommand{\thefigure}{\arabic{figure}\alph{subfigure}}
\setcounter{subfigure}{1}
\begin{figure}
\includegraphics[width=3.45in,height=4.5in]{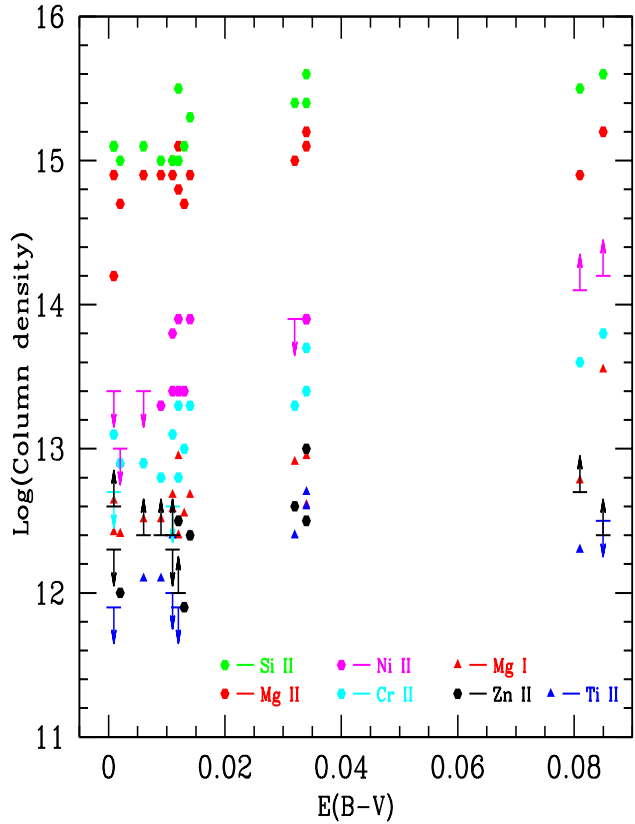}
\caption{Column densities (dex) for Mg II, Mg I, Si II, Ti II, Cr II, Ni
II and Zn II for all samples from Table A4 as a function of their derived
$E(B-V)$, assuming the SMC extinction curve applies. Mean values are
plotted for Mg I. Upper and lower limits are indicated by error bars
pointing down and up respectively. Upper limits are from non-detections,
lower limits are from indications of saturation. While the derived
reddening differs by a factor of 80, the column densities do not increase
that much from the left to the right. Variations at a given $E(B-V)$ are
only a factor of three or less. This figure and Figure 7b show that
sub-samples defined with different selection criterion but having similar
values of $E(B-V)$ yield similar column densities, a fact we use to
justify presenting representative samples in Tables 6, 7 and 8.}
\end{figure}
\addtocounter{figure}{-1}
\addtocounter{subfigure}{1}
\begin{figure}
\includegraphics[width=3.45in,height=4.5in]{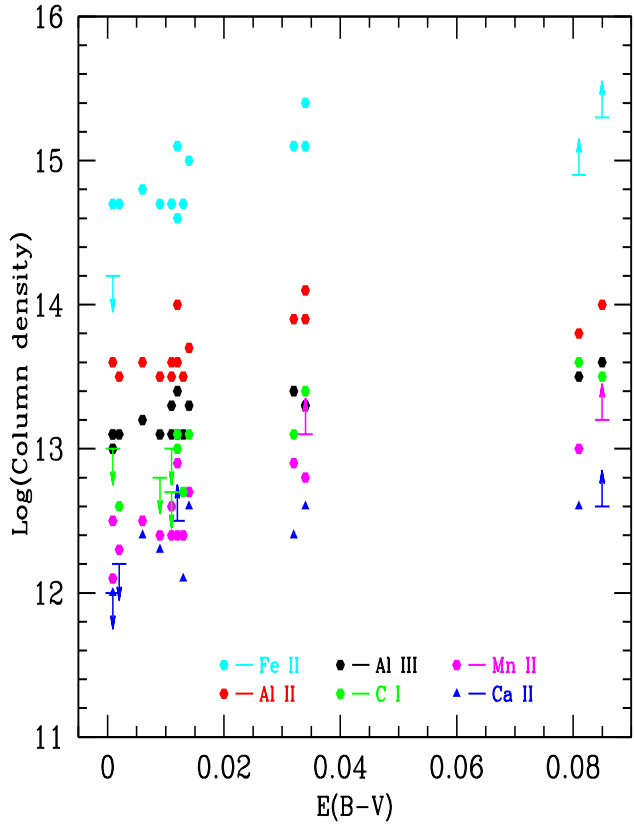}
\caption{Column densities (dex) for C I, Al II, Al III, Ca II, Mn II and
Fe II for all samples from Table A4 as a function of their derived
$E(B-V)$, assuming the SMC extinction curve applies. Upper and lower
limits are indicated by error bars pointing down and up respectively.
Upper limits are from non-detections, lower limits are from indications
of saturation.}
\end{figure}
\renewcommand{\thefigure}{\arabic{figure}}
\section{Abundances and depletions}
Using the column densities discussed in section 7 (Table A4), and the
values of H I (also given in Table A4, derived assuming an SMC relationship
between $E(B-V)$ and H I), we proceed to derive abundances, relative to H I,
and depletions, relative to zinc. The justification for this assumption is that
extinction curves for all of our samples are of the SMC type. This result may
well be a consequence of a selection effect associated with the observability
of QSOALS. There may be other samples of QSOALS for which the assumption is not
valid. For instance, for the very few systems that are known to have the 2175
{\AA} bump in their extinction curves, the assumption is probably not valid.

A direct determination of H I for our samples would be preferable, but as we
discuss in section 8.3, this was not possible.  We have already established on
the basis of column densities that our most reddened systems can be identified
with DLAs at low redshift. Here we will show that abundances for sub-samples of
smaller extinction (sub-DLAs) are close to solar and could only be higher (in
case the dust-to-gas ratio is higher than the SMC value).
\subsection{Abundances}
The abundances for four sub-samples of Table A4, covering $E(B-V)$ from
0.002 to 0.085, derived as explained above, are listed in Table 7. The
errors quoted are based on propagating the errors on the equivalent widths in
Table A3. The column densities plotted in Figure 7 show that these four
systems are a fair sample over this range and not special in any particular
way.

In Table 7, it is evident that the abundances, based on Table A4 and the SMC
dust-to-H I relationship, generally decrease from the lowest extinction
sub-sample (24) to the highest extinction sub-sample (27). The decrease is a
factor of  10 to 20. The two elements Ca II and C I are trace elements and are not
interesting for this particular discussion.  Comparing the lowest extinction
sub-sample 24 (column 3) with solar system abundances (column 2), we see
that Zn has cosmic abundances in sub-sample 24. The other elements 
are three to ten times lower than the solar system values, Si being the least
deficient and Mn being the most deficient.  On the high extinction side, we
make an empirical connection to the SMC via the star Sk 155 (Welty  et al.
2001). The color excesses, H I column density and abundances for sample 27
are consistent with the values of the Sk 155 and both have an SMC extinction
curve. 

Thus, the pattern from our composite spectra is consistent with lower total
abundances in the more reddened QSOALSs, in reference to solar system
values, similar to the local, SMC dwarf galaxy with sub-solar abundances.
Titanium stands out as a possible key indicator of different dust depletion
patterns that has not yet been fully utilized in studies of QSOALSs. It
has been detected in a few DLAs (Dessauges-Zavadsky, Prochaska \& D'Odorico
2002; Ledoux, Bergeron \& Petitjean 2002). Note that we have not made
ionization corrections, which are formally required except for Ti II, which
has the same ionization potential as neutral hydrogen. While we argue on the
basis of the ratio N$_{\rm Al\; III}$ to N$_{\rm Al\; II}$ that such
corrections are unimportant, the sense of the corrections would be to enhance
the relative abundance of Ti noted above, leaving the Ti abundances
unchanged while lowering the abundances of Zn II, Cr II, Ni II, etc.

\subsection{Relative abundances (apparent depletions)} 
In Table 8, we compare relative abundances, [X/Zn], a parameter most
universally available, even when N$_{\rm H\;I}$ is not obtainable
\footnote[1]{We note that the comparison should be with N$_{\rm H}$=N$_{\rm H
\;I}$ + N$_{\rm H_2}$ + N$_{\rm H\; II}$, but the sum is essentially the same
as N$_{\rm H\; I}$ in all known QSOALSs and in the SMC. Ignoring N$_{\rm  H\;
II}$ amounts to our assumption that there are no ionization corrections for the
dominant first ions.}. The solar system values are all zero, in this
notation. Also listed are values for the ``hot", ``warm" and ``cold" gas in the Milky
Way, inferred by Jenkins and Savage (1996) and updated by Welty et al. (1999b);
values for the four samples used in Table 7; values for the SMC star Sk 155
(Welty et al. 2001; Welty et al. 2005, in preparation); and the average values
for five, low redshift DLAs from Khare et al. (2004). Nothing in this table
depends on the assumption of an SMC dust to gas ratio for the sub-samples of
this paper. 

The presence of various types of incompleteness: uncertainties in
Mg II and Al II, limits for sample 27 and species that are not dominant
(C I and Ca II) preclude definite conclusions. However, it seems that
relative abundances with respect to Zn for sub-samples 24, 26, and 8
($E(B-V)<\sim$0.032) are most similar to those found in MW
sightlines that contain gas with ``halo" or ``warm" gas depletion
patterns (columns 2 and 3 of Table 8). Only for sub-sample 27
($E(B-V)\sim$0.085), Sk 155 and the DLA sample do the relative
abundances fit into the warm gas category for the MW. In none of these
sub-samples is there any evidence for ``cold" clouds, though there well
could be ``cold components" mixed into the composite equivalent widths
from which we derive the column densities. The similarities in these
last three columns of Table 8 confirm our identification of the most
reddened sub-samples with the general class of QSOALSs known as DLAs,
as well as our identification of sub-sample 24 with the standard
category of sub-DLAs (N$_{\rm H\; I} < 2 \times 10^{20}$ cm$^{-2}$).
Typical values of H I in the MW halo sightlines are N$_{\rm H\; I}\sim
10^{19}-10^{20}$ cm$^{-2}$.

The empirical comparison of relative abundances given in the last
paragraph does not distinguish the effects of dust depletion from those of
nucleosynthetic signatures. In regard to the latter, the ratio of silicon to
iron is the classic pattern found in old Galactic stars (Lauroesch et al. 1996
and references therein) and the value of Mn compared to Fe could be partly due
to the odd-even effect. However, we do not know that grains are the same
everywhere: Welty et al. (2001) show that the grains in SMC, toward Sk 155,
seem to be mainly Fe and do not contain much Si. The ratio [Si/Fe]then mimicks
the value found in old halo stars of the Milky Way, but two are not related.

While it is clear from Table 7 that the abundances of first ions
decrease with increasing $E(B-V)$, the corresponding trend is not evident in
Table 8, for relative abundances with respect to Zn. In particular, the
relative abundances of Mg II, Al II, Si II, Cr II, Mn II, Fe II and Ni II
are lower in the sample with the least extinction (sample 24). It will be
interesting to look for this trend in high resolution studies of large
numbers of low $z_{abs}$, sub-DLA systems.

We return now to the issue of our assumption of an SMC ratio of $E(B-V)$ to H I
column density. If, as might be indicated by the relative abundance commented
on in the previous paragraph, our sub-DLA sample includes some MW extinction
curves in the averaged QSOALS, then there might be less H I per unit of
$E(B-V)$. In this case, our abundances as given in Table 7 are underestimates.
The dust to gas ratio can be affected by both grain size and grain content.
Clayton et al. (2000) show that there are low reddening sightlines in the MW
that have extinction curves closer to the SMC curves as opposed to the high
reddening regions of the MW for which extinction curves are normally measured.
So as long as the dust-to-gas ratio in the QSOALS is not lower than the SMC
value our conclusion about zinc being solar in sub-DLAs appears to be robust.

We conclude that properties that can be known from selected QSOALSs from
studies of individual QSOs observed at high resolution, are generic to a
large sample of QSOALSs with color excesses near 0.1. A much larger
fraction of QSOALSs than previously thought may have solar abundances,
judging from the large sample size of the lowest extinction sub-sample
(number 24) in Table 8.

\begin{table}
\tiny
\caption{Abundances$^a$ for selected sub-samples}
\begin{tabular}
{|c|c|c|c|c|c|c|}
\hline
\multicolumn{2}{|c|}{Sub-sample Number}&24&26&8&27&--\\
\hline
\multicolumn{2}{|c|}{Number of systems} &698&97&251&48&--\\
\hline
\multicolumn{2}{|c|}{Criteria} W$_{\rm Mg\;II}$&All&$>$2.5
{\AA}&$<$2 {\AA}&$>$2.5 {\AA}&\\
\multicolumn{2}{|r|}{$\Delta(g-i)$}&$<$0.2& $<$0.2&All&$>$0.2&--\\
\hline
\multicolumn{2}{|c|}{Other}&&&&&Sk 155$^b$\\
\hline
\multicolumn{2}{|c|}{$E(B-V)$$^c$}&0.002&0.012&0.032&0.085&0.11\\
\hline
\hline
\multicolumn{2}{|c|}{Log(N$_{\rm H\; I}$)}&20.0&20.7&21.2&21.6&21.4\\
\hline
Species$^d$&Solar$^e$&\multicolumn{5}{|c|}{Log(N$_{\rm X}$/N$_{\rm H\; I}$)}\\
&System&&&&&\\
\hline
C (I) &-3.6 & -7.4 & -7.6 & -8.1 & -8.1 &-8.1 \\
Mg (II) & -4.4 & -5.1: & -5.6: & -6.2: & -6.4:&-5.2$^f$ \\
Al (II+III) &-5.5 & -6.4: & -6.6: & -7.3: &-7.8:&-7.3:\\ 
Si (II) &-4.5  &-5.0 & -5.2 & -5.8 & -6.1&-5.8 \\
Ca (II) & -5.7  & -7.8 & $>$-8.2 &-8.8&-9.0 & -8.8 \\
Ti (II) &   -7.1 & -8.1 & -8.2 & -8.7&-9.3& -8.7 \\
Cr (II)  & -6.4  & -7.1 & -7.4 & -7.9&$>$-8.0 & -7.9 \\
Mn (II) &-6.5   & -7.7 & -7.8 & -8.3&-8.6 & -8.3 \\
Fe (II) &-4.5  &-5.3 & -5.6 & -6.1&$>$-6.7 & -6.1 \\
Co (II) & -7.1  & $<$-7.1 & $<$-7.4 &$<$-8.1& $<$-8.2 & $<$-8.1  \\
Ni (II) &  -5.8  & (-6.6) & (-6.8) &$<$-7.3& $>$-7.5 & -7.3  \\
Zn (II) &-7.4   & -7.6 & -8.2 &-8.6& $>$-8.9 & -8.6 \\
\hline
\end{tabular}

\noindent $^a$ Abundances are formally better than 0.1-0.2 dex in
precision, assuming the derived values of $E(B-V)$ can be used to infer
N$_{\rm H\; I}$ for our samples as described in the text. Values followed
by colons are based on saturated lines (Mg II and Al II) and the
assumption that the well determined curve of growth of Fe II can be used
to infer N$_{\rm Mg\; II}$ and N$_{\rm Al\; II}$. Values in parentheses
are formally good values but are uncertain because of inconsistencies
with other measurements of lines of Ni II.\\ 
\noindent $^b$ Welty et al. (2001). $E(B-V)$ is the inferred value for
the star, corrected for 0.04 mag of extinction due to the MW. N$_{\rm H\;
I}$ was derived from a profile fit to the blended Lyman alpha line (from
the HST and FUSE data) with SMC and MW components. The value given is
only for the SMC, as are the abundances.\\
\noindent $^c$ colour excess for an SMC extinction curve as derived for
these sub-samples, as described in the text.\\
\noindent $^d$ The ion stage of the listed species from which the
abundance is derived is in parentheses. No ionization corrections are
made, though N$_{\rm Al\; II}$ and N$_{\rm Al\; III}$ are summed.
Ionization corrections are believed not to be important except for C I
and ca II
because N$_{\rm Al\; II}$ is greater than N$_{\rm Al\; III}$ in all of
our samples.\\
\noindent $^e$ Solar system abundances from Lodders. K. 2003, ApJ, 591, 1220\\ 
\noindent $^f$ The N$_{\rm Mg\;II}$ for Sk155 comes from measurements of
the Mg II $\lambda\lambda$1238, 1242, with $f$ value 0.0001. It is thus
much more accurate (formally), but lacking in wings comparable to the
2800 {\AA} doublet.
\end{table}
\begin{table*}
\centering
\begin{minipage}{140mm}
\caption{Comparison depletion in selected sub-samples with other gas}
\begin{tabular}
{|l|r|r|r|r|r|r|r|r|r|}
\hline
Gas&H$^a$&W$^a$&C$^a$&S24$^b$&S26$^b$&S8$^b$&S27$^b$&SMC$^c$&DLA$^d$\\
\hline
[Zn/H]&&&&-0.2&-0.6&-1.2&$>$-1.3&-0.9&-1.1\\
\hline
E(B-V)&&&&0.002&0.012&0.032&0.085&0.11&\\
\hline
\hline
Species&\multicolumn{9}{c}{[X/Zn]}\\
\hline
C&-0.2&-0.1&0.1&[-0.4]$^e$&[-0.6]&[-0.5]&[$>$-0.8]&---&---\\
Mg&-0.2&-0.4&-0.8&-0.5:&-0.2:&-0.6:&$<$-0.5:&0.0&---\\
Al&--&-0.9&-2.0&-0.7:&-0.3&-0.6:&$<$-0.8:&---&-0.9\\
Si&-0.2&-0.3&-0.9&-0.3&+0.1&-0.1&$<$-0.1&+0.1&-0.2\\
Ca&--&-1.8&-3.2&[$<$-2.9]&[$>$-1.7]&[-1.9]&[$<$-1.6]&-2.2&---\\
Ti&-0.6&-1.0&-2.4&-0.5&-0.3&-0.4&$<$-0.7&-0.9&-1.0\\
Cr&-0.5&-1.0&-1.7&-0.5&-0.2&-0.3&---&-0.8&-0.7\\
Mn&-0.6&-0.8&-0.9&-1.0&-0.3&-0.6&$<$-0.6&-1.0&-0.8\\
Fe&-0.5&-1.2&-1.8&-0.6&-0.3&-0.4&---&-0.9&-1.2\\
Co&---&-0.7&-1.7&$<$0&$<$+0.5&$<$+0.2&$<$+0.4&$<$-0.8&---\\
Ni&-0.5&-1.2&-1.8&-0.6&-0.2&$<$-0.3&----&-1.1&-1.2\\
Zn&0.0&0.0&0.0&0.0&0.0&0.0&0.0&0.0&0.0\\
\hline
\end{tabular}

$^a$ Representative values for halo, warm and cold Galactic interstellar medium
components, from ongoing work noted in the text.  These values are [X/H], taken
here to be [X/Zn] since Zn is solar and undepleted in the local interstellar
gas.\\ 
$^b$  These columns are the same sub-samples, in the same order, as
in Table 7.\\
$^c$  Values for the star Sk 155, as in Table 7.\\
$^d$  Depletions for five damped Lyman alpha systems between $z_{abs}$
of 0.8 and 1.5 (Khare et al. 2004).\\
$^e$  Values in brackets are for species for which there are large
ionization corrections, C I and Ca II. Errors for all values are as in
Table 7.\\
\end{minipage}
\end{table*}

\subsection{Using known DLA samples for determination of dust-to-gas ratio} We
initially considered making use of the H I column densities derived for a
sample of 197 Mg II absorbers by Rao et al. (2005) with HST UV spectroscopy,
for the analysis of abundances and depletions.  However, we found that their
sample did not have significant overlap with our high $E(B-V)$ sub-samples.
Specifically we divided their sample into N$_{\rm H\;I} > 10^{20}$ cm$^{-2}$
and N$_{\rm H\;I} < 10^{20}$ cm$^{-2}$ sub-samples and made matched
non-absorber comparison samples using the method described in Section 5.
Relative to the comparison samples, neither of their sub-samples showed any
signature of extinction, as both gave $E(B-V)$ slightly less than zero.
However, this result is consistent with the present work in that it is clearly
possible to select a sample with a range of W$_{\rm Mg\;II\lambda2796}$ that
has low $\Delta(g-i)$, as can be seen from Figures 3b and 5a. When Rao et al.
(2005) selected their sample for HST observation, they chose absorbers which
had a range of W$_{\rm Mg\;II\lambda2796}$ but also required the $g$ magnitudes
of the QSOs to be brighter than 19.1, thus ensuring that there would be
sufficient UV flux to obtain the needed signal-to-noise ratios to measure
N$_{\rm H\;I}$ values from the Ly$\alpha$ absorption lines.  By employing this
criterion, the most reddened QSOs may have been moved out of the their sample.
Any small extinction that may exist in their sample cannot be easily measured
with our technique for the small number of objects (15 and 17) in the two
sub-samples we can check here.

With regard to the Rao et al. (2005) sample, it should be noted that they do
find a relationship that can be used to predict the mean N$_{\rm H\;I}$
value as a function of W$_{\rm Mg\;II\lambda2796}$. Specifically, they find
that for W$_{\rm Mg\;II\lambda2796} < 0.6$\AA\ the mean N$_{\rm H\;I} \approx
1\times10^{19}$ cm$^{-2}$, but for W$_{\rm Mg\;II\lambda2796} > 0.6$\AA\ the
mean N$_{\rm H\;I} \approx 4\times10^{20}$ cm$^{-2}$. If we were to simply use
this relationship between N$_{\rm H\;I}$ and W$_{\rm Mg\;II\lambda2796}$ we
would find results similar to those described by Nestor et al. (2003) and
Turnshek et al. (2005), i.e., a trend of increasing mean metallicity with
increasing W$_{\rm Mg\;II\lambda2796}$, where increasing W$_{\rm
Mg\;II\lambda2796}$ is primarily an indication of the increasing velocity
spread of the absorbing gas. Since neither of these studies separated the
effects of extinction from those of the kinematics, the result can be thought
of as a separate effect from depletion, related to our comments in section 9.3
on the Routly-Spitzer effect. 
\section{Discussion}
\subsection{Extinction and uncertainties} It is clear that the extinction
derived here is due to intervening absorbers as it is obtained from the
comparison of QSOs with absorbers and those without absorbers along their
line of sight (see the top panels of Figures 2, 5a, 5b and 5c).
Additional support for this conclusion comes from the correlation of
absorber properties with the $E(B-V)$ in the following. (i) Extinction
depends on the strengths of absorption lines e.g. extinction depends on
the mean W$_{\rm Mg\;II\lambda2796}$. It is very small for QSOALS with
W$_{\rm Mg\;II\lambda2796}$ $<$1.53 {\AA} (sub-samples 2-4) and increases
with increase in W$_{\rm Mg\;II\lambda2796}$ beyond 1.53 {\AA}
(sub-samples 5, 6, 7 and also 8 and 23). (ii) The strengths of the
weakest lines, least affected by saturation, increase directly with the
derived values of $E(B-V)$ from the extinction curves (see Tables 2, 3, 5
and A3). (iii) In sub-samples defined by properties not expected to be
related to gas absorption lines, the average extinctions and the
absorption line strengths show little change from sub-sample to
sub-sample (see Table 4).

Consistency of the spectroscopic and photometric measurements of extinction
further supports the view that the extinction arises in the intervening
absorbers in our sample. Both the average $\Delta(g-i)$ and average
$E(B-V)_{(g-i)}$ are correlated tightly to the $E(B-V)$ values determined from
the composite spectra (the linear least squares fits are $<E(B-V)_{(g-i)}>$ =
0.98$\times E(B-V)$-0.002 (rms=0.006) and ${<\Delta(g-i)>\over 4.0}$ =
1.01$\times E(B-V)$ - 0.002 (rms=0.006)).  The average value of $\Delta(g-i)$
for the absorber sample is positive (0.013) while that for the non-absorber
sample is negative (-0.013). Also, a higher fraction of QSOs with absorbers
have large $\Delta(g-i)$ values as compared to the QSOs without absorbers. 

To quantify the degree of association of extinction with gas absorption,
notice that in the absorber sample, roughly 14\% of the systems have
$\Delta(g-i)>$0.2 as compared to 5\% of the systems of the non-absorber
sample.  It thus appears that for 65\% of such highly reddened QSOs in
our sample, intervening absorbers are the cause of the reddening. In
other words, QSOs with grade A absorption systems in their spectra appear
to be three times as likely to be reddened as those without any
absorption system in their spectra. 

As expected the reddening seems to be independent of $\beta$ (sub-samples
13 \& 14) which is consistent with all the systems with $\beta>0.01$
being of intervening origin in our BAL-free sample. The reddening is also
very similar for the sub-samples of bright and faint QSOs (sub-samples 11
\& 12). Surprisingly, the sub-samples selected on the basis of $z_{abs}$
(sub-samples 9 and 10) show very different values of $E(B-V)$; the
sub-sample consisting of lower $z_{abs}$ has half the value of
$E(B-V)$ as compared to the higher $z_{abs}$ sub-sample. This could in
principle, be due to the fact that we have only considered systems with
grade A in choosing our sample and it is possible that there are other
systems of lower grade as well as systems beyond the range of $z_{abs}$
considered here, along the lines of sight. 

However, the only systems not seen by us are those at $z_{abs} <$ 0.5,
which could only be picked up in Ca II. The numbers of these systems
would be the same in both high $z_{abs}$ and low $z_{abs}$ subsamples,
since all of our QSOs have redshifts above 1. They would also be the
same, statistically in our matched QSOs, so the effect of such systems
should not be seen in our extinction curves. As for lower grade systems,
we have estimated the effect of their presence from the redshift
dependence of d{\it N}/dz, the number of Mg II systems per unit redshift
interval, per line of sight, values calculated from the SDSS data base
(York, Van den Berk et al. 2005, in preparation) and find the effect to
be $\le$ 19\%. In reality, the effect is likely to be much smaller as our
d{\it N}/dz estimates show that the total number of systems with grades
A, B and C, having W$_{\rm Mg\;II\lambda2796}>1.5$ {\AA} (which are
likely to cause reddening) is expected to be less than $<150$ in our
sample of 809 QSOs if these QSOs were chosen randomly without any
knowledge about the presence of absorption systems in their spectra. We
have, however, chosen these QSOs for the presence of absorption systems
in their spectra and the probability of their having other absorption
systems is thus very small. The reason for the large difference between
the $E(B-V)$ values for the sub-samples with different $z_{abs}$ ranges
is thus not understood. 
\subsection{Comparisons with other searches for
2175 {\AA} feature and dust} We have established that extinction is from
the intervening absorption systems and have shown that for none of the
sub-samples does the extinction curves have a discernible bump at 2175
{\AA}, though up to 30\% of the lines of sight being reddened by MW type
of dust can not be ruled out.

There are less than ten known extinction curves for QSOALSs that show the
2175 {\AA} bump. The best case is AO 0235+164 (Junkkarinen et al. 2004).
Wang et al. (2004) detect the bump in three QSOs with $z_{em}$ above 1.6
(out of 6350 in the DR1 sample searched). One of these is included in our
sample (SDSS J144612.98+035154.1). The other two are in QSOs at higher
redshift than we include (SDSS J145907.19+002401.2 and SDSS
J012147.73+002718.7). The common object has the fourth highest value of
$\Delta(g-i)$ in our sample. The three QSOs with the highest values of
$\Delta(g-i)$ show no sign of the bump, under visual inspection
(J011117.36+142653.6, J144008.75+001630.4, J085042.24+515911.7). Wang et
al. (2004) review previous studies, including indications that some
lensed QSOs show the bump in the spectra of at least one of the multiple
images (but, see McGough et al. 2005).

Malhotra (1997) used a subset of spectra from Steidel and Sargent (1992) to
obtain an average extinction curve and to search for the 2175 {\AA} bump,
evidence for which showed up in the composite spectra. Among the sample are at
least two BAL QSOs (UM485 (SDSS J121549.8-003422.1) and SBS0932+524 (SDSS 
J093552.92+495314.3)) and several strong, associated absorption line
systems. Therefore, the sample is not selected in the same way as the
samples we use here. In addition, some of the spectra have strong Fe II
emission complexes.

None of the 19 spectra from Malhotra's sample which have been observed by
the SDSS, shows the 2175 {\AA} absorption feature in the SDSS spectra. We
did notice that in one of the BAL QSOs (UM485), the C IV and Si IV troughs
coincide with 2175 {\AA} in the rest frame of two, narrow Mg II
absorption systems. It is possible that these BAL troughs contribute to
the strength of the feature found in the composite spectra.  That BAL
QSOs may be contributing is also indicated by the fact that the composite
spectrum obtained by Malhotra is quite red, a common feature of BAL
spectra (Reichard et al. 2003b). Thus we believe that the strength of the
feature found by Malhotra is likely to be caused by artifacts of the QSOs
chosen, or by a few strong 2175 {\AA} features in a few of the spectra
(ones not yet found in SDSS data). As the SDSS survey is completed, a
larger number of matches with Malhotra's sample will be found and the
experiment can be repeated.  

Thus it appears that the carrier of the 2175 {\AA} bump is not ubiquitous
throughout the Universe, a fact that may go with the absence of heavy
depletion of heavy elements that we find here in our sub-samples. It may
be that the bump is not caused by carbonaceous material as is commonly
assumed (Draine 2003) but rather depends for its existence on heavy
depletion of elements such as Fe, Ti, Ca and/or other refractory species.

The ubiquity of the SMC extinction curve also suggests that grains in 
the Galaxy are less completely processed (relatively more large 
grains) than elsewhere in the Universe. Whether the difference has to 
do with selection effects in the QSOALS sample or represents global 
differences between galaxies is not clear. For instance, dense clouds 
sampled for MW extinction curves may be selectively avoided in 
QSO spectra and there are low density lines of sight in the MW 
with very weak bumps and very steep UV extinction curves (Clayton et 
al. 2000).

Ellison et al. (2005) have studied DLAs in a radio selected sample of
QSOs to overcome any dust bias that may be favoring the selection of
dust-free absorbers in the magnitude limited QSO studies. They obtain a
best fit value of $E(B-V)$ = 0.02(0.05) for SMC (MW) type dust which is
similar, and, in particular, not larger than, the values we find here.
Their dust-to-gas ratio is however, 30-50\% higher than that for the SMC.  
\subsection{The Routly Spitzer effect} In section 6.3, we defined two
sub-samples specifically to test for the Routly Spitzer effect (Section
6.3, sub-samples 15 and 16). Since the Fe II $\lambda$2382 line is weaker
than the Mg II $\lambda$2796 line, the release of Fe II from grains into
gas phase at high velocity should yield a high ratio of Fe II to Mg II
equivalent widths (using the noted lines) if the effect is important.
Dividing the sample in half (one half with ratios less than 0.58 and the
other with ratios greater than 0.58) should select for the effect and the
equivalent widths of other lines may reveal confirming evidence.
Comparing sub-sample 16, with the higher ratios, with the full
sample, there is a small effect of enhanced Ca II and Ti II (see Tables
A3 and A4). However, we have noted elsewhere that several lines seem to
be partaking in the same wide profiles noted for Mg II.
 Whether the higher velocity components in the
broad profiles are selectively enhanced cannot be discerned without
higher resolution spectra.

\subsection{Interpretation of trends in abundance variations} Our results
show that the gas-phase abundances decrease with increasing $E(B-V)$ for
systems with $E(B-V)>$ 0.002 (the smallest measurable value in our
sub-samples). Sub-sample 24, having 698 systems, thus shows the highest
abundances. This sub-sample has the smallest measurable $E(B-V)$ and
hence the smallest estimated average H I column density of 10$^{20}$
cm$^{-2}$ and is thus composed of sub-DLAs. Peroux et
al. (2003) have observed sub-DLAs having abundances higher than typical DLAs.
Super-solar abundance has been observed in two sub-DLAs (Pettini et al. 2000;
Khare et al. 2004). There are reasons to believe that the DLAs may not in fact,
be the progenitors of the present day disk galaxies as has been suggested
earlier (Wolfe et al. 1995) : \hfill\break (i) Deep imaging shows that DLAs are
often associated with dwarf galaxies.\hfill\break (ii) The rate of metallicity
evolution in DLAs is found to be very low and DLAs have been found to have
highly sub-solar abundances even at very low redshifts (Kulkarni et al. 2005a).
\hfill\break (iii) Star formation is usually associated with molecular clouds,
but DLAs have very low molecular fraction (Ledoux, Petitjean \& Srianand
2003)\hfill\break (iv) The matching of $\Omega_{\rm DLAs}$ at high redshifts
with $\Omega_{\rm stars}$ at present may not be correct (Peroux et al. 2005b)
or may just be a coincidence. Also, sub-DLAs may contribute a significant
fraction of $\Omega_{\rm H\;I}$ (Peroux et al.  2005b).\hfill\break  (v) Zwaan
et al. (2005) have shown that the properties and incidence rates of DLAs are in
good agreement with DLAs originating in galaxies of various morphological types
like the local galaxies.

On the other hand, Mg II systems, which are likely to be non-DLA, Lyman
limit systems (LLSs), are very often found to be associated with bright
spiral galaxies (Bergeron \& Boisse 1991; Churchill et al. 2005 and
references therein).  Near-solar abundances have been observed in one
LLS (Jenkins, et al. 2005).  Mg II systems also show evidence of rotation
(Churchill \& Vogt 2001). Thus it seems likely that non-DLA, Mg II
systems, and not the DLAs, are the progenitors of present day disk
galaxies. The lower H I columns in these systems could be a result of
higher star formation rates as compared to those in DLAs which seem to
have very low rates of star formation (Kulkarni et al. 2005b).

\subsection{Uncertainties in abundance measurements} In section 7 and 8
we have discussed several reasons for being cautious about interpreting
the abundances derived from the composite spectra. These include the
effects of saturation, which in some cases cannot be certain and the
effects of using the faintest QSOs, which have lower equivalent width
sensitivity and lead to slight overestimates of column densities. Other
uncertainties are associated with the possible presence of circum-QSO
extinction without detectable absorption lines and with the assumption of
an SMC dust to gas ratio. Here we discuss these reasons which could
affect the abundance measurements for some sub-samples. 
\subsubsection{The effect of dust around the QSO, with no absorption
lines} First consider the sub-samples based on high $\Delta(g-i)$ values.
These sub-samples are particularly chosen to contain reddened
(sub-samples 25 and 27) QSOs. No such criterion is applied for the
corresponding non-absorber sub-samples. It is of course likely, as is
clear from the above discussion, that the high values of $\Delta(g-i)$
are caused by the presence of absorbers, however, some of the high
$\Delta(g-i)$ values could have other origins. These sub-samples are,
therefore, likely to be contaminated (up to 35\%) by systems which happen
to have high $\Delta(g-i)$ values due to reasons unrelated to the
absorbers. Therefore, even though 65\% of the systems in these
sub-samples may have high dust abundance in the absorbers, the remaining
35\% may not. Thus, though the $E(B-V)$ for these sub-samples will be
correct, the equivalent widths will be diluted (assuming for the moment,
that high amount of dust in the absorbers corresponds to high metallicity
and large equivalent widths) due to the presence of these stray systems.
This will result in getting lower calculated abundances for such
sub-samples. Since our abundances (except for Al II and Mg II) are based
on weak lines, this effect is of the order of 30\% (in the sense of an
underestimate) for unsaturated lines, but cannot be evaluated for
saturated lines (denoted in tables by giving lower limits). For the worst
cases, sub-samples 25 and 27, there are enough reliable determinations
that the error cannot be large, providing the relative abundances of the
heavy elements do not change drastically at high extinction (see Figure
7). The effect is similar in amplitude, but opposite in sign to the
effect of including faint QSOs in the composite spectra, as noted at the
beginning of section 9.5 and in section 7.2.2.
\subsubsection{The assumption of an SMC dust-to-gas ratio} The largest
uncertainties in abundances have to do with the derivation of N$_{\rm H\;
I}$ from $E(B-V)$ using the SMC conversion factor. Since the SMC is the
extreme curve in all cases known and since the less steep extinction
curves (such as the standard MW curve) have lower values of N$_{\rm
H\;I}$ for unit $E(B-V)$, the existence of some non-SMC curves amidst our
sample overestimates the value of N$_{\rm H\; I}$. The true, average
N$_{\rm H\; I}$ would be lower and the resulting abundances would be
higher than we derive. If anything, our abundances could be
underestimates, though not by too much.

As argued earlier, there is little evidence for MW extinction curves in
our sample. Furthermore, Tables 6 and 7 show that sub-sample 25 (one of
our two most reddened) has column densities and abundances (derived by
using our technique) that can be related to the classic case of damped
Lyman alpha absorbers, for which the abundances are overwhelmingly
sub-solar (see section 1). 

Hence, our conclusion that there is a real difference between the low
extinction sightlines and our highest extinction sightlines: there are
solar abundance sightlines (in Zn II) in our lightly reddened, largest
sub-samples. The relative depletion of Fe and other refractories
resembles the halo or warm clouds of the MW, in any case, indicating that
the Zn effect is due to overall metallicity, not to depletion differences
on grains in the different sub-samples.

In the above analysis we have assumed a constant dust-to-gas ratio for
determining the abundances, and, based on our result of finding
predominantly SMC extinction curves, used the dust-to-gas ratio for the
SMC. An examination of all relevant data on the SMC reveals little
evidence of wide spread variations in the dust-to-gas ratio for the SMC
(or, for that matter, in the LMC or MW), though there are notable
exceptions. For 40 known cases in the SMC, for instance, the rms scatter
in the dust-to-gas ratio is 0.4 dex. If such differences existed between
a large sample of galaxies with SMC curves, they would average out in our
composite spectra. It should be noted that the stars used in such studies
generally have higher extinction than our samples with the lowest color
excesses and that some of the variation for the SMC comes from
uncertainties in the Galactic foreground extinction. As already noted,
low reddening sightlines within our Galaxy have curves more similar to
the SMC curves than to the MW curves derived for higher amounts of
extinction (Clayton et al. 2000). This variation is consistent with our
assumptions here.

In the analysis presented in this paper, we have assumed the SMC $E(B-V)$
- N$_{\rm H\;I}$ relationship. However, it should be kept in mind that if
one does not adopt the SMC $E(B-V)$ - $N_{\rm H\;I}$ relation, but
instead takes an approach similar to Nestor et al. (2003) and Turnshek et
al (2005), i.e., inferring mean N$_{\rm H\;I}$ from W$_{\rm
Mg\;II\lambda2796}$, completely different abundance trends are inferred.
Of course, a variety of trends may be present, representative of
different sub-populations of absorbers as detailed in section 8 and,
below, in section 9.6. Future work with the full SDSS sample may reveal
these.
\subsection{Nature of the defining Mg II doublet}  
It is believed that large values of W$_{\rm Mg\;II\lambda2796}$ are
caused by high velocity gas. This is also evident from the line
profiles in our composite spectra for various sub-samples. In
particular, the strongest Mg II lines in sample 7 (2.52 $<$ W$_{\rm
Mg\;II\lambda2796} <$ 5.0 {\AA}) have average widths of 450 km
s$^{-1}$. The correction for the instrumental profile, which is
typically has FWHM of about 170 km s$^{-1}$, is negligible.  In all
probability, some of that width is represented by components that
are saturated and some by profiles in the wings that are not
saturated, as seen in the Mg II profile of Peroux et al. (2005a),
for instance, or as is common in SMC and LMC sightlines.

The widths of lines of Mg I, Fe II and to a lesser extent, Mn II, grow
with that of lines of Mg II and it is clear from the line profiles that
the growth is in added equivalent width over a wider velocity range.
This fact explains why the absorption lines seldom saturate in the
sub-samples we measured and why the Mg I $\lambda$2852 line is not
saturated, despite its large equivalent width, in most of the
sub-samples measured and presented in Table 2, 3, 4, 5 and A4. 
\section{Conclusions}
We have studied a sample of 809 absorption systems, in the spectra of
SDSS non-BAL QSOs, with W$_{\rm Mg\;II\lambda2796}>$0.3 {\AA}, 1$ \le
z_{abs}<$1.86 and $\beta>$0.01 to determine the average extinction and
the average strengths of absorption lines in QSOALSs. 

The extinction curves, obtained by comparing the geometric mean average
spectra of the QSOs in 27 sub-samples of absorbers (constructed on the
basis of the absorber as well as QSO properties) with those of matching
samples of QSOs without any absorbers in their spectra, show clear
evidence of extinction by the intervening absorbers. The absorber rest
frame colour excess $E(B-V)$ for the sub-samples varies from $<$0.001 to
0.085. The extinction curves uniformly share the main properties of the
typical SMC extinction curve: no 2175 {\AA} bump and rising extinction
below 2200 {\AA}. An average MW curve, similar to those of local
interstellar clouds, is excluded. A mix of up to 30\% of MW
curves cannot, however, be ruled out, but the MW curves are not a
dominant contributor.

Adapting the SMC extinction curve, the observer frame colours excesses
($\Delta(g-i)$) were  used, with the absorption redshift, to derive
average absorber frame colour excesses, ($<E(B-V)_{\Delta(g-i)}>$), which
are consistent with the values of $E(B-V)$ derived from the extinction
curves. 

While about 14\% of the absorber sample has high values of $\Delta(g-i)$
($>$0.2), 5\% of the sample of QSOs with no absorbers also shows such
high values. These high values, for the non-absorber QSOs (and some of
the absorber QSOs as well) could be caused by strong absorbers with
$z_{abs}<$0.5, where SDSS can not detect Mg II doublets or else, there is
a class of QSOs with anomalously red continua or there are regions with
large amounts of dust associated with gas that is too hot to produce the
near UV absorption lines.  

The average strengths of absorption lines between 1800 and 3300 {\AA} in
the absorber rest frame, are found to increase as the value of W$_{\rm
Mg\;II\lambda2796}$ increases. While Mg II line strengths are
generally correlated with reddening, there is large scatter: in
particular, of 145 systems with W$_{\rm Mg\; II\lambda 2796}$ above 2.5
{\AA} in rest frame equivalent width, 2/3 show little reddening ($\sim$
0.01 in $E(B-V)$, on average), while the reminder have an average
$E(B-V)$ that is eight times higher.

The column densities for Si II, Ca II, Ti II, Cr II, Mn II, Fe II, Ni II
and Zn II show a strong trend, increasing slowly with our derived values
of $E(B-V)$. If we adopt the SMC, $E(B-V)$ - N$_{\rm H\; I}$ relation to
derive the mean N$_{\rm H\;I}$ for our sub-samples, the abundances
derived from the average column densities and average N$_{\rm H\;I}$ show
a consistent trend of lower abundances with increasing $E(B-V)$. Our
largest sub-sample, with $E(B-V)=0.002$ and N$_{\rm H\;I} \sim 10^{20}$
cm$^{-2}$, has near solar zinc and titanium. Systems in this sub-sample
(which contains 85\% of our full sample) are thus sub-DLAs or Lyman-limit
systems. For samples with $E(B-V)\sim$0.08, we find gas depletions with
respect to Zn of up to 10 compared to solar abundances. Depletions of Fe,
Mn and Cr measured with respect to Zn are similar to those for DLAs as
published in the literature. The depletion patterns are similar to those
found in the halo gas in the MW.

In general, the ionization corrections appear to be small. The ratio of
the column densities of Al III to Al II is smaller than 0.5. The Al III
does not seem to be a part of the Al II gas and C IV appears to be
separated from both Al II and Al III.

Typical QSOALS have wide blends of velocity components in Mg II
$\lambda\lambda$2796, 2802, from below our instrumental width of 170 km
s$^{-1}$ up to 450 km s$^{-1}$ or more, independent of extinction. These
wide systems contribute to the apparent fact that W$_{\rm Mg\; I\lambda
2852}$ is near the linear portion of the curve of growth even for
equivalent width $\sim$ 500 m{\AA}. Many elements seem to participate in
these wide flows and the effects of gas depletion seem to be selective,
with Ti mildly depleted amounts in low reddening samples and highly
depleted in samples of modest reddening. Since Ti is not yet well
observed in studies of individual systems, this effect needs to be
explored in more detail.

Studies of systems with low but well defined N$_{\rm H\; I}$ are needed,
using high $S/N$ and high resolution, to determine what fraction of
QSOALS have solar or greater abundances of Zn.
\section*{acknowledgment}
DGY acknowledges student support from Kavli Institute of Cosmological
Physics and the Department of Astronomy and Astrophysics at the
University of Chicago. PK acknowledges support from the University of
South Carolina Research Foundation, travel support from the Department of
Astronomy \& Astrophysics, University of Chicago and use of computer and
Internet facilities at the Institute of Physics, Bhubaneswar. VPK
acknowledges support from the NSF grant AST-0206197. DVB and DPS are
supported by NSF grant AST-0307582.
We thank an anonymous referee for a quick and very helpful report
which improved the presentation of the paper considerably.

Funding for the creation and distribution of the SDSS Archive has 
been provided by the Alfred P. Sloan Foundation, the Participating 
Institutions, the National Aeronautics and Space Administration, the 
National Science Foundation, the U.S. Department of Energy, the 
Japanese Monbukagakusho, and the Max Planck Society. The SDSS Web 
site is http://www.sdss.org/.

The SDSS is managed by the Astrophysical Research Consortium (ARC) 
for the Participating Institutions. The Participating Institutions 
are The University of Chicago, Fermilab, the Institute for Advanced 
Study, the Japan Participation Group, The Johns Hopkins University, 
the Korean Scientist Group, Los Alamos National Laboratory, the 
Max-Planck-Institute for Astronomy (MPIA), the Max-Planck-Institute 
for Astrophysics (MPA), New Mexico State University, University of 
Pittsburgh, University of Portsmouth, Princeton University, the 
United States Naval Observatory, and the University of Washington.\\

\newpage
\appendix
\section{Figures and tables}
In this appendix, we collect various figures and tables that support 
the main text, but are illustrative in nature, or  not needed in full 
detail in the text but contain important information derived in the 
course of this study.

Figure A1 gives the composite spectrum of the full sample of 809 
absorbers, but with an expanded relative intensity scale compared to 
Figure 1, to show the weakest absorption lines more clearly.

Figure A2 shows the difference in $i$ magnitude and in QSO emission 
line redshift for the pairs of QSOs matched in the derivation of 
extinction curves. The details are discussed in the text (section 5).

\begin{figure*}
\includegraphics[width=7.00in,height=8.8in]{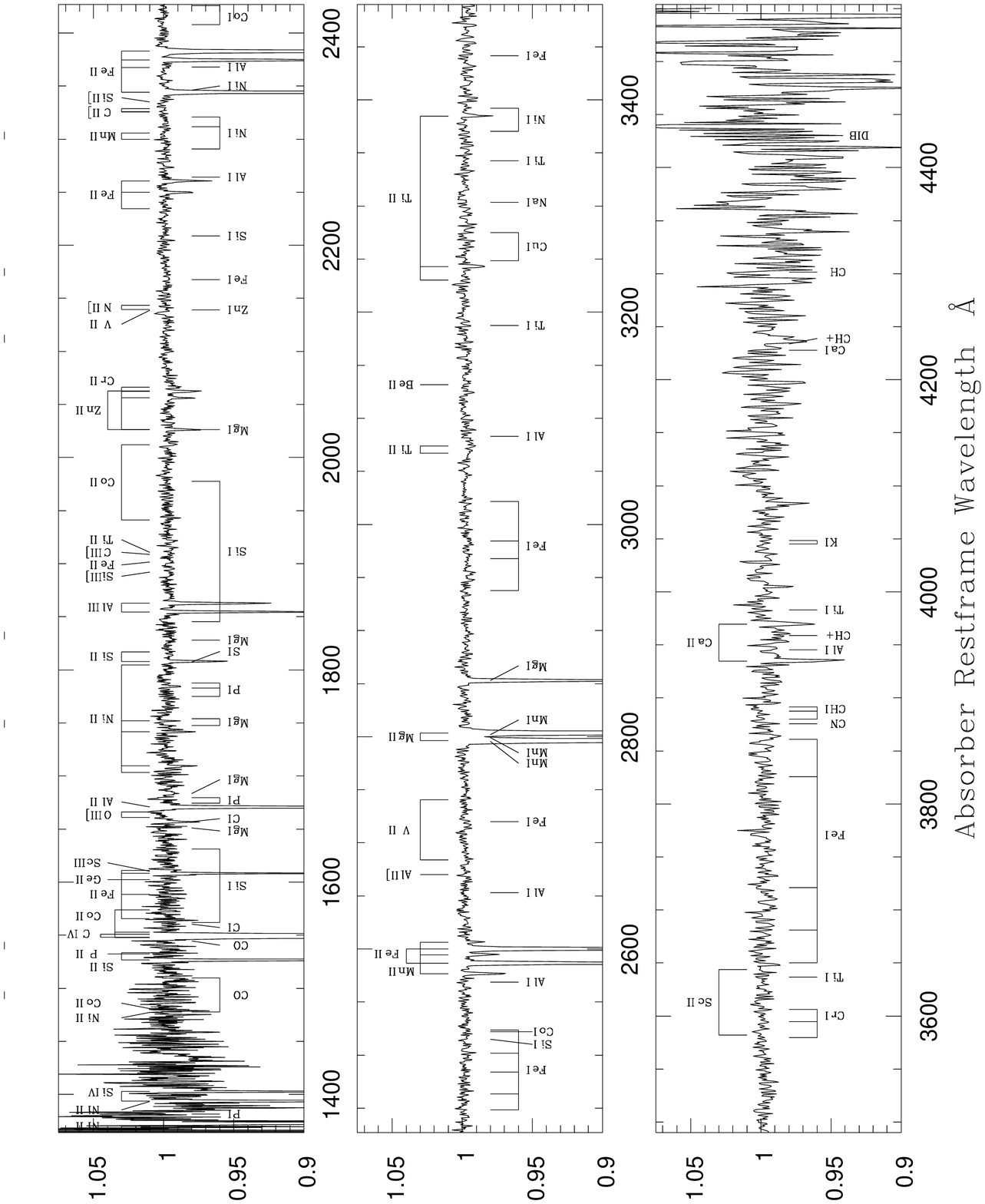}
\caption{Composite spectra of the full sample; same as figure 1 but the y
axis is expanded to show weak lines}
\end{figure*}
Table A1 contains the list of QSOs used in this study. In both Table A1a
and A1b, the data on absorber QSOs is on the left and that for the
matching (unabsorbed) QSO is on the right. Use of the pairs is discussed
in section 5. Table A1a, includes for QSOs with class A absorbers, the
plate, fiber and mean Julian date (MJD) of the recording of the spectrum
for each object; the QSO coordinates,  in decimal degrees; the emission
and absorption redshifts; the $i$ magnitude; and the relative velocity of
the absorber with respect to the background QSO. For the matched QSO, the
listed items are the plate, fiber, MJD, emission redshift, $i$ magnitude
and the match radius (in units of Figure A2). SDSS data for all objects
can be found using this information, on the website, http://www.sdss.org.
Any additional information can be obtained from DGY. Note that the plate
and fiber are unique only when the MJD is specified: the same QSO can be
observed on multiple plates and the same plate can be used on different
dates.

Table A1b includes data on the absorbers in the QSOs of Table A1a, and on
the observer frame reddening of the absorber and non-absorber
$\Delta(g-i)$, defined in section 1.3. The columns include, for the QSO
with the absorber; the plate; fiber; MJD; absorption line redshift; the
value of $E(B-V)_{(g-i)}$ inferred from the observer frame
reddening and redshift, according to the prescription given in section 5
($E(B-V)_{(g-i)}$ = $\Delta (g-i)$(1+$z_{abs})^{-1.2}$/1.506); and the
rest frame equivalent widths of 5 absorption lines listed in the
footnotes. For the non-absorber, matched QSO, we list plate, fiber, MJD
and observer frame reddening.

Table A2 gives a finding list of absorption lines that may be found 
in spectra such as ours from intervening QSOALSs.  Listed are the 
vacuum wavelength, the species, the oscillator strength, and the 
reference from which the data are drawn. The table includes only 
lines between 1370 and 4430 {\AA}, and includes all ions of elements that 
can be observed, the main diatomic molecules and the main diffuse 
interstellar band (4430 {\AA}) that fall in our spectral region. The lines 
included are chosen to include all lines ever detected in QSOALSs as 
well as additional weak lines that could be used, in much better 
spectra, to confirm single detected lines or to obtain lines on the 
linear portion of the curve of growth (e.g., Si II] $\lambda$2335.123).

The main numerical results of our study are listed in Table A3, as the
equivalent widths of the weakest absorption lines, useful for determining
column densities. Sixteen samples of the 27 described in section 6 are
included, chosen to cover the full range of $E(B-V)$ values we derive and
to provide redundancy in sub-samples with the same reddening drawn from
different selection criteria. Four rows across the top include sample
number (see Table 1), the number of absorption line systems in each
sample, the criteria used to select absorbers in the sub-sample and the
derived colour excess for an SMC extinction curve (section 5, Figure 2
and Table 1). The vacuum wavelengths, the species and the oscillator
strengths are given on the left of page 1 of Table A3 and the wavelengths
and species are redundantly listed on the left side of page two of Table
A3, for ease of use. Equivalent widths are given in milli Angstroms.
Errors are derived from the empirical noise in the continuum adjacent to
the lines and take into account the width of the lines. Since the line
widths are comparable to the resolution element and the
spectra are normalized, using principal components for each QSO (section
4), no allowance is made for errors in drawing the continuum. Errors are
1 $\sigma$, upper limits are 3 $\sigma$. Strong lines not found in
Table A3 may be found, for some systems, in Tables 2, 3, 4 and 5.

Table A4 contains the column densities derived using the weak lines in
Table A3 or, in the case of Al II and Mg II, from the strong lines that
are available with saturation corrections derived from the empirical
curves of growth (Figure 6). The column densities tabulated here are
plotted as a function of $E(B-V)$ in Figure 7 of the text.

The banner rows include the same information as listed for Table A3, with
the addition of the column density of N$_{\rm H\; I}$,  as derived from
the relation between $E(B-V)$ and H I for the SMC. In the second part of
the table, the species are redundantly listed down the left hand side.

For each of the sixteen samples, the column densities and upper limits
are given in logarithmic form, based in most cases on assuming that the
weakest line of a species is on the linear portion of the curve of
growth, for cases where the line ratios justify such an assumption. The
errors are derived in different ways for different species. For weak
lines and limits, the errors come directly from conversion of the
equivalent width errors or limits in Table A3. The errors are indicated by
letters a,b,c etc, each letter indicating the logarithmic error, 0-0.049,
0.05-0.099, etc. For Mg II and Al II, the additive correction derived
from the appropriate Fe II curve of growth (Figure 6) is given in
brackets. Parentheses, for Ni II, indicate that there is some uncertainty
beyond that listed because different transitions observed do not give
consistent values. When the assumption of the weakest line being on the
linear portion of the curve of growth is not justified, the entries are
given as lower limits (or listed with a colon, if the situation is
uncertain). 
\begin{figure*}
\includegraphics[width=5.15in,height=5.2in]{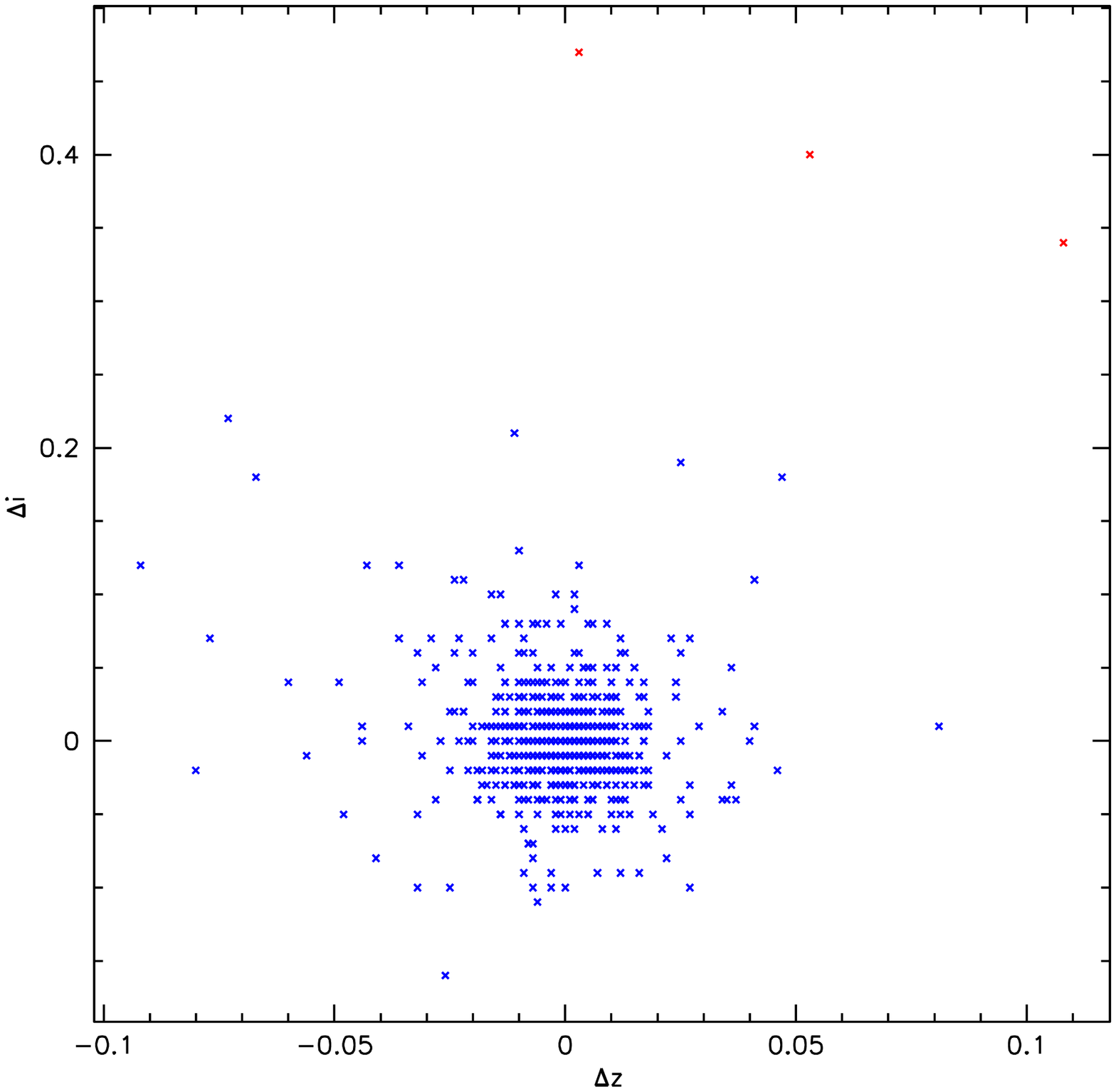}
\caption{Plot between difference in $z_{em}$ and $i$ magnitude of the
absorber QSOs and corresponding non-absorber QSOs. Points marked in red
are excluded from the sample as the differences are large.}
\end{figure*}
\begin{table*}
\begin{minipage}{140mm}
\rotatebox{90}{
\footnotesize
\begin{tabular}
{|llllrllllllllll|}
\multicolumn{15}{l}{Table A1a: Full sample of the absorber and non-absorber QSOs}\\
\hline
\multicolumn{9}{|c|}{Absorber sample}&\multicolumn{6}{|c|}{Non-absorber
sample}\\
\hline
Plate&Fiber&MJD&RA&Dec&$z_{em}$&$z_{abs}$&$i$ mag&
$\beta$&\multicolumn{1}{|c}{Plate}&Fiber&MJD&$z_{em}$&$i$ mag&Match \\
&&&&&&&&&\multicolumn{1}{|c}{}&&&&&radius\\ 
\hline
0267& 264& 51608& 147.554270&  -0.477538& 1.894& 1.6734&  19.440&  0.079&0386& 610&51788 &  1.886& 19.440& 0.0364 \\
0267& 287& 51608& 146.933580&  -1.245374& 1.363& 1.2122&  19.460&  0.066&0591& 018& 52022&  1.369& 19.480& 0.0430 \\
0267& 294& 51608& 147.139390&  -0.850967& 1.604& 1.0103&  18.550&  0.253&0468& 356& 51912&  1.607& 18.570& 0.0359 \\
0268& 059& 51633& 149.909530&  -0.583584& 1.875& 1.5983&  18.870&  0.101&0619& 377& 52056&  1.869& 18.860& 0.0319 \\
0269& 091& 51910& 151.131670&  -0.821981& 1.637& 1.3359&  18.380&  0.121&0413& 038& 51929&  1.628& 18.350& 0.0644 \\
0269& 613& 51910& 151.814730&   0.716218& 1.676& 1.0377&  19.010&  0.266&0404& 554& 51812&  1.679& 19.020& 0.0215 \\
0270& 213& 51909& 152.639380&  -0.790152& 1.667& 1.3270&  18.010&  0.136&0296& 417& 51984&  1.636& 18.000& 0.1418 \\
0270& 297& 51909& 151.983930&  -1.220578& 1.943& 1.5603&  19.880&  0.138&0549& 238& 51981&  1.930& 19.880& 0.0591 \\
0270& 462& 51909& 152.575820&   0.064227& 1.401& 1.2649&  18.420&  0.058&0552& 153& 51992&  1.416& 18.430& 0.0702 \\
0271& 154& 51883& 154.762230&  -0.340854& 1.477& 1.2240&  18.930&  0.107&0430& 072& 51877&  1.481& 18.910& 0.0379 \\
0273& 197& 51957& 157.951720&  -0.602809& 1.257& 1.0760&  18.860&  0.083&0283& 560& 51959&  1.257& 18.850& 0.0166 \\
0273& 494& 51957& 158.163390&   0.065052& 1.189& 1.0170&  19.260&  0.082&0351& 480& 51780&  1.197& 19.270& 0.0400 \\
0273& 496& 51957& 158.001160&   0.574553& 1.279& 1.1936&  19.350&  0.038&0650& 185& 52143&  1.278& 19.350& 0.0045 \\
0274& 357& 51913& 159.435190&   0.469256& 1.733& 1.4243&  18.150&  0.119&0533& 029& 51994&  1.717& 18.220& 0.1371 \\
0275& 357& 51910& 160.163580&   0.226905& 1.234& 1.0112&  19.060&  0.105&0552& 232& 51992&  1.234& 19.050& 0.0166 \\
0276& 068& 51909& 164.083610&  -0.147638& 1.439& 1.2846&  18.340&  0.065&0496& 338& 51988&  1.414& 18.360& 0.1184 \\
0276& 579& 51909& 163.833330&   0.966411& 1.817& 1.0785&  19.090&  0.295&0315& 598& 51663&  1.810& 19.100& 0.0359 \\
0276& 579& 51909& 163.833330&   0.966411& 1.817& 1.7858&  19.090&  0.011&0315& 598& 51663&  1.810& 19.100& 0.0359 \\
0277& 059& 51908& 165.936700&  -0.850738& 1.723& 1.6405&  18.800&  0.031&0484& 119& 51907&  1.726& 18.800& 0.0136 \\
0277& 246& 51908& 164.707500&  -1.167496& 1.725& 1.1623&  19.230&  0.227&0463& 076& 51908&  1.743& 19.250& 0.0883 \\
0277& 417& 51908& 165.041450&   1.245120& 1.748& 1.6811&  19.060&  0.025&0657& 161& 52177&  1.745& 19.060& 0.0136 \\
0277& 628& 51908& 166.527180&   0.536786& 1.530& 1.3352&  18.920&  0.080&0382& 024& 51816&  1.515& 18.950& 0.0844 \\
0277& 629& 51908& 166.514230&   0.368860& 1.660& 1.0198&  18.830&  0.269&0282& 488& 51658&  1.649& 18.840& 0.0527 \\
0278& 009& 51900& 168.083660&  -0.858161& 1.514& 1.0134&  19.070&  0.218&0539& 484& 52017&  1.505& 19.070& 0.0409 \\
0278& 446& 51900& 167.496260&   1.032133& 1.493& 1.3274&  18.970&  0.069&0513& 125& 51989&  1.508& 18.950& 0.0758 \\
0278& 595& 51900& 168.233890&   0.228728& 1.431& 1.2419&  18.890&  0.081&0300& 379& 51943&  1.432& 18.910& 0.0335 \\
0280& 130& 51612& 170.545400&  -1.190975& 1.765& 1.3065&  19.370&  0.179&0656& 548& 52148&  1.774& 19.340& 0.0644 \\
0280& 208& 51612& 170.288990&  -1.214862& 1.675& 1.4973&  19.190&  0.069&0583& 061& 52055&  1.661& 19.180& 0.0657 \\
0280& 302& 51612& 169.153720&   0.243500& 1.833& 1.3476&  19.730&  0.186&0416& 442& 51811&  1.840& 19.710& 0.0460 \\
\hline
\end{tabular}
}
\end{minipage}
\end{table*}
\begin{table*}
\begin{minipage}{140mm}
\rotatebox{90}{
\footnotesize
\begin{tabular}
{|llllrrlllll|lllr|}
\multicolumn{15}{l}{Table A1b: Full sample of the absorber and non-absorber QSOs}\\
\hline
\multicolumn{11}{|c|}{Absorber sample}&\multicolumn{4}{|c|}{Non-absorber
sample}\\
\hline
Plate&Fiber&MJD&$z_{abs}$&$\Delta(g-i)$&$E(B-V)_{(g-i)}$&$W_{\rm
Mg\;I}^a$&W$_{\rm Mg\;II}^b$&W$_{\rm Mg\;II}^c$&W$_{\rm Al\;II}^d$&W$_{\rm
Fe\;II}^e$&
\multicolumn{1}{|c}{Plate}&Fiber&MJD&$\Delta(g-i)$ \\
&&&&&&&&&&&\multicolumn{1}{|c}{}&&&\\
\hline
0267& 264& 51608 & 1.6734 &  0.2007 & 0.0409 &  0.96 &  2.94  & 2.66  & 1.62  &   2.90 & 0386& 610& 51788&  -0.0904\\ 
0267& 287& 51608 & 1.2122 &  0.2067 & 0.0529 &     0 &  1.37  & 1.01  &    0  &   0.73 & 0591& 018& 52022&  -0.1138\\
0267& 294& 51608 & 1.0103 &  0.0093 & 0.0027 &  0.54 &  1.14  & 0.67  &    0  &   0.29 & 0467& 560& 51912&   0.1045\\
0268& 059& 51633 & 1.5983 & -0.0402 &-0.0085 &  0.34 &  2.49  & 1.94  & 0.87  &   1.57 & 0619& 377& 52056&  -0.0407\\
0269& 091& 51910 & 1.3359 &  0.1388 & 0.0333 &  0.35 &  0.67  & 0.49  &    0  &   0.35 & 0413& 038& 51929&   0.1269\\
0269& 613& 51910 & 1.0377 &  0.0554 & 0.0157 &  0.72 &  3.51  & 3.70  &    0  &   1.37 & 0404& 554& 51812&  -0.1323\\
0270& 213& 51909 & 1.3270 &  0.2288 & 0.0551 &  0.28 &  1.99  & 1.75  & 0.95  &   1.38 & 0296& 417& 51984&  -0.1346\\
0270& 297& 51909 & 1.5603 &  0.2862 & 0.0615 &  2.21 &  6.60  & 6.55  & 2.18  &   2.59 & 0549& 238& 51981&  -0.1720\\
0270& 462& 51909 & 1.2649 & -0.0832 &-0.0207 &  0.40 &  1.20  & 1.08  &    0  &   0.78 & 0552& 153& 51992&  -0.2173\\
0271& 154& 51883 & 1.2240 &  0.0703 & 0.0179 &  0.50 &  1.61  & 1.33  &    0  &   0.84 & 0430& 072& 51877&  -0.0808\\
0273& 197& 51957 & 1.0760 &  0.2709 & 0.0749 &  0.91 &  2.46  & 2.13  &    0  &   1.42 & 0283& 560& 51959&  -0.0009\\
0273& 494& 51957 & 1.0170 & -0.1036 &-0.0296 &     0 &  1.92  & 1.56  &    0  &   0.77 & 0351& 480& 51780&  -0.0636\\
0273& 496& 51957 & 1.1936 &  0.0883 & 0.0228 &     0 &  0.70  & 0.97  &    0  &   0.62 & 0650& 185& 52143&  -0.0339\\
0274& 357& 51913 & 1.4243 &  0.0822 & 0.0189 &  0.38 &  2.52  & 2.10  & 0.63  &   1.41 & 0533& 029& 51994&  -0.2390\\
0275& 357& 51910 & 1.0112 &  0.0209 & 0.0060 &     0 &  0.65  & 0.72  &    0  &   0.44 & 0552& 232& 51992&  -0.0348\\
0276& 068& 51909 & 1.2846 & -0.0697 &-0.0172 &  0.19 &  0.84  & 0.83  & 0.43  &   0.46 & 0496& 338& 51988&  -0.1248\\
0276& 579& 51909 & 1.0785 &  0.0784 & 0.0216 &     0 &  1.14  & 0.97  &    0  &   0.39 & 0315& 598& 51663&  -0.1361\\
0276& 579& 51909 & 1.7858 &  0.0784 & 0.0152 &  0.27 &  1.87  & 1.70  & 0.73  &   0.99 & 0315& 598& 51663&  -0.1361\\
0277& 059& 51908 & 1.6405 & -0.0901 &-0.0187 &     0 &  0.58  & 0.80  &    0  &   0.40 & 0484& 119& 51907&   0.0623\\
0277& 246& 51908 & 1.1623 &  0.0283 & 0.0074 &  0.75 &  1.31  & 1.18  &    0  &   0.74 & 0463& 076& 51908&   0.0301\\
0277& 417& 51908 & 1.6811 &  0.1806 & 0.0367 &     0 &  1.02  & 0.91  &    0  &   0.47 & 0657& 161& 52177&   0.1214\\
0277& 628& 51908 & 1.3352 & -0.0000 & 0.0000 &     0 &  1.49  & 0.74  &    0  &   0.58 & 0382& 024& 51816&   0.0858\\
0277& 629& 51908 & 1.0198 & -0.1291 &-0.0369 &     0 &  1.08  & 0.77  &    0  &   0.70 & 0282& 488& 51658&  -0.0653\\
0278& 009& 51900 & 1.0134 & -0.0521 &-0.0149 &     0 &  0.88  & 0.78  &    0  &   0.56 & 0539& 484& 52017&  -0.0836\\
0278& 446& 51900 & 1.3274 & -0.1470 &-0.0354 &  0.22 &  1.34  & 1.01  & 0.50  &   0.52 & 0513& 125& 51989&   0.0555\\
\hline
\multicolumn{5}{l}{$^a$ Equivalent width of Mg I $\lambda$2583 {\AA}}\\
\multicolumn{5}{l}{$^b$ Equivalent width of Mg II $\lambda$2796 {\AA}}\\
\multicolumn{5}{l}{$^c$ Equivalent width of Mg II $\lambda$2803 {\AA}}\\
\multicolumn{5}{l}{$^d$ Equivalent width of Al II $\lambda$1670 {\AA}}\\
\multicolumn{5}{l}{$^e$ Equivalent width of Fe II $\lambda$2382 {\AA}}\\
\end{tabular}
}
\end{minipage}
\end{table*}
\newpage
\begin{table*}
\begin{minipage}{140mm}
\footnotesize
\begin{tabular}
{|llll|llll|}
\multicolumn{8}{l}{Table A2: Finding list for interstellar lines in
spectra of QSOs: 1370-4500 {\AA} }\\
\hline
Vacuum&Species&Oscillator&Reference&Vacuum&Species&Oscillator&Reference\\
Wavelength&&strength&&Wavelength&&strength&\\
\hline
1370.132 & Ni II   & 7.690E-2& M03   &2325.4029& C II]     & 4.780e-8& M03   \\ 
1381.476 & P I     & 3.160E-1& M03   &2326.1126& C II]*    & 5.520e-8& M03   \\
1393.324 & Ni II   & 1.010E-2& M03   &2328.8374& C II]*    & 2.720e-8& M03   \\
1393.7602& Si IV   & 5.130e-1& M03   &2335.123 & Si II]    & 4.250e-6& M03   \\ 
1402.7729& Si IV   & 2.540e-1& M03   &2344.2139& Fe II     & 1.140e-1& M03   \\
1477.222 & Ni II   & 9.720e-4& M03   &2346.2616& Ni I      & 1.430e-1& M03   \\ 
1477.57  & CO APXS2& 4.013e-2& MN    &2367.5905& Fe II     & 0.216e-4& M03   \\ 
1480.9546& Co II   & 1.190e-2& M03   &2367.7750& Al I      & 1.060e-1& M03   \\ 
1509.75  & CO APXS1& 2.911e-2& MN    &2374.4612& Fe II     & 3.130e-2& M03   \\ 
1526.7070& Si II   & 1.330e-1& M03   &2382.7652& Fe II     & 3.200e-1& M03   \\ 
1532.5330& P II    & 3.030e-3& M03   &2407.9875& Co I      & 3.800e-1& M03   \\ 
1533.4316& Si II*  & 1.320e-1& M03   &2425.6708& Co I      & 2.820e-1& M03   \\ 
1544.45  & CO APXS0& 1.578e-2& MN    &2448.4509& Fe I      & 2.310e-3& M03   \\ 
1547.9451& Co II   & 1.110e-2& M03   &2463.3922& Fe I      & 5.320e-2& M03   \\ 
1548.204 & C IV    & 1.899e-1& M03   &2484.0209& Fe I      & 5.440e-1& M03   \\ 
1550.781 & C IV    & 9.475e-2& M03   &2501.8858& Fe I      & 4.930e-2& M03   \\ 
1552.7624& Co II   & 1.160e-2& M03   &2515.0725& Si I      & 2.110e-1& M03   \\ 
1560.31  & C I     & 7.740e-2& M03   &2522.1238& Co I      & 2.290e-1& M03   \\ 
1562.001 & Si I    & 3.758e-1& M01,M03&2523.6083& Fe I      & 2.030e-1& M03   \\ 
1565.5255& Fe II   & 0.766e-5& M03   &2568.7518& Al I      & 3.760e-2& M03   \\ 
1573.92  & Co II   & 3.700e-2& M03   &2576.877 & Mn II     & 3.610e-1& M03   \\ 
1588.6876& Fe II   & 1.480e-4& M03   &2586.6500& Fe II     & 6.910e-2& M03   \\ 
1602.4863& Ge II   & 1.440e-1& M99   &2594.499 & Mn II     & 2.800e-1& M03   \\ 
1608.4511& Fe II   & 5.770e-2& M03   &2600.1729& Fe II     & 2.390e-1& M03   \\ 
1610.194 & Sc III  & 8.660e-2& M03   &2606.462 & Mn II     & 1.980e-1& M03   \\ 
1611.2005& Fe II   & 1.380e-3& M03   &2653.2654& Al I      & 1.500e-2& M03   \\ 
1631.170 & Si I    & 2.051e-1& M01,M03 &2669.948 & Al II]    & 1.060e-5& M03   \\ 
1651.164 & Mg I    & 6.577e-4& M01,M03 &2683.887 & V II      & 1.090e-1& M03   \\ 
1656.93  & C I     & 1.490e-1& M03   &2719.8329& Fe I      & 1.220e-1& M03   \\ 
1660.809 & O III]* & 1.590e-7& M03   &2740.525 & V II      & 1.010e-1& M03   \\ 
1666.150 & O III]**& 2.090e-7& M03   &2795.641 & Mn I      & 5.650e-1& M03   \\ 
1670.7886& Al II   & 1.740e00& M03   &2799.094 & Mn I      & 4.190e-1& M03   \\ 
1674.5953& P I     & 3.250e-2& M03   &2801.907 & Mn I      & 2.900e-1& M03   \\ 
1679.6969& P I     & 5.360e-2& M03   &2796.3543& Mg II     & 6.155e-1& M03   \\ 
1683.4116& Mg I    & 2.490e-3& M03   &2803.5315& Mg II     & 3.058e-1& M03   \\ 
1703.4119& Ni II   & 6.000e-3& M03   &2852.9631& Mg I      & 1.830e00& M03   \\ 
1709.6042& Ni II   & 3.240e-2& M03   &2937.7623& Fe I      & 1.810e-2& M03   \\ 
1741.5531& Ni II   & 4.270e-2& M03   &2967.7646& Fe I      & 4.380e-2& M03   \\ 
1747.7937& Mg I    & 9.080e-3& M03   &2984.4402& Fe I      & 2.900e-2& M03   \\ 
1751.9157& Ni II   & 2.770e-2& M03   &3021.5187& Fe I      & 1.040e-1& M03   \\ 
1753.8406& Mg I    &         & M03   &3067.2379& Ti II     & 4.890e-2& M03   \\ 
1774.9487& P I     & 1.690e-1& M03   &3073.8633& Ti II     & 1.210e-1& M03   \\ 
1782.8291& P I     & 1.130e-1& M03   &3083.0462& Al I      & 1.770e-1& M03   \\
1787.6481& P I     & 6.040e-2& M03   &3131.54  & Be II     & 4.981e-1& M03   \\ 
1804.473 & Ni II   & 3.800e-3& W99   &3187.372 & Ti I      & 1.950e-1& M03   \\ 
1807.3113& S I     & 9.050e-2& M03   &3230.1221& Ti II     & 6.870e-2& M03   \\ 
1808.0129& Si II   & 2.080e-3& M03   &3242.9184& Ti II     & 2.320e-1& M03   \\ 
1816.93  & Si II*  & 1.790e-3& M03   &3248.4739& Cu I      & 4.340e-1& M03   \\ 
1827.9351& Mg I    & 2.420e-2& M03   &3274.8980& Cu I      & 2.180e-1& M03   \\ 
1845.5205& Si I    & 2.700e-1& M03   &3303.529 & Na I      & 1.345e-2& M03   \\ 
1854.7184& Al III  & 5.590e-1& M03   &3342.834 & Ti I      & 1.740e-1& M03   \\ 
1862.7910& Al III  & 2.780e-1& M03   &3370.5336& Ni I      & 2.430e-2& M03   \\ 
1892.030 & Si III] & 2.690e-5& M03   &3384.7304& Ti II     & 3.580e-1& M03   \\ 
1901.773 & Fe II   & 7.000e-5& M03   &3392.0165& Ni I      & 0.999e-2& M03   \\ 
1908.734 & C III]  & 1.687e-7& M03   &3441.5918& Fe I      & 2.360e-2& M03   \\ 
1910.78  & Ti II   & 2.020e-1& M03   & 3579.705 & Cr I      & 3.660e-1& M03  \\ 
1941.2852& Co II     & 3.400e-2& M03 & 3581.947 & Sc II     & 2.370e-1& M03  \\ 
1977.5972& Si I      & 4.910e-2& M03 & 3594.507 & Cr I      & 2.910e-1& M03  \\ 
2012.1664& Co II     & 3.680e-2& M03 & 3606.350 & Cr I      & 2.260e-1& M03  \\ 
\hline
\end{tabular}
\end{minipage}
\end{table*}
\begin{table*}
\begin{minipage}{140mm}
\footnotesize
\begin{tabular}
{|llll|llll|}
\multicolumn{8}{l}{Table A2(continued): Finding list for interstellar lines
in spectra of QSOs: 1370-4500 {\AA}}\\
\hline
2026.1370& Zn II     & 5.010e-1& M03 & 3636.498 & Ti I      & 2.520e-1& M03  \\ 
2026.269 & Cr II     & 1.300e-3& M03 & 3643.822 & Sc II     & 3.750e-1& M03  \\ 
2026.4768& Mg I      & 1.130e-1& M03 & 3650.3426& Fe I      & 0.699e-4& M03  \\ 
2056.2569& Cr II     & 1.030e-1& M03 & 3680.9611& Fe I      & 2.800e-3& M03  \\ 
2062.2361& Cr II     & 7.590e-2& M03 & 3720.9928& Fe I      & 4.110e-2& M03  \\ 
2062.6604& Zn II     & 2.460e-1& M03 & 3825.5288& Fe I      & 4.830e-3& M03  \\ 
2066.1640& Cr II     & 5.120e-2& M03 & 3861.0058& Fe I      & 2.170e-2& M03  \\ 
2138.827 & V II      & 8.200e-2& M03 & 3875.705 & CN BSXS0  & 3.380e-2& BvD \\  
2139.2477& Zn I      & 1.470e00& M03 & 3879.873 & CH BSXP0  & 1.070e-3& GvDB \\ 
2139.683 & N II]*    & 5.860e-8& M03 & 3887.510 & CH BSXP0  & 3.200e-3& GvDB \\ 
2143.450 & N II]**   & 8.190e-8& M03 & 3891.319 & CH BSXP0  & 2.130e-3& GvDB \\
2167.4534& Fe I      & 1.500e-1& M03 & 3934.7750& Ca II     & 6.267e-1& M03  \\ 
2208.6666& Si I      & 5.750e-2& M03 & 3945.1224& Al I      & 1.170e-1& M03  \\ 
2234.4472& Fe II     & 2.520e-5& M03 & 3958.812 & CH+APXS1  & 3.400e-3& GvDB \\ 
2249.8768& Fe II     & 1.820e-3& M03 & 3969.5901& Ca II     & 3.116e-1& M03  \\ 
2260.7805& Fe II     & 2.440e-3& M03 & 3982.888 & Ti I      & 1.050e-1& M03  \\ 
2264.1647& Al I      & 8.920e-2& M03 & 4045.2847& K I       & 5.680e-3& M03  \\ 
2290.6934& Ni I      & 1.280e-1& M03 & 4048.3565& K I       & 2.640e-3& M03  \\ 
2299.663 & Mn II     & 4.810e-4& M03 & 4227.918 & Ca I      & 1.770e00& M03  \\ 
2305.714 & Mn II     & 1.150e-3& M03 &4233.740 & CH+APXS0   & 5.500e-3& GvDB \\ 
2311.6723& Ni I      & 3.765e-1& M01,M03&4301.5229& CH ADXP0  & 5.060e-3& GvDB\\
2320.7468& Ni I      & 6.850e-1& M03 &4430.12  & DB 4428   & 1.000e00& dib  \\ 
\hline
\end{tabular}

References: BvD: Black, J. H. \& van Dishoeck, E. F. 1988, ApJ, 331,
986\hfill\break
GvBD: Gredel, R., van Dishoeck, E. F., \& Black, J. H. 1993, A\&A, 269,
477\hfill\break M91: Morton, D.C. 1991, ApJS, 77, 119;  M03: Morton, D.C.
2003, ApJS, 149, 205\hfill\break MN: Morton, D. C. \& Noreau 1994, ApJS, 95, 301\\
\end{minipage}
\end{table*}
\begin{table*}
\centering
\begin{minipage}{140mm}
\tiny
\begin{tabular}
{|c|l|c|c|c|c|c|c|c|c|c|c|}
\multicolumn{12}{l}{Table A3: Equivalent widths of weak lines for derivation of
abundances}\\
\hline
\multicolumn{3}{|c|}{Sub-sample Number} & 3& 4& 24& 9& 11$^a$ & 13$^a$ &
12$^a$ &10$^a$&26 \\
\hline
\multicolumn{3}{|c|}{Number of systems} & 135 & 133 & 698 & 404 & 398 &
405 & 411&405 & 97\\
\hline 
\multicolumn{3}{|c|}{Criteria}  & W$_{\rm Mg\;II}$& W$_{\rm Mg\;II}$&
$\Delta^b$ & $z_{abs}$ & $i$ & $\beta$ & $i$&$z_{abs}$&W$_{\rm Mg\;II}$\\
\multicolumn{3}{|c|}{}  & 0.9-1.2 & 1.2-1.5 &$<$ 0.2  & $<$1.3 &
$<$19.1&$<0.1$&$>19.1$&$>1.3$ & $>$2.5\\
\multicolumn{3}{|c|}{}  & &&&&&& &&$\Delta<0.2$ \\
\hline
\multicolumn{3}{|c|}{$E(B-V)$} & $<$0.001 & $<$0.001 & 0.002 & 0.006 & 0.009 &
0.011 & 0.011 &0.012 &0.012\\
\hline
\hline
$\lambda$ & Species& $f$ & \multicolumn{9}{|c|}{Equivalent Width in m{\AA}}  \\
\hline			
1656.9&C I &	0.1490&	$<$48&	$<$78&	16$\pm$4&	NA&	$<$24&	$<$40& $<$39&  39$\pm$13&49$\pm$18\\
3303.6&Na I&	0.0130&	$<$29&	$<$24&	$<$13&	$<$15&	$<$15&	$<$19& $<$32&$<$33&$<$33\\
2853.0&Mg I&	1.8300&	171$\pm$22&	281$\pm$16&	258$\pm$4&	341$\pm$20&	298$\pm$13&	279$\pm$15& 379$\pm$21&300$\pm$17&596$\pm$16\\
2026.5&Mg I&	0.1130& $<$23($>$7)&$<$39($>$11)&$<$14($>$10)&$<$7($>$14)&$<$21($>$12)&$<$29($>$11)&$<$37($>$15)&   $<$15($>$12)&$<$82($>$24)\\ 
1827.9&Mg I&	0.0242&	$<$17&	$<$21&	$<$8&	$<$17&	$<$9&	$<$10& $<$21&   $<$10&	$<$30  \\ 
3083.1&Al I&	0.1770&	$<$24&	$<$28&	$<$16&	$<$13&	$<$13&	$<$17& $<$26&    $<$26&	$<$38  \\
2367.8&Al I&	0.1060&	$<$54&	$<$45&	$<$12&	$<$21&	$<$24&	$<$27& $<$54&   $<$33&	$<$43   \\
1845.5&Si I&	0.2700&	$<$17&	$<$20&	$<$8&	$<$17&	$<$10&	$<$10& $<$20&  $<$10&$<$30 \\
2515.1&Si I&	0.2110&	$<$18& 	$<$19&	$<$8&	$<$11&	$<$9&	$<$11& $<$20& $<$12&	$<$29 \\
2335.1&Si II]&	4.25E-6&$<$19&	$<$17&	$<$8&	$<$8&	$<$8&	$<$10& $<$19& $<$11&	$<$26\\
1808.0&Si II&	0.0021&	63$\pm$18&	79$\pm$18&	62$\pm$5&	76$\pm$14&	59$\pm$7&	59$\pm$7& 83$\pm$24& 58$\pm$7&	187$\pm$19\\
1807.3&S I&	0.0905&	23$\pm$6&$<$24& 29$\pm$3&36$\pm$6&30$\pm$3&26$\pm$3&29$\pm$8&  27$\pm$4&95$\pm$10 \\
3934.8&Ca II&	0.6270&	92$\pm$36&	$<$147&	102$\pm$13&	98$\pm$26&	98$\pm$36&	199$\pm$84& 226$\pm$75&  NA&	253$\pm$56 \\
3969.6&Ca II&	0.3120&	$<$77&	$<$106&	53$\pm$13&	102$\pm$26&
86$\pm$26&	156$\pm$51& $<$124& NA&$<67$ \\
3643.8&Sc II&	0.3750&	$<$39&	$<$38&	$<$16&	$<$21&	$<$19&	$<$25& $<$48&  $<$41&	$<$66\\
3582.0&Sc II&	0.2370&	$<$37&	$<$39&	$<$16&	$<$17&	$<$17&	$<$24& $<$42&  $<$39&	$<$63 \\
3384.7&Ti II&	0.3580&	$<$37&	$<$51&	26$\pm$5&	44$\pm$10&	34$\pm$12&	41$\pm$12& $<$44&  $<$39&	100$\pm$28 \\
3242.9&Ti II&	0.2320&	$<$29&	$<$49&	$<$16&	27$\pm$8&	29$\pm$7&	51$\pm$14& $<$33&  $<$32&	78$\pm$19\\
1910.8&	Ti II&	0.1980&	$<$24&	$<$21&	$<$9&	$<$15&	$<$9&	$<$10& $<$18&    $<$10&$<$26 \\
3073.9&Ti II&	0.1210&	$<$24&	$<$28&	$<$12&	17$\pm$4&	$<$13&	$<$18& $<$25&  $<$24&	$<$38 \\
3230.1&Ti II&	0.0690&	$<$25&	$<$31&	$<$12&	$<$14&	$<$13&	$<$16& $<$30& $<$28&	$<$39 \\
2683.9&V II&	0.1100&	$<$17&	$<$18&	$<$8&	$<$8&	$<$8&	$<$9& $<$17&$<$13&	$<$29\\
2740.5&V II&	0.1000&	$<$20&	$<$20&	$<$8&	$<$9&	$<$10&	$<$11& $<$18&$<$14&	$<$27\\
2056.3&Cr II&	0.1030&	$<$23&	49$\pm$15&	35$\pm$5&	30$\pm$9&	23$\pm$6&	$<$17& 45$\pm$15&22$\pm$8&	93$\pm$16\\
2062.2&Cr II&	0.0760&	$<$23&	$<$36&	24&	$<$22&	$<$17&	$<$17& $<$34&$<$18&	61\\
2066.2&Cr II&	0.0510&	$<$23&	$<$24&	12(4)&	$<$14&	$<$11&	$<$17& $<$23&$<$15&	28$\pm$9\\
2576.9&Mn II&	0.3610&	27$\pm$11&	61$\pm$14&	47$\pm$4&	61$\pm$8&	51$\pm$6&	43$\pm$7& 76$\pm$17&53$\pm$9&	143$\pm$14\\
2594.5&Mn II&	0.2800&	39$\pm$14&	43$\pm$13&	42$\pm$4&	54$\pm$7&	54$\pm$9&	47$\pm$8& 60$\pm$12&42$\pm$9&	108$\pm$22\\
2606.5&Mn II&	0.1980&	$<$23&	24$\pm$11 &	21$\pm$4&	24$\pm$7&	23$\pm$8&	15$\pm$7& $<$23&$<$14&	60$\pm$18\\
2484.0&Fe I&	0.5440&	$<$17&	$<$19&	$<$8&	$<$11&	$<$9&	$<$11& $<$21&$<$13&	$<$28\\
2523.6&Fe I&	0.2030&	$<$19&	$<$19&	$<$8&	$<$11&	$<$9&	$<$10& $<$20&$<$12&	$<$31\\
2374.5&Fe II&	0.0313&	180$\pm$18&	281$\pm$15&	274$\pm$4&	307$\pm$7&	252$\pm$8&	264$\pm$9& 416$\pm$18&253$\pm$11&	761$\pm$14\\
2260.8&Fe II&	0.0024&	$<$22&	58$\pm$16&	43$\pm$4&	61$\pm$11&	48$\pm$7&	53$\pm$10& 59$\pm$17&44$\pm$10&	137$\pm$16\\
2249.9&Fe II&	0.0018&	$<$22&	36$\pm$11 &	31$\pm$4&	42$\pm$8&	32$\pm$7&	30$\pm$8& 39$\pm$12&19$\pm$9&	123$\pm$16\\
2234.2&Fe II&	2.52E-5&$<$18& 	$<$18&	$<$8&	$<$10&	$<$8&	$<$10& $<$17&$<$12&	$<$27\\
2367.6&Fe II&	2.16E-5&$<$18& 	20$\pm$6&	$<$9&	$<$11&	$<$9&	$<$10& $<$20&$<$12&	$<$26\\
2012.2&Co II&
0.0370&$<18$&$<23$&13$\pm$4&$<12$&$<9$&$<12$&$<19$&$<12$&$<26$\\
1941.3&Co II&	0.0340&	$<$22&	$<$21&	$<$8&	$<$13&	$<$10&	$<$12& $<$20&$<$11&	$<$24\\
1741.6&Ni II&	0.0430&	$<$32&	$<$36&	25$\pm$4&$<$31&	25$\pm$9&29$\pm$7& 71$\pm$22&30$\pm$9&	108$\pm$12\\
1709.6&Ni II&	0.0320&	$<$28&	$<$37&	20$\pm$6&$<$25&	$<$14&	$<$14& 58$\pm$14&$<$14&	62$\pm$13\\
1751.9&Ni II&	0.0280&	$<$22&	$<$32&	$<$10&	$<$26&	17$\pm$4& 23$\pm$5&$<$23&$<27$&40$\pm$9\\
1703.4&Ni II&	0.0060&	$<$30&	$<$37&	$<$12&	$<$25&	$<$15&	$<$15 &$<$24&$<$14&	$<$39\\
1804.5&Ni II&	0.0038&	$<$54&	$<$54&	$<$14&	$<$42&	$<$21&	$<$21 &$<$72&$<$21&	$<$58\\
2139.3&Zn I&	1.4700&	$<$21&	$<$26&	$<$8&	$<$12&	$<$10&	$<$11 &$<$20&$<$12&	$<$26\\
2062.7&Zn II&	0.2460&	$<$23&	$>$35&	21&	$>$33&	$>$26&	$<$17 &$>$25&$>$9&	30\\
2026.3&Zn II&	bl$^c$&	$<$23&	74$\pm$26&	35$\pm$4&	40$\pm$17&	21$\pm$6&	29$\pm$9 &62$\pm$19&$<$15&	112$\pm$14\\
2062.4&Zn II&	bl$^c$&	$<$23&	71$\pm$2&	45$\pm$4&	55$\pm$9&	43$\pm$10&	$<$17 &59$\pm$14&28$\pm$8&	91$\pm$13\\
\hline 
\end{tabular}

$^a$ The numbers in the samples for measuring equivalent width were greater 
by 1,1, 1 and 2, for sub-samples 11, 13, 12 and 10, than what is listed 
here. The numbers shown  for number of objects in the sub-sample and for 
the extinction of the sub-sample are from Table 1.\\
$^b$ $\Delta$ is the observer frame colour excess $\Delta(g-i)$\\
$^c$ The equivalent widths for these lines are the actual measured values
for the blended lines.\\
\end{minipage}
\end{table*}
\begin{table*}
\centering
\begin{minipage}{140mm}
\tiny
\begin{tabular}
{|c|c|c|c|c|c|c|c|c|}
\multicolumn{9}{l}{Table A3(continued): Equivalent widths of weak lines for derivation of
abundances}\\
\hline
Sample Number&&	1&	16&	8&	17&	19&	25&	27\\
\hline
Number of systems&&	809&	369&	251&	85&	58&	111&	48\\
\hline
Criteria&&W$_{\rm Mg\;II}$&W$_{\rm Fe\;II}$/W$_{\rm Mg\;II}$&W$_{\rm
Mg\;II}$&W$_{\rm Al\;II}$/W$_{\rm Mg
\;I}$&Fe II$\lambda$2260&$\Delta>$0.2&W$_{\rm Mg\;II}$\\
&&$>$0.3&$>$0.58&$>$2&$<$1.54&present&&$>$2.5\\
&&&&&&&&$\Delta$$>$0.2\\
\hline
$E(B-V)$&&	0.013&	0.014&	0.0320&	0.034&	0.034&	0.081&	0.085\\
\hline
\hline
$\lambda$&Species&\multicolumn{7}{|c|}{Equivalent Width in m{\AA}}\\
\hline
1656.9&C I &19$\pm$4&46$\pm$13&	50$\pm$12&79$\pm$21&$<57$&57$\pm$19&	114$\pm$32\\
3303.6&Na I&$<$12&$<$17&$<$21&$<$81&$<$37&31&$<$67\\
2853.0&Mg I&282$\pm$4&397$\pm$10&585$\pm$10&691$\pm$35&	689$\pm$16&536$\pm$15&795$\pm$26\\
2026.5&Mg I&	$<$32($>$11)&$<$34($>$16)&$<$77($>$23)&$<$37($>$28) &$<$76($>$28)&$>$48($>$21)&$<$240($>$32)\\
1827.9&Mg I&	$<$8&	$<$14&	$<$17&	$<$22&	$<$32&	$<$28&	$<$48\\
3083.1&Al I&	$<$14&	$<$14&	$<$24&	$<$46&	$<$35&	$<$34&	$<$58\\
2367.8&Al I&	$<$11&  $<14$&  $<$27&	$<$93&	$<$41&	$<$48&	$<$85\\
1845.5&Si I&	$<$8&	$<$13&	$<$17&	$<$22&	$<$31&	$<$29&	$<$49\\
2515.1&Si I&	$<$8&   $<$12&	$<$16&	$<$29&	$<$33&	$<$25&	$<$48\\
2335.1&Si II]&	$<$7&	$<$12&	$<$16&	$<$27&	$<$25&	$<$30&	$<$48\\
1808.0&Si II&	68$\pm$5&121$\pm$10&	167$\pm$11&	145$\pm$22&	240$\pm$15&	175$\pm$24&	212$\pm$37\\
1807.3&S I&	32$\pm$3&61$\pm$4&87$\pm$5&59$\pm$9&124$\pm$11&91$\pm$10&115$\pm$18\\
3934.8&Ca II&	110$\pm$16&	254$\pm$64&	201$\pm$40&	NA&	254$\pm$47&	307$\pm$86&	752$\pm$139\\
3969.6&Ca II&	57$\pm$13&	156$\pm$42&	101$\pm$26&	NA&
141$\pm$36&99$\pm$21&$<163$\\
3643.8&Sc II&	$<$16&	$<$26&	$<$31&	$<$106&	$<$50&	$<$59&	$<$82\\
3582.0&Sc II&	$<$16&	$<$24&	$<$30&	$<$103&	$<$52&	$<$62&	$<$86\\
3384.7&Ti II&	31$\pm$5&	67$\pm$16&	111$\pm$18&	130$\pm$51&	160$\pm$16&	120$\pm$29&	193$\pm$56\\
3242.9&Ti II&	32$\pm$5&	49$\pm$10&	64$\pm$11&	$<$94&	120$\pm$16&	51$\pm$13&	$<$432?\\
1910.8&	Ti II&	$<$8&	$<$11&	$<$15&	$<$20&	$<$29&	$<$27&	$<$43\\
3073.9&Ti II&	$<$11&	$<$15&	$<$24&	$<$46&	$<$39&	$<$33&	$<$53\\
3230.1&Ti II&	$<$12&	$<$17&	$<$22&	$<$61&	$<$32&	$<$35&	$<111$\\
2683.9&V II&	$<$8&	$<$11&	$<$17&	$<$29&	$<$29&	$<$28&	$<$42\\
2740.5&V II&	$<$8&	$<$14&	$<$18&	$<$29&	$<$27&	$<$27&	$<$43\\
2056.3&Cr II&	39$\pm$4&	72$\pm$11&	101$\pm$10&	89$\pm$17&	147$\pm$14&	77$\pm$13&	154$\pm$27\\
2062.2&Cr II&	40&	56&	78&	70&	123&	75&	132\\
2066.2&Cr II&	14$\pm$4&	39$\pm$10&	54$\pm$10&	50$\pm$16&	100$\pm$13&	72$\pm$18&	109$\pm$27\\
2576.9&Mn II&	54$\pm$4&	114$\pm$11&	140$\pm$9&	146$\pm$21&	231$\pm$14&	156$\pm$17&	237$\pm$26\\
2594.5&Mn II&	50$\pm$4&	78$\pm$10&	129$\pm$11&	103$\pm$16&	167$\pm$12&	162$\pm$19&	277$\pm$32\\
2606.5&Mn II&	26$\pm$4&	49$\pm$9&	79$\pm$10&	64$\pm$14&	142$\pm$13 &96$\pm$18&	147$\pm$30\\
2484.0&Fe I&	$<$8&	$<$13&	$<$18&	$<$26&	$<$33&	$<$28&	$<$50\\
2523.6&Fe I&	$<$8&	$<$12&	$<$16&	$<$30&	$<$34&	$<$25&	$<$48\\
2374.5&Fe II&	298$\pm$4&503$\pm$12& 702$\pm$9&476$\pm$31&894$\pm$14&624$\pm$16&958$\pm$28\\
2260.8&Fe II&	46$\pm$4&99$\pm$11&113$\pm$9&125$\pm$23&280$\pm$10&90$\pm$14&145$\pm$18\\
2249.9&Fe II&	35$\pm$4&63$\pm$10&105$\pm$10&58$\pm$17&163$\pm$15&113$\pm$20&215$\pm$32\\
2234.2&Fe II&	$<$8&	$<$12&	$<$17&	$<$27&	$<$24&	$<$30&	$<$51\\
2367.6&Fe II&	$<$8&	$<$13&	$<$16&	$<$27&	$<$25&$<30$&$<$50\\
2012.2&Co II&	15$\pm$4&$<11$&	56$\pm$13&$<20$&57$\pm$17&$<30$&$<47$\\
1941.3&Co II&	$<$8&	$<$10&	$<$16&	$<$19&	$<$27&	$<$26&	$<$43\\
1741.6&Ni II&	28$\pm$4&84$\pm$24&90$\pm$8&80$\pm$20&124$\pm$16&87$\pm$19&100$\pm$23\\
1709.6&Ni II&	16$\pm$4&22$\pm$5&83$\pm$17&$<$29&117$\pm$24&	$<$41&$<$63\\
1751.9&Ni II&	$<$10&	$<$16&	51$\pm$17&52$\pm$13&$<$40&76$\pm$20&93$\pm$30\\
1703.4&Ni II&	$<$11&	$<$18&	$<$38&	$<$28&	$<$45&	$<$42&	$<$66\\
1804.5&Ni II&	$<$14&	$<$30&	$<$32&	$<$66&	$<$36&	$<$72&	$<$110\\
2139.3&Zn I&	$<$8&	$<$12&	$<$17&	$<$25&	$<$28&	$<$25&	$<$47\\
2062.7&Zn II&	8&	21&	36&	31&102	&	47&	26\\
2026.2&Zn II&	40$\pm$4&	55$\pm$8&	111$\pm$8&	68$\pm$127&	176$\pm$14&	95$\pm$15&	266$\pm$29\\
2062.7&Zn II&	48$\pm$4&	77$\pm$10&	114$\pm$8&	101$\pm$15&	225$\pm$14&	122$\pm$14&	158$\pm$20\\
\hline
\end{tabular}
\end{minipage}
\end{table*}
\begin{table*}
\centering
\begin{minipage}{140mm}
\scriptsize
\begin{tabular}
{|c|c|c|c|c|c|c|c|c|c|}
\multicolumn{9}{l}{Table A4: Column densities of weak lines }\\
\hline
{Sample Number}&3&4&24&9&11&13&12&10\\
\hline
{Number of systems}&135&133&698&404&398&405&411&405\\
\hline
{Criteria}&W$_{\rm Mg\;II}$&W$_{\rm
Mg\;II}$&$\Delta^a$&$z_{abs}$&$i$&$\beta$&$i$&$z_{abs}$\\
&0.9-1.2&1.2-1.5&$<$0.2&$<$1.3&$<$19.1&$<$0.1&$>$19.1&$>1.3$\\
\hline
{$E(B-V)$}&$<$0.001&$<$0.001&0.002&0.006&0.009&0.011&0.011&0.012\\
\hline
\hline
Species&\multicolumn{8}{|c|}{Column densities$^{b}$ in cm$^{-1}$}\\
\hline
N$_{\rm H\; I}^{c}$/10$^{20}$&$<$0.5&$<$0.5&0.9&2.8&4.2&5.0&5.0&5.4\\
\hline                                                                      
log N$_{\rm H\; I}$&$<$19.7&$<$19.7&20.0&20.4&20.6&20.7&20.7&20.7\\
\hline                                                                       
Mg II&14.2a[0.5]&14.9[1]&14.7a[1]&14.9[1]&14.9[1]&14.9[1]&15.0[0.9]&14.8[0.9] \\
Al
II&13.1a[0.2]&13.6a[0.5]&13.5a[0.5]&13.6b[0.5]&13.5a[0.5]&13.5a[0.5]&13.6a[0.5]&13.6a[0.5]
\\
Al III&13.0b&13.1a&13.1a&13.2a&13.1a&13.1a&13.3a&13.1a\\
Si II $\lambda$1808&15.1c&15.1b&15.0a&15.1b&15.0a&15.0a&15.0a&15.0a\\
Si II $\lambda$1526$^d$&14.0a&14.1b&14.1a&NA&14.1e&14.1a&14.2a&14.1a\\
Ca II&12.0c&$<$12.0&$<$12.2&12.4b&12.3b&12.6b&12.4c&NA\\
Sc II&$<$12&$<$11.9&$<$11.6&$<$11.7&$<$11.6&$<$11.8&$<$12.0 &$<$12.0\\
Ti II&$<$11.9&$<$11.9&11.9b&12.1b&12.1b&12.4c&$<$12.0&$<$11.9\\
V II& $<$12.4&$<$12.4&$<$12.1&$<$12.1&$<$12.1&$<$12.1&$<$12.4&$<$12.3\\
Cr II&$<$12.7&13.1c&12.9b&12.9b&12.8b&$<$12.6&13.1c&12.8c\\
Mn II&12.1c&12.5c&12.3a&12.5a&12.4a&12.4b&12.6b&12.4b\\
Fe II&$<$14.2&14.7b&14.7b&14.8b&14.7b&14.7b&14.7b&14.6b\\
Co II&$<$13.3&$<$13.3&$<$12.9&$<$13.1&$<$13.0&$<$13.1&$<$13.3&$<$13.0\\
Ni II&$<$13.4&$<$13.4&(13.4)&  $<$13.4&13.3c&13.4b&13.8c&13.4c\\
Zn II&$<$12.3&$>$12.6&12.4&$>$12.4&$>$12.4&$<$12.3&$>$12.4&$>$12.0\\
C I&$<$13&$<$13&12.6c&NA&$<$12.8&$<$12.7&$<$13&13.0c\\
Mg I&12.1-12.6&12.4-12.8&12.3-12.5&12.4-12.6&12.4-12.6&12.4-12.7&12.5-12.8&12.4b\\
Al I& $<$12.2&$<$12.2&$<$11.9&$<$11.9&$<$12.0&$<$12.1&$<$12.2&$<$12.3\\
Si I& $<$12.2&$<$12.2&$<$11.8&$<$12.0&$<$12.0&$<$12.0&$<$12.2&$<$12.0\\
Fe I& $<$11.8&$<$11.8&$<$11.5&$<$11.6&$<$11.5&$<$11.6&$<$11.9&$<$11.6\\
\hline
\end{tabular}

$^a$ $\Delta$ is the observer frame colour excess $\Delta(g-i)$\\
$^b$ Letters (lower case) indicate errors, dex, a=0-0.049, b=0.05-0.099,
c=0.1-0.149, etc. () indicates uncertainties or inconsistent raw
data.\\ 
$^{c}$ N$_{\rm H\; I}$ comes from using the SMC extinction curve to derive
$E(B-V)$ and the using the SMC ratio of N$_{\rm H\; I}$ to $E(B-V)$ from Welty et
al. 2005 (which is similar to previous values).\\
$^d$ We used the column densities from Si II $\lambda1808$. The values
from Si II $\lambda1526$ are to show that the saturation corrections for
Si II are consistent with the saturation corrections for Al II and Mg II
derived from Fe II (in brackets for the entries for these two species.)\\
\end{minipage}
\end{table*}

\newpage

\begin{table*}
\centering
\begin{minipage}{140mm}
\scriptsize
\begin{tabular}
{|c|c|c|c|c|c|c|c|c|}
\multicolumn{9}{l}{Table A4 (Continued): Column densities of weak lines }\\
\hline
Sample Number&26&1&16&8&17&19&25&27\\
\hline
Number of systems&97&809&369&251&85&58&111&48\\
\hline
Criteria&W$_{\rm Mg\;II}$&W$_{\rm Mg\;II}$&$W_{\rm Fe\;II}$/$W_{\rm
Mg\;II}$&W$_{\rm Mg\;II}$&W$_{\rm Al\; II}$/W$_{\rm Mg
\;I}$&FeII$\lambda$2260&$\Delta>$ 0.2&W$_{\rm Mg\;II}$\\
&$>$2.5&$>$0.3&$>$0.58&$>$2.0&$<$1.54 &present&&$>$2.5\\
&$\Delta<$0.2&&&&&&&$\Delta>$0.2\\
\hline
$E(B-V)$&0.012&0.013&0.014&0.032&0.034&0.034&0.081&0.085\\
\hline
\hline
Species&\multicolumn{8}{|c|}{Column densities in cm$^{-1}$}\\
\hline
N$_{\rm H\; I}$/10$^{20}$&5.4 &5.9&6.3&14.4&15.3&15.3&36.0&38.0\\
\hline
log N$_{\rm H\; I}$&20.7&20.8&20.8&21.2&21.2&21.2&21.6&21.6\\
\hline
Mg II&15.1[1.0]&14.7[0.9]&14.9[0.9]&15.0a[0.9]&15.1[1.0]&15.2[1.2]&14.9[0.9]&15.2[1.0]\\
Al II&14.0a[0.6]&13.5a[0.5]&13.7a[0.5]&13.9a[0.5]&13.9a[0.5]&14.1a[0.7]&13.8[0.5]&14.0a[0.5]\\
Al III&13.4a&13.1a&13.3a&13.4a&13.3a&13.3a&13.5a&13.6a\\
Si II$\lambda$1808&15.5a&15.1a&15.3a&15.4a&15.4b&15.6a&15.5a&15.6b\\	
Si II$\lambda$1526&14.5a&14.1a&14.4a&14.5a&14.4a&14.5a&14.6a&14.8a\\	
Ca II&$>$12.5 &12.1&12.6c&12.4c&NA&12.6b&12.6c&$>$12.6\\
Sc II&$<$12.2&$<$11.6&$<$11.8&$<$11.9&$<$12.4&$<$12.1&$<$12.1c&$<$12.3\\
Ti II&12.5c&11.9b&12.4b&12.4b&12.6e&12.7b&12.3c&$<$12.5\\
V II&$<$12.6&$<$12.1&$<$12.2&$<$12.4&$<$12.6&$<$12.6&$<$12.6&$<$12.8\\
Cr II&13.3b&13.0b&13.3b&13.3b&13.4c&13.7b&$>$13.6c&13.8b\\
Mn II&12.9b&12.4a&12.7a&12.9a&12.8a&$>$13.1&13.0a&$>$13.2\\
Fe II&15.1b&14.7a&15.0a&15.1b&15.1b&15.4a&$>$14.9&$>$15.3c\\
Co II&$<$13.3&$<$12.8&$<$13.0&$<$13.1&$<$13.2&$<$13.4&$<$13.4&$<$13.6\\
Ni II&(13.9c)&(13.4c)&13.9c&$<$(13.9c)&13.9b&(13.9c)&$>$14.1&($>$14.2)\\
Zn II&12.5&11.9& 12.4&12.6&12.5&13&$>$12.7&$>$12.4\\
C I&13.1e&12.7c&13.1b&13.1c&13.4b&$<13.2$ &13.6:&13.5d\\
Mg
I&12.7-13.1&12.3-12.7&12.5-12.8&12.6-13.1&12.5-12.7&12.7-13.1&12.6-12.9&12.9-13.8\\
Al I&$<$12.4&$<$11.9&$<$12.0&$<$12.2&$<$12.5&$<$12.4&$<$12.4&$<$12.6\\
Si I&$<$12.4&$<$11.8&$<$12.0&$<$12.1&$<$12.4&$<$12.5&$<$12.3&$<$12.6\\
Fe I&$<$12.0&$<$11.4&$<$11.7&$<$11.8&$<$12.0&$<$12.1&$<$12.0&$<$12.2\\
\hline
\end{tabular}

\end{minipage}
\end{table*}
\section{The special case of blended lines: Zn II, Cr II and Mg I}
The interpretation of the lines that are blends of lines of separate
species (mainly Mg I and Zn II near $\lambda$2026 and Zn II and Cr II
near $\lambda$2062) is complicated and we follow procedures similar to
those outlined by Khare et al. (2004) as follows. (a) W$_{\rm Cr\;
II\lambda2066}$ and W$_{\rm Cr \;II\lambda2056}$ were averaged, to
estimate W$_{\rm Cr\; II\lambda2062}$, as the oscillator strength of
$\lambda$2062 is very nearly equal to the average of the oscillator
strengths of the $\lambda$2056 and $\lambda$2066 transitions of Cr II.
(b) The measured equivalent width for $\lambda$2062 (last row, Table 2 as
well as Tables 5 and A4) was reduced by this estimate to get W$_{\rm Zn\;
II\lambda2062}$. This value was multiplied by 2, the ratio of $f$ values
for the two lines, to get W$_{\rm Zn\; II\lambda2026}$. The values are in
parentheses when the assumption that the Zn II lines are on the linear
portion of the curve of growth was used. (c) To estimate the
contamination of the Zn II $\lambda$2026 line by Mg I, we calculated the
value for W$_{\rm Mg\; I\lambda2026}$ from the linear curve of growth
assumption, using W$_{\rm Mg\; I\lambda2026}$=W$_{\rm Mg\;
I\lambda2852}$/42\footnote[1]{The errors propagated in these steps are as
follows: (a) The error in W$_{\rm Cr\;II\lambda2062}$ =
1/$\sqrt{2}\times~$the error in W$_{\rm Cr\;II\lambda2056}$; (b) The
error in W$_{\rm Zn\;II\lambda2062} = \sqrt{2}\times~$the error in
W$_{\rm Cr\;II\lambda2062}$; (c) The error in W$_{\rm Zn\;II\lambda2026}
= 2\times~$the error in W$_{\rm Cr\;II\lambda2056}$. The error in W$_{\rm
Mg\;I\lambda2026}$ should be very small, if the Mg I $\lambda2852$ line
is on the linear portion of the curve growth}. We compared this with the
residual left for Mg I after subtracting the Zn II $\lambda$2026
estimate. We found that the calculated contribution of Mg I is remarkably
close to the lower limit assuming the linear part of the curve of growth
applies for Zn II $\lambda$2026. This interesting result, discussed
later, is consistent with the failure to detect Mg I $\lambda$1827 which
would be detectable if Mg ${\rm I}$ $\lambda$2026 were highly saturated.
To be conservative (see discussion below about blends of saturated
lines), we actually enter the upper limit to Mg I $\lambda$2026 by
showing, on the left for Mg I $\lambda2026$ in Table 2, the maximum value
of W$_{\rm Mg\; I \lambda2026}$ assuming that W$_{\rm Zn\;
II\lambda2026}$=W$_{\rm Zn \;II\lambda2062}$, that is, that the zinc is
fully saturated. The entry to the right (in parenthesis) for Mg I is the
minimum value given the $\lambda$2852 equivalent width. Only for
sub-sample 27 is Mg I the dominant contribution to the strength of the
$\lambda$2026 line; this sub-sample is the only one with the equivalent
width of $\lambda$2852 $>$ 600 m{\AA}, in the table.

Any conclusions on abundances are strongly affected by our technique to
unscramble the Zn II/Cr II/Mg I equivalent widths. If the Cr II
$\lambda$2062 line is saturated on a component-by component basis, so
probably is the Zn II $\lambda$2062 line. At any rate,
component-by-component saturation of one species in a complicated blend
means there is no way to separately extract information on the second
species. Thus, in the last two sub-samples, the equivalent widths of Zn II
lines do not grow with $E(B-V)$, possibly because we have assigned most
of the equivalent width to Cr II using our procedure and Zn II is
thereby, underestimated. This problem is accentuated by the large
probable contribution of the Mg I $\lambda$2026 line, where neither 
W$_{\rm Zn\; II\lambda2026}$, as we predicted by using our simple rule,
nor the linear curve of growth prediction for W$_{\rm Mg\; I\lambda2026}$
comes close to explaining the 266 m{\AA} measured equivalent width of the
blend: Mg I must be saturated, consistent with the much larger value of
W$_{\rm Mg\; I\lambda2852}$ compared to the rest of our sub-samples: in
this case the Zn II/Mg I blend can not be separated either.
\section{Use of both weak and saturated lines in deriving column
densities}
It follows from Appendix B that both Zn II lines are near the linear
portion of the curve of growth in a number of cases and that we can
derive N$_{\rm Zn\;II}$ from the corrected value for W$_{\rm Zn\;
II\lambda2062}$. This assumption should be verified by getting high
resolution observations of the four lines involved here for a number of
sources in our sub-samples.

Turning to other weak lines, Cr II is consistent with no saturation in
the first four sub-samples, through sub-sample 8, with $E(B-V)$=0.032,
but strong saturation is evident in the sub-samples in the last two
columns ($E(B-V)\sim $0.08). Mn II is consistent with little saturation
for the first three sub-samples, but shows saturation in the sub-samples
in the last two columns. The weak lines of Fe II show similar effects.
Note that when the weakest lines of Fe II, the lines of Cr II, or the Mn II
triplet become saturated (sub-samples 25 and 27), the large errors in
equivalent width for the sub-samples allow values that seem to be
inverted.

Finally, a pattern related to saturation is discernible assuming the
first ions lie roughly on the same curve of growth in a given average
spectrum (Spitzer \& Jenkins 1976).  Inspection of Table 2 shows that
lines can be assumed to be unsaturated when their equivalent width is
$<$50 m{\AA}.  Lines of higher equivalent widths may or may not be
saturated.

We note the close tracking of Al II $\lambda$1670 and Al III
$\lambda$1854. This implies that the average ionization correction does
not vary much between the sub-samples as defined, though the two ions may
well be in different physical regions, proximate to each other but not
mixed. The fact that Al II $\lambda$1670 tracks Fe II $\lambda$2586 shows
that it may be possible to estimate Al II column densities by using the
Fe II curve of growth, which is well defined because of the large number
of Fe II lines. Interestingly, Al III shows modest saturation at all
apparent values of $E(B-V)$, consistent with its being in separate
regions that do not contribute much to reddening but do have appreciable
column densities. Thus, ratios of N$_{\rm Al\; III}$/N$_{\rm Al \;II}$
are upper limits to the ratios in the H I regions. Our samples indicate
N$_{\rm Al\; III}$/N$_{\rm Al\; II}$ $<0.5$ (see below), consistent  with
the findings of Howk \& Sembach (1999) and Vladilo et al. (2001). Our
large sample makes this result very general.

\end{document}